\documentclass[a4paper,prb,superscriptaddress,showpacs,twocolumn]{revtex4}
\usepackage[dvips]{graphicx}
\usepackage{amssymb}
\usepackage{natbib}

\begin{document}

\title{Probing multiband superconductivity by point-contact spectroscopy}

\author{D.Daghero \email{E-mail:dario.daghero@polito.it}}
\author{R.S. Gonnelli}
\affiliation{Dipartimento di Fisica and CNISM, Politecnico di
Torino, 10129 Torino, Italy}

\begin{abstract}
Point-contact spectroscopy was originally developed for the
determination of the electron-phonon spectral function in normal
metals. However, in the past 20 years it has become an important
tool in the investigation of superconductors. As a matter of fact,
point contacts between a normal metal and a superconductor can
provide information on the amplitude and symmetry of the energy gap
that, in the superconducting state, opens up at the Fermi level. In
this paper we review the experimental and theoretical aspects of
point-contact spectroscopy in superconductors, and we give an
experimental survey of the most recent applications of this
technique to anisotropic and multiband superconductors.
\end{abstract}

\maketitle

\section{Introduction: point-contact spectroscopy}\label{sect:intro}
Point-contact spectroscopy (PCS) was developed more than 35 years
ago as an experimental tool to investigate the interaction
mechanisms between electrons and phonons in metals. Yanson
\cite{yanson74} was the first to observe that small
microconstrictions between two metals show non-linearities in the
$I$-$V$ characteristic (and in the second derivative $d^2 V/dI^2$)
that are the hallmark of inelastic scattering of electrons by
phonons.  The point-contact technique was later used to study all
kinds of scattering of electrons by elementary excitation in metals,
like magnons and so on \cite{duif89,naidyuklibro}. When one of the
sides of a point contact is a superconductor, quantum phenomena such
as quasiparticle tunneling or Andreev reflection (see
Sect.\ref{sect:andreev}) occur at the interface, depending on the
height of the potential barrier between the two electrodes. As a
result, the $I-V$ shows -- in addition to the features related to
inelastic electron scattering -- much stronger non-linearities that
give rise to particular structures in the \emph{first} derivative
$dI/dV$ (that is, in the differential conductance) which contain
fundamental information on the excitation spectrum of the
quasiparticles, i.e. on the superconducting energy gap and its
properties in the direct and reciprocal space. For this reason, and
apparently in spite of its simplicity, point-contact spectroscopy
has become an important, sometimes unique, tool for the
investigation of superconducting materials. In some recent cases,
PCS has provided precious spectroscopic information on newly
discovered superconductors when more complex, technologically
demanding techniques such as scanning tunneling microscopy (STM) and
angle-resolved photoemission spectroscopy (ARPES) were still
hindered by the absence of single-crystal samples of sufficient
size. There is a number of excellent reviews that deal with the
theoretical and experimental aspects of point-contact spectroscopy
in normal metals and superconductors
\cite{jansen80,duif89,naidyuklibro}. An extensive and comprehensive
review was dedicated especially to point-contact results in cuprates
\cite{deutscher05}. The present review is therefore focused on the
most recent applications of point contact spectroscopy to the study
of multiband superconductors. A general theoretical introduction is
provided, whose aim is to explain in a simple, experimental-oriented
way, and with a consistent notation, theoretical models of
increasing complexity for the interpretation of point-contact data
in superconductors.

\section{Fabrication of point contacts}\label{sect:fabrication}
\begin{figure}
\begin{center}
\includegraphics[width=0.9\columnwidth]{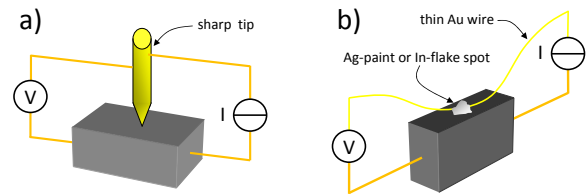}
\end{center}
\caption{(a) Experimental arrangement for point-contact measurements
with the typical needle-anvil technique.  A metallic tip (Au, Pt-Ir,
Pt, Ag) is gently pressed against the surface   of the sample. (b)
The ``soft'' point-contact technique where a tiny spot of Ag paste
(or a tiny flake of In) replaces the tip.}\label{fig:needle-anvil}
\end{figure}

A point contact is simply a contact between two metals, or a metal
and a superconductor, whose radius is smaller than the electron mean
free path, and this in most cases means that the contact is
nanometric. Historically, point contacts were fabricated in a number
of ways \cite{naidyuklibro}. The pioneering technique exploited by
Yanson \cite{yanson74} for PCS was based on the realization of
microshorts in the dielectric layer of a tunnel junction between two
metals. Another technique widely used especially in superconductors
(but that allows only the creation of homocontacts between two
electrodes of the same material) is the break-junction technique in
which a single sample is broken at low temperature into two pieces
that are then brought back in contact. More recently, point contacts
have been made by lithographical creation of a small hole in a thin
membrane on both sides of which a metal film is then deposited. But
the most used technique simply consists in bringing the two
electrodes in contact by using a micromechanical apparatus. In the
most common configuration, often called ``needle-anvil'', the sample
to be studied is one of the electrodes, and the other is a metallic
tip, electrochemically or mechanically sharpened, which is gently
pressed against the sample surface (figure \ref{fig:needle-anvil}a).
Typically, the tip has an ending diameter of some tens of
micrometers  and it is easily deformed during the contact
\cite{blonder83}. This means that, except in very special cases
\cite{srikanth92}, parallel contacts are very likely to form between
sample and tip \cite{baltz09}. In general this is not detrimental to
spectroscopy, unless the sample is highly inhomogeneous on a length
scale comparable with the tip end \cite{baltz09}. The needle-anvil
technique has several advantages: i) it is non-destructive and
several measurements can be carried out in the same samples; ii) the
resistance of the contact can be controlled to some extent by fine
tuning of the pressure applied by the tip. Its main drawbacks are
the poor thermal and mechanical stability of the junction and the
fact that, if the sample is very small (tens of micrometers, as it
can happen with single crystals), the whole procedure becomes
extremely difficult. For these reasons, since 2001 we adopted the
so-called ``soft'' point-contact technique, in which the contact is
made between the clean sample surface and a small drop (about 50
$\mu$m in diameter) of Ag paste or a small In flake. The Ag or In
counterelectrode is connected to current and voltage leads through a
thin Au wire (10 - 25 $\mu$m in diameter) stretched over the sample,
as depicted in Fig \ref{fig:needle-anvil}(b). Despite the large
``footprint'' of the counterelectrode (in particular in the case of
Ag paste) if compared to the electronic mean free path, these
contacts very often provide spectroscopic information. This clearly
means that, on a microscopic scale,  the real electrical contact
occurs only here and there through parallel nanometric channels
connecting the sample surface with the In flake or with individual
grains in the Ag paste, whose size is 2-10 $\mu$m. With respect to
the needle-anvil technique, the ``soft'' one does not involve any
pressure applied to the sample and this can be sometimes very
useful, as we will show in Sect. \ref{sect:FeAs}. The resistance of
the as-made contacts is usually already in the suitable range for
Andreev reflection to occur. If needed, it can be tuned by applying
short ($\approx50$ $ms$) voltage or current pulses until a
spectroscopic contact is achieved. This effect (sometimes called
``fritting'' \cite{holm58}) is well known in standard
electrotechnics. The pulses have the effect of destroying some of
the existing microjunctions and/or creating new ones by piercing a
small oxide layer on the surface of either electrode. The contacts
are mechanically and thermally very stable so that, for example, PCS
measurements can be performed even in a cryocooler. Moreover, they
can be made also on the thin side of small single crystals allowing
directional point-contact spectroscopy even in samples too small for
the needle-anvil technique. Often (but not always) the conductance
curves of ``soft'' point contacts are more broadened than those
obtained by the needle-anvil technique. As we will show, this is
probably related to inelastic scattering near the interface,
possibly by an oxide layer on the surface of Ag grains or of the
sample. As a matter of fact, the same holds for contacts made with
the Au wire alone, or even with a tip, whenever the pressure applied
by the tip on the sample is small.

\section{Point-contact spectroscopy (PCS) in the normal state}
\label{sect:PCSnormal} The uniqueness of point-contact spectroscopy
in the normal state is due to its ability to provide
\emph{spectroscopic}, energy-resolved information on the inelastic
scattering of quasiparticles with elementary excitations like
phonons, magnons and so on by using a very simple and cheap
experimental setup. To do so, however, some important experimental
requirements must be fulfilled. The relevant quantity is the Knudsen
ratio $K=\ell/\textit{a}$, where $\ell$ is the electron mean free
path ($1/\ell=1/\ell_e+1/\ell_i$ where $\ell_{e,i}$ are the elastic
and inelastic mean free paths) and $\textit{a}$ is the contact
radius. From now on it will be assumed that the shape of the contact
is a circular orifice with radius $\textit{a}$ in an otherwise
completely reflecting barrier. Unless otherwise specified, we will
specially refer to homocontacts (i.e. contacts between two
electrodes made of the same metal). Depending on the value of the
Knudsen ratio, different regimes of conduction are possible, as
described in the following.
\subsubsection{Ballistic regime}\label{sect:ballistic} 
\begin{figure}
%\vspace{1cm}
  \centering
  \includegraphics[width=\columnwidth]{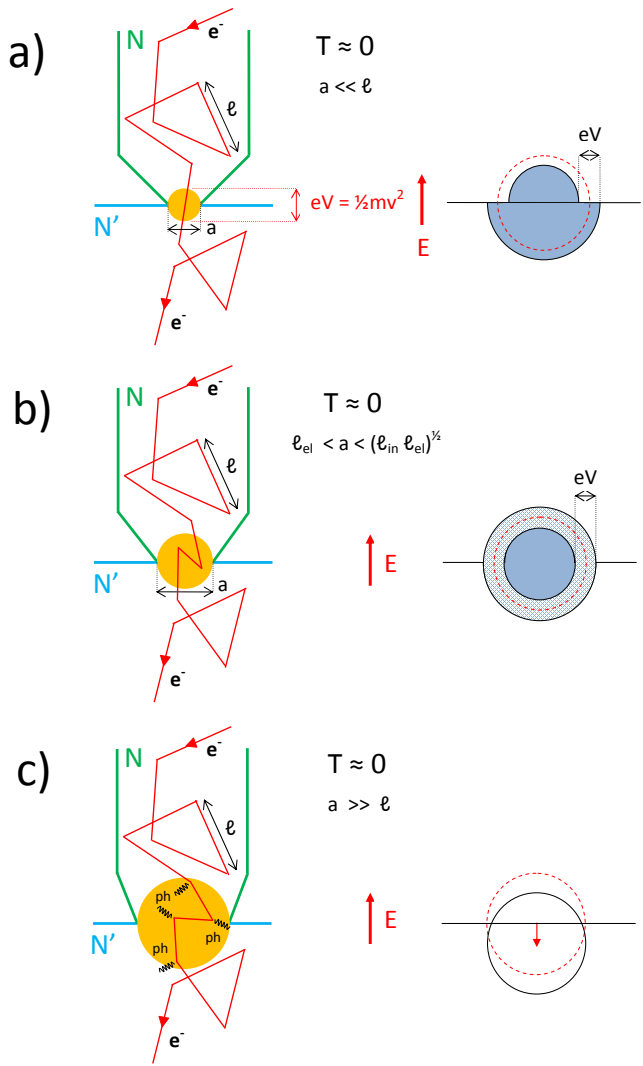}
  \caption{Schematic diagram of the contributions to the current in a
  point contact obtained by solving the Boltzmann equation. (a):
  zeroth-order Sharvin current (no scattering).
  (b): Diffusive regime (only elastic scattering in the contact area). (c): Thermal
  regime with inelastic scattering in the contact region.
  The right sides of the figures
  show electron distribution functions at the center of the
  contact for the three main regimes. (a): ballistic regime. The FS
  is formed by two half-spheres with different radius, i.e. defined by
  the surfaces at energy $E$ and $E+eV$. (b): diffusive regime.
  The elastic scattering redistributes the electrons over
  the sphere but only in an energy shell with a width given by
  $eV$. (c): thermal regime. The inelastic scattering
  reduces the shift in energy space as it is usual for normal transport
  in a conductor.}\label{fig:collisions}
\end{figure}

In the
ballistic regime the electron mean free path $\ell$ is much larger
than the contact radius $\textit{a}$ ($\ell\gg\textit{a}$ or $K \gg
1$). The applied voltage $V$ accelerates electrons within the
distance of a mean free path. The electrons will then flow through
the contact ballistically (with \emph{no} scattering) gaining a
kinetic energy equal to $eV$ (see Fig. \ref{fig:collisions} (a)). In
this way, the energy of the injected electrons is perfectly known
and corresponds to the voltage applied to the junction. The
resistance of the contact in such a situation was calculated by
Sharvin \cite{sharvin65} and is equal to
\begin{equation}\label{eq:Sharv_Res}
R_{S}=\frac{4\rho\ell}{3\pi a^2}
\end{equation}
where $\rho$ is the resistivity of the material under study. Since
in metals $\rho \varpropto \ell^{-1}$, $R_{S}$ is independent of the
electron mean free path, and depends only on the contact geometry.
As a matter of fact, it can also be written as
\begin{equation}\label{eq:Sharv_Res2}
R_{S}=\frac{2h}{e^2 k_F^2 a^2}
\end{equation}
being $k_F$ the Fermi momentum \cite{deJong94}. In the $k$ space,
the (supposed spherical) Fermi surface (FS) expands for forward
electrons by a quantity $eV$ (see Fig.\ref{fig:collisions}(a)).
Inelastic scattering events taking place in the bottom electrode
give rise to a measurable (negative) corrections to the current only
if they cause the backflow of carriers through the orifice. The
backscattered electron must jump back onto the shrunk FS and this is
possible only if it can lose an energy $eV$ in the scattering
process. This explains why in the ballistic regime the applied
voltage sets the energy scale of the spectroscopic investigation.

The first-order correction to the current due to the backscattered
electrons is \cite{duif89,naidyuklibro}:
\begin{equation}\label{eq:I_1}
\delta I=-\frac{2\pi  e}{\hbar}\Omega_{eff}N(0)\int_{0}^{eV}d
E\int_{0}^{E}dE^{\prime}S(E-E^{\prime})
\end{equation}
where $\Omega_{eff}= 8a^3/3$ is the effective volume in which the
inelastic scattering of electrons contributing to $\delta I$ occurs,
$N(0)$ is the density of states at the Fermi level and
\begin{equation}\label{eq:S(e)}
 S(E)=\frac{N(0)}{32\pi^{2}}\int\frac{d^{2}\textbf{k}}{k^{2}}\int\frac{d^{2}\textbf{k}^{\prime}}{k^{\prime
 2}}|g_{\textbf{k}\textbf{k}^{\prime}}|^{2}K(\textbf{k},\textbf{k}^{\prime})\delta(E-E_{\textbf{k}}+E_{\textbf{k}^{\prime}})
\end{equation}
is the spectral function for the relevant interaction, which results
from an integration over all the initial and final electron states
of the scattering matrix elements
$|g_{\textbf{k}\textbf{k}^{\prime}}|$ weighted by an efficiency
function $K(\textbf{k},\textbf{k}^{\prime})$ which accounts for the
direction of the incoming and the inelastically scattered electron.
It can be shown \cite{duif89,naidyuklibro} that
\begin{equation}\label{eq:2nd_deriv}
  \frac{d^{2}I}{dV^{2}}=-\frac{2\pi
  e^{3}}{\hbar}\Omega_{eff}N(0)S(eV)
\end{equation}
A direct determination of the spectral function by means of PCS
$I-V$ measurements is thus possible. If the elementary excitations
are phonons, $S(eV)$ is the so called ``point-contact
electron-phonon spectral function'' $\alpha^2_{PC}F(eV)$ which
differs only slightly (due to the efficiency function
$K(\textbf{k},\textbf{k}^{\prime})$) from the thermodynamic
Eliashberg function $\alpha^{2}F(eV)$. In this case, using the
formulas of the free electron model, one obtains:
\begin{equation}\label{eq:2nd_deriv2}
\frac{d^{2}I}{dV^{2}}=-\frac{16 e a}{3 \hbar v_F}\alpha^2_{PC}F(eV)
\end{equation}
It is worth mentioning that, according to eq.\ref{eq:2nd_deriv2},
one expects the experimental $-\frac{d^{2}I}{dV^{2}}$ to rapidly
fall to zero above the Debye energy. Very often this is not the case
\cite{jansen80,duif89,naidyuklibro} and a considerable background is
found, which has been attributed to the presence of non-equilibrium
phonons. It is however possible to correct for the background and to
determine the $\alpha^2_{PC}F(eV)$ function \cite{duif89}. This
method has allowed extracting the electron-phonon spectral function
in many normal metals \cite{naidyuklibro}, but can be applied also
to superconductors above the critical temperature or driven normal
by means of a magnetic field. Some examples will be discussed in
Sect. \ref{sect:MgB2_phonons} for the case of MgB$_2$ and in Sect.
\ref{sect:borocarbides} for the case of borocarbides.
\subsubsection{Thermal regime}\label{sect:thermal}
As opposed to the ballistic regime is the thermal (or Maxwell) one
in which $\ell\ll a$ [see figure \ref{fig:collisions} (c)]. Some
authors\cite{naidyuklibro} prefer to identify this regime by the
condition $\ell_i \ll a$ to make it explicit that electrons can
undergo inelastic scattering in the contact region as they normally
do in the bulk. In this case, the resistance of the junction
(already calculated by Maxwell) depends on the resistivity of the
metal \cite{duif89}:
\begin{equation}\label{eq:R_Maxwell}
  R_{M}=\frac{\rho}{2\textit{a}}.
\end{equation}
Joule heating occurs in the contact region and causes a local
increase in temperature. The maximum temperature $T_{max}$ at the
center of the contact can be estimated by using the following
expression \cite{duif89}
\begin{equation}\label{eq:T_max_thermal}
T_{max}^{2}=T_{bath}^{2}+V^{2}/4L
\end{equation}
where $T_{bath}$ is the bath temperature and L is the Lorenz number.
In this case, at any finite bias the contact resistance is related
to the resistivity of the material at $T_{max}>T_{bath}$. Since in
metals $\rho$ increases with temperature, the $I-V$ curves become
S-shaped and the conductance decreases with bias \cite{baltz09}. Any
spectroscopic information on the electron inelastic scattering is
lost. Since the standard transport theory for bulk materials applies
also to the contact, the FS is only slightly shifted in the
direction of the electric field, as in Fig. \ref{fig:collisions}(c).
\subsubsection{Intermediate regime}\label{sect:diffusive}
Between the two aforementioned extreme regimes, the resistance of
the contact can be expressed by a simple interpolation formula
derived by Wexler \cite{wexler66}:
\begin{eqnarray}\label{eq:R_diff}
R&=&\frac{4\rho\ell}{3\pi a^2}+
\Gamma(K)\frac{\rho}{2a}\\
&=&\frac{2h}{e^2 k_F^2 a^2}+ \Gamma(K)\frac{\rho}{2a}.\nonumber
\end{eqnarray}
Here the first term is the Sharvin resistance and the second is the
Maxwell resistance, multiplied by a function of the Knudsen ratio K.
$\Gamma$ is always of the order of unity. If the two metals are
different (i.e. for a heterocontact), the resistance of the contact
can be written as \cite{baranger85,deutscher02}
\begin{equation}\label{eq:R_diff_hetero}
R=\frac{2h}{e^2 a^2 k_{F,min}^2 T}+
\Gamma(K)\frac{\rho_1+\rho_2}{4a}
\end{equation}
assuming a spherical Fermi surface for both metals 1 and 2. Here
$k_{F,min}=\min[k_{F,1},k_{F,2}]$ and \footnote{If the effective
electron mass can be supposed to be the same for metals 1 and 2, the
Fermi velocities in equation \ref{eq:T} can be replaced by the Fermi
wavevectors.}
\begin{equation}
T=\frac{4v_{F,1}v_{F,2}}{(v_{F,1}+v_{F,2})^2}.\label{eq:T}
\end{equation}
In both eqs. \ref{eq:R_diff} and \ref{eq:R_diff_hetero} the
prevalence of the Sharvin or Maxwell term depends only on the size
of the contact. For a junction between given materials, the Maxwell
contribution dominates in large contacts, while the Sharvin one
becomes more and more important on decreasing $a$.

Between the thermal and ballistic regime one can also define the
so-called diffusive regime in which the elastic mean free path
$\ell_{e}$ of the electrons is small compared with the contact
radius $\textit{a}$ but the diffusion length
$\Lambda=\sqrt{\ell_{i}\ell_{e}}$ for inelastic scattering is still
bigger than $\textit{a}$ ($\textit{a}\ll\Lambda$). The
quasiparticles can now experience \emph{elastic} scattering
processes inside the contact region but not inelastic ones, as shown
in the left panel of figure \ref{fig:collisions}(b). The elastic
scattering redistributes the quasiparticles isotropically over the
FS, in an energy shell of width $eV$ (right panel of Fig.
\ref{fig:collisions}(b)). Though energy-resolved information is
still available, the effective volume $\Omega_{eff}$ in which the
inelastic scattering of electrons gives rise to the backflow current
is now reduced by a factor of the order of $\textit{a}/\ell$ with
respect to the ballistic regime (see eq. \ref{eq:I_1}). This is due
to the fact that the probability for an electron to cross the
contact, undergo an inelastic scattering event and then flow back
through the orifice is reduced by elastic scattering in the contact
region. The intensity of the spectroscopic signal (proportional to
$-d^2I/dV^2$) is thus strongly reduced. Moreover, a different
efficiency function must be used in the spectral function $S(E)$
(see eq. \ref{eq:S(e)}), since the elastic scattering relaxes the
requirement of momentum conservation.

\subsection{Determination of the conduction regime of a real point contact}
The radius of a real point contact (for example made by pressing a
metallic tip against the sample surface) is unknown and, in general,
experimentally inaccessible. As a matter of fact, the size of the
actual contact is not related to the apparent contact area or to the
footprint of the tip \cite{baltz09}. So the problem arises of how to
check whether the contact is ballistic or not. One possibility is to
admit that the resistance of the contact $R_N$ \emph{can} be written
as in the Sharvin formula, i.e. $R_N=(4\rho \ell)/(3 \pi a^2)$ where
the product $\rho \ell$ refers to the bank with the smaller Fermi
energy (see eq.\ref{eq:R_diff_hetero}) and thus, generally, to the
superconductor. The condition $a \ll \ell$ can then be turned in a
condition on the contact resistance:
\begin{equation}\label{eq:test_resistance}
R_{N} \gg \frac{4\rho }{3\pi \ell}
\end{equation}
Alternatively, one can (very crudely) evaluate the contact radius
$a$ by means of
\begin{equation}\label{eq:radius}
a=\sqrt{\frac{4\rho \ell}{3 \pi R_N}}
\end{equation}
and then compare it to $\ell$. This estimation is based on the
assumption that only one contact is present. In almost all real
cases, because of the rather likely formation of parallel contacts,
the value of $a$ obtained in this way is nothing but an upper limit
to the size of the contacts (whose number is unknown). As a matter
of fact, in this case $R_N$ is the resistance of the parallel as a
whole and the resistance of individual contacts is necessarily
larger than that. This means that, if $a$ estimated from eq.
\ref{eq:radius} is smaller than $\ell$, the contact (either single
or multiple) is necessarily ballistic. If instead $a \geq \ell$,
this does not necessarily mean that the contact is not ballistic. In
these cases, the conductance curves ($dI/dV$ vs. $V$) can help
understanding what is the regime of conduction. If the conductance
shows a downward curvature, for example, heating may occur in the
contact. If the conductance shifts on increasing temperature, this
may mean that a Maxwell term (proportional to the resistivity) is
playing a role.

In the case of point contacts on superconductors, as we will see
later on, some specific features show up in the conductance curves
when the contact is not ballistic (see sect.\ref{sect:dips}).
Moreover, a critical temperature of the junction smaller than the
bulk $T_c$ can be due to a surface degraded layer but also, more
banally, to Joule heating in the contact (so that the actual
temperature of the contact $T_{max}$ is higher than that of the
bath).

\section{Point-contact Andreev-reflection spectroscopy (PCAR) in the superconducting
state}\label{PCARsuperconducting}

\subsection{Andreev reflection}\label{sect:andreev}
\begin{figure}
  \centering
  \includegraphics[width=0.7\columnwidth]{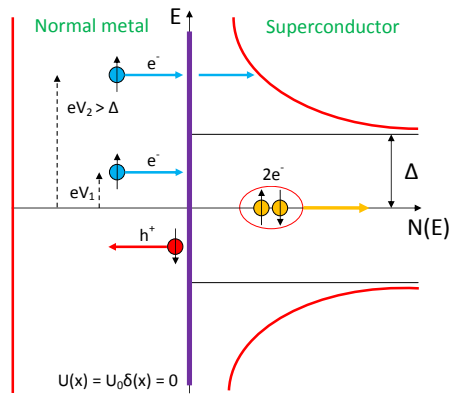}
  \caption{Electrical transport at an ideal (barrierless) $N/S$
  interface at $T=0$. Incoming electrons with $eV<\Delta$
  are reflected as holes and, for each electron, a Cooper pair
  is transmitted (Andreev reflection). Electrons approaching
  the interface with $eV>\Delta$ are normally transmitted as
  electron-like quasiparticles.}\label{fig:Andreev}
\end{figure}

Let us consider a normal metal (N) brought in direct contact with a
superconductor (S), with no potential barrier between them. Let's
apply to this junction a voltage $V<\Delta/e$ being $\Delta$ the
energy gap in the S side. If the contact is ballistic, the whole
voltage drop occurs at the interface. An electron coming from the N
side will not be able to propagate through the interface because
only Cooper pairs exist in this energy range in S. But if a hole is
reflected and two electrons are transmitted in S as a Cooper pair
(Fig. \ref{fig:Andreev}) the total charge and momentum are
conserved. This phenomenon is called Andreev reflection
\cite{Andreev64} and can be theoretically described by solving the
Bogoliubov-de Gennes equations \cite{degennes66} at a N/S interface.
The reflected hole has opposite wave vector and (if the Cooper pairs
are singlets, as in all the cases analyzed here) opposite spin with
respect to the incoming electron, so it traces back the trajectory
of the incoming electron until a scattering event occurs.  %%%%%%%%% controllare

If the applied voltage is much greater than the gap ($eV\gg\Delta$),
all the electrons whose energy is lower that the gap still undergo
Andreev reflection, but now their contribution to the current is
constant and does no longer depend on the applied voltage. Instead,
the electrons with energy higher than the gap are transmitted
through the interface (see fig. \ref{fig:Andreev}) giving a
voltage-dependent current. The total current for $eV\gg\Delta$ is
thus \cite{blonder83}
\begin{equation}\label{eq:excess_current}
I\propto
ev_{F}(eV-\Delta)+2ev_{F}\Delta\approx\frac{V}{R_{S}}+\frac{\Delta}{eR_{S}}.
\end{equation}
The second term of the right-hand side of eq.
\ref{eq:excess_current} is called ``excess current'' and is the
hallmark of the superconducting state even at energies much higher
than the gap. This result is exact only if the gap rises from zero
up to the bulk value over a distance larger than the superconducting
coherence length $\xi$. If the gap is instead modeled as a sharp
barrier at the interface an additional term equal to
$\Delta/3eR_{S}$ must be included.

Because of Andreev reflection, the conductance of the junction turns
out to be doubled for $V<\Delta/e$. This clearly suggests a simple
way to determine the energy gap in the S side by point contact
spectroscopy. This technique is often referred to as point-contact
Andreev-reflection spectroscopy (PCAR).
From the solution of the Bogoliubov-de Gennes equations near a $N/S$
interface \cite{saint-james64} it is possible to note that Andreev
reflection does not occur abruptly at the interface but over a
length scale of the order of $\xi$. In general $\xi$ is also the
length over which $\Delta$ is depressed due to the proximity effect
generated by N on S. However, if the contact size is smaller than
$\xi$ this effect can be neglected.

\begin{figure}
  \centering
  \includegraphics[width=\columnwidth]{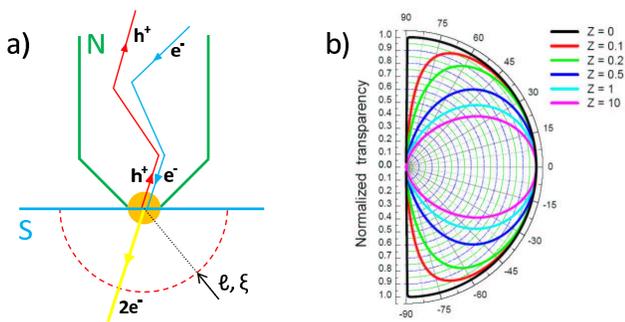}
  \caption{(a) Sketch of a ballistic Andreev-reflection point
  contact whose radius $\textit{a}$ is much smaller that the
  electron mean free path $\ell$ and the coherence length $\xi$;
  (b) A polar plot of the normalized transparency of an
  $NS$ junction as function of the angle of injection of the
  current for different values of the Z parameter.}\label{fig:PCS_TIP}
\end{figure}

As already mentioned, PCAR requires that the gain in energy of the
electrons crossing the junction is well defined. This is true in the
ballistic regime but, also, in the diffusive one. If one wants to
measure the gap by PCAR, it is clear that the voltage across the
junction will reach values of the order of, and even greater than,
the gap $\Delta$. If the contact is ballistic, using the value for
the carrier density in the free-electron model, it is possible to
show \cite{deutscher05} that the velocity of electrons across the
contact, at $V \simeq \Delta/e$, is on the order of the depairing
velocity in the superconductor. In other words, the current density
becomes overcritical in the contact. Just outside the contact the
current spreads out, its density decreases and will reach the
critical value a short distance away from the actual junction
\cite{waldram96,daghero06c}, as shown in figure \ref{fig:PCS_TIP}
(a). If the size of the overcritical region is smaller than the
coherence length $\xi$ the spectroscopy is still possible
\cite{deutscher05}, because superconductivity cannot be quenched
over distances smaller than $\xi$. Therefore it is necessary to
adopt contacts (see figure \ref{fig:PCS_TIP} (a)) which are smaller
than the electron mean free path (to avoid heating effects) and
smaller than the coherence length (to avoid proximity effect and
destruction of superconductivity in the contact region)
\cite{deutscher05}.

\subsection{The Blonder, Thinkam and Klapwijk (BTK)
model}\label{sect:BTK}

Even if Andreev reflection was discovered in the early 60s, it was
only in 1982 that Blonder, Thinkam and Klapwijk \cite{BTK} (from now
on referred to as BTK) gave a complete, even though simplified,
theoretical discussion of the phenomenon, including the effect of a
finite transparency of the interface. The most noticeable
simplification is that the model is 1D, i.e. all the involved
momenta are normal to the interface and parallel to the $x$ axis.
The barrier is represented by a repulsive potential $U_0\delta(x)$
located at the interface, which enters in the calculations through
the dimensionless parameter
\begin{equation}\label{eq:Z}
Z=\frac{U_0}{\hbar v_{F}}
\end{equation}

Of course, the smaller is $Z$, the more transparent is the barrier.
The parameter $Z$ is originally meant to represent the effect of the
typical oxide layer in a point contact, the localized disorder in
the neck of a short microbridge or the intentional oxide barrier in
a tunnel junction. According to the BTK model, calculated at $T=0$,
the electron coming from the N side can undergo four processes whose
probabilities are:
\begin{description}
  \item[\emph{A}] $\Rightarrow$ probability of Andreev reflection. The probability
  decreases with increasing $Z$ for $eV<\Delta$ and is always small for
  $eV>\Delta$;
  \item[\emph{B}] $\Rightarrow$ probability of normal specular reflection.
  This probability increases with $Z$, i.e. on decreasing the barrier
  transparency;
  \item[\emph{C}] $\Rightarrow$ probability of transmission in S as electronlike quasiparticle (ELQ).
  The probability decreases if $Z$ increases but it is always zero for $eV<\Delta$;
  \item[\emph{D}] $\Rightarrow$ probability of transmission with FS
  crossing (i.e. as holelike quasiparticle, HLQ). The probability is small for $eV>\Delta$
  and always zero for $eV<\Delta$.
\end{description}
Of course the sum of the four probabilities must be equal to 1.
Fig.\ref{fig:Andreev} shows the particular case of a barrierless
($Z=0$) N/S junction at $T=0$, where only the terms A and C are
present.
\begin{figure}
\begin{center}
\includegraphics[width=\columnwidth]{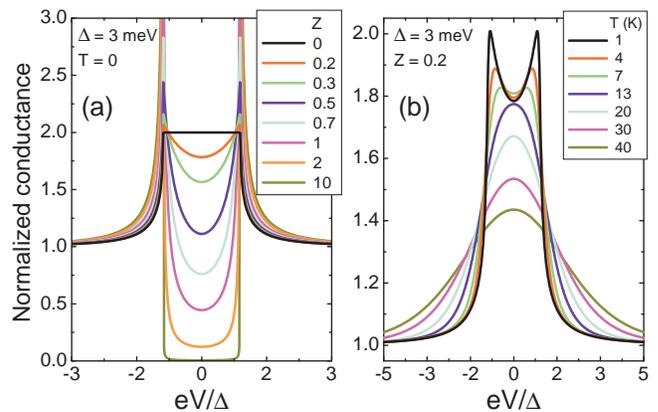}
\end{center}
\caption{(a) Normalized conductance curves $\sigma(E)/\tau_N$ of a
N/S interface at $T=0$ calculated within the BTK model
(eq.\ref{eq:sigma}) as a function of the barrier parameter $Z$, from
pure Andreev ($Z=0$) to pure tunneling ($Z=10$) regimes. (b) Effect
of the thermal smearing on the normalized conductance (note that we
used $\Delta=3$ meV and $Z=0.2$ in \emph{all} the
curves).}\label{fig:BTK_Z=0}
\end{figure}

In can be shown that the expression of the total current across the
junction, at T=0, is given by \cite{blonder83}
\begin{equation}\label{I_BTK}
I_{NS}=I_{0}\int_{-\infty}^{\infty}[f(E-eV)-f(E)][1+A(E)-B(E)]dE
\end{equation}
where $f(E)$ is the Fermi distribution function, $A(E)$ and $B(E)$
are the coefficients giving the probability of Andreev and ordinary
reflection, and the quantity $[1+A(E)-B(E)]$ (which is the
transmission probability) is often indicated by $\sigma(E)$. Note
that, although $\sigma(E)$ is formally written only as a function of
$A$ and $B$, the contribution of $C$ an $D$ has been taken into
account in the calculations. $I_0$ is a constant which depends on
the area of the junction, on the density of states and on the Fermi
velocity. The derivative of the current with respect to the bias,
$dI_{NS}/dV$, provides the conductance of the junction. When divided
by the conductance of the same junction \emph{when the
superconductor is in the normal state}, $dI_{NN}/dV$, this gives the
\emph{normalized} conductance of the junction, $G$ (which is the
outcome of PCAR experiments).

Here, instead of giving the explicit expressions for the
probabilities $A$, $B$, $C$ and $D$ that can be found easily in
literature \cite{BTK,deutscher05,naidyuklibro}, we prefer to show in
detail the results of a different approach \cite{kashiwaya96} that
allows writing the AR normalized conductance at $T=0$,
$G=(dI/dV)_{NS}/(dI/dV)_{NN}$ as a function of the quantities
$N_q(E)=E/\sqrt{E^2-\Delta^2}$ and
$N_p(E)=\Delta/\sqrt{E^2-\Delta^2}$ whose real parts are the BCS
quasiparticle and pair density of states, respectively.

We can start from the definition of the transparency $\tau_N$ of the
barrier in the BTK approximation of current injection totally
perpendicular to the N/S interface:
\begin{equation}\label{eq:tau}
\tau_N=\frac{1}{1+Z^2}
\end{equation}
and then we introduce the function:
\begin{equation}\label{eq:gamma}
\gamma(E)=\sqrt{\frac{E-\sqrt{E^{2}-\Delta^{2}}}{E+\sqrt{E^{2}-\Delta^{2}}}}=\frac{E-\sqrt{E^{2}-\Delta^{2}}}{\Delta}.
\end{equation}

It is trivial to show that:
\begin{equation}\label{eq:gamma1}
\gamma(E)=\frac{N_{q}(E)-1}{N_{p}(E)}.
\end{equation}

Note that $\gamma(E)$ is a complex function even if the gap $\Delta$
is real, as in the BCS case, since $N_p(E)$ and $N_q(E)$ become
imaginary for $E<\Delta$. By using these definitions it is possible
to demonstrate that the BTK conductance at $T=0$ is given by:

\begin{equation}\label{eq:sigma}
\sigma(E)=\tau_N\cdot\frac{1+\tau_N|\gamma(E)|^{2}+(\tau_N-1)|\gamma(E)^{2}|^{2}}{|1+(\tau_N-1)\gamma(E)^{2}|^{2}}.
\end{equation}

The calculated normalized conductance $G(E)=\sigma(E)/\tau_N$ is
shown in Fig. \ref{fig:BTK_Z=0}(a) for various values of $Z$ and for
$\Delta=$ 3 meV. In a perfectly transparent junction ($Z=0$, pure
Andreev regime) the conductance within the gap ($|eV|\leq\Delta$) is
doubled with respect to the normal-state one. When $Z>0$, two peaks
appear at $|eV|\approx\Delta$ and their amplitude increases on
increasing $Z$ while the zero-bias conductance (ZBC) is depressed.
Finally, at $Z\gtrsim10$, the normalized conductance at $T=0$
coincides with the BCS quasiparticle density of states, i.e. the
real part of $N_q(E)$. Indeed, it can be demonstrated that the
results of the BTK model for $Z\rightarrow\infty$ coincide with the
standard results of the theory for NIS (I=insulator) tunnel
junctions. Hence, the BTK model can reproduce, by simply changing a
parameter, all the different experimental situations corresponding
to different transparencies at the N/S interface, from zero to
infinity.

Equation \ref{eq:sigma} is particularly useful to discuss the
extensions of the simple BTK formalism we will present in the
following sections. As a matter of fact it should be borne in mind
that, even if widely used as a simple tool for fitting the
experimental PCAR spectra, the original BTK model is based on a
large number of approximations and simplifications, i.e.:
\begin{itemize}
\item[(1)] All the calculations are made at $T=0$;
\item[(2)] The problem is 1D, i.e. the current injection
is only perpendicular to the plane interface;
\item[(3)] The barrier is ideal and presents a null thickness;
\item[(4)] The Fermi surfaces of both materials in N
and S sides are spherical;
\item[(5)] The Fermi velocities are the same in both sides;
\item[(6)] The superconductor is supposed homogeneous and
isotropic. Because of the mono-dimensionality, the gap $\Delta$
entering the equations is actually the gap in one single direction
and represents ``the'' gap only if the order parameter is isotropic
(i.e. it has a $s-$wave symmetry).
\item[(7)] The N/S interface is atomically flat (somehow implicit
in the 1D current injection).
\end{itemize}
In the following we will show that most of these restrictions can be
easily relaxed giving a more realistic tool for the analysis of PCAR
experiments in a variety of unconventional superconductors.

\section{Beyond the BTK model} \label{sect:beyondBTK}
\subsubsection{Finite temperature}\label{sect:finiteT} The
calculation of the differential conductance of a N/S junction at
finite temperature is a quite easy task. It can be simply
accomplished by introducing in the equation for the current the
standard convolution with the Fermi function at finite $T$,
$f(E,T)$, and then taking the derivative of the current with respect
to the bias voltage, i.e.:
\begin{equation}\label{eq:dIns}
\frac{dI_{NS}}{dV}(V)=I_{0}\frac{d}{dV}\intop_{-\infty}^{+\infty}\left[f(E-eV,T)-f(E,T)\right]\sigma(E)dE
\end{equation}
where $\sigma(E)$ is given by eq. \ref{eq:sigma}. In figure
\ref{fig:BTK_Z=0}(b) the effect of the thermal broadening on the
normalized conductance is calculated by using a
temperature-independent gap $\Delta=3$ meV and $Z=0.2$. At the
increase of $T$ the two peaks typical of the AR at $Z\neq 0$ are
smeared out finally leaving a single zero-bias maximum at $V>10$
meV.
If the (supposed BCS) temperature dependence of the gap $\Delta(T)$
is taken into account in the expressions of $N_q(E)$ and $N_c(E)$,
i.e. in the $\sigma(E)$, the curves become as shown in
Fig.\ref{fig:BTK_3D} (b).  The AR features now correctly disappear
at the critical temperature of the contact (usually equal or very
close to the $T_c$ of the superconductor).

The pre-factor $I_0$ of eq.\ref{eq:dIns} is expressed in terms of
the normal density of states of the two materials and thus could, at
least in principle, depend on temperature and on energy: in this
case it should be brought inside the integral, and would no longer
simplify when normalizing. This could be the case when the normal
state conductance is found experimentally to change with temperature
or to be voltage-dependent, as it is in cuprates \cite{deutscher05}
and in the recently discovered Fe-based superconductors (see
sect.\ref{sect:FeAs}). However, one usually assumes for simplicity
that $I_0$ is constant and uses the expression for the normalized
conductance to fit the experimental PCAR spectra. From the
experimental point of view, however, these cases present the extra
problem of defining \emph{what} is the normal-state conductance to
be used for the normalization, as we will show in
sect.\ref{sect:FeAs}.

\subsubsection{2D or 3D BTK model}\label{sect:2DBTK}
If the current injection was really only perpendicular to the
interface as the BTK model assumes, one could in principle probe the
$\textbf{k}$ dependence of the gap by making directional PCAR
(DPCAR) measurements on the different crystallographic planes of
high-quality superconducting single crystals. Actually, charge
carriers can approach the interface from any direction and the only
condition set by the AR theory is that the component of the
$\textbf{k}$ vector parallel to the interface is conserved in all
processes. This implies, for example, that the reflected hole comes
back in $N$ with $\textbf{k}$ opposite to that of the incident
electron and traces back its trajectory until the first scattering
event in N occurs (see Fig. \ref{fig:PCS_TIP}(a)). In the S side a
Cooper pair propagates essentially in the same direction as the
incident electron (neglecting the small refraction due to the
expansion of the FS). Calling $\theta_N$ the angle between the
direction of the incident electron and the normal to the interface,
the conservation of transverse momenta leads to the following
dependence of the transparency $\tau_N$ on $\theta_N$:
\begin{equation}\label{eq:tau1}
\tau_N(\theta_N)=\frac{cos(\theta_N)^2}{cos(\theta_N)^2+Z^2}.
\end{equation}
Of course eq.\ref{eq:tau1} coincides with eq.\ref{eq:tau} for
$\theta_N=0$. In figure \ref{fig:PCS_TIP}(b) the angular dependence
of the normalized transparency (i.e. $\tau_N(\theta_N)/\tau_N(0)$)
is shown for different values of $Z$. When $Z=0$ all the
quasiparticles are transmitted with the same unitary probability in
the whole half-space $-\pi/2 \leq \theta_N \leq \pi/2$, but at the
increase of $Z$ the transmission becomes progressively weaker and
more directional around the perpendicular to the interface. Strictly
speaking, the injection is always in the whole half-space but one
can decide to conventionally fix a threshold (e.g. 75 \% of the
maximum transparency) to determine an equivalent injection angle
$\theta^*$. In the limit $Z\geq10$ (tunnel regime) one gets
$\theta^* \approx\pm30^\circ$, i.e. the tunneling process is
certainly highly directional. For the typical $Z$ values observed in
real PCAR experiments ($\sim 0.2-0.5$), $\theta^*$  ranges between
$\pm 70^\circ$ and $\pm 52^\circ$ thus evidencing the reduced
directionality of the PCAR technique. In addition to these
``theoretical'' limitations, some practical problems have to be
taken into account. Irrespective of the way the PC are realized
(needle-anvil or ``soft'' technique) the contact footprint has a
relatively large area (some hundreds of square microns). If this
area contains crystal-growth terraces, defects, pits or cracks the
probability to have some contacts along a different crystallographic
direction becomes high. Directional PCAR (DPCAR) spectroscopy can
give reliable results only if very high-quality single crystals with
highly regular (and large) surfaces parallel to the crystallographic
planes are used. Despite these limitations, we will show in the
experimental survey (sect. \ref{sect:Experiments}) that recent DPCAR
experiments were able to precisely determine the anisotropic
properties of the gap in several unconventional superconductors.

As shown in eq. \ref{eq:tau1} the barrier transparency depends on
the direction of the incoming electron in the N side. By introducing
this expression in eq.\ref{eq:sigma}, integrating over the whole
half-plane and properly normalizing, we get the normalized
conductance at $T=0$ \cite{kashiwaya96}:
\begin{equation}\label{eq:sigma2D}
G_{2D}(E)=\frac{\intop_{-\frac{\pi}{2}}^{+\frac{\pi}{2}}\sigma(E,\theta_N)\cos\theta_N
d\theta_N}{\intop_{-\frac{\pi}{2}}^{+\frac{\pi}{2}}\tau_N(\theta_N)\cos\theta_N
d\theta_N}
\end{equation}

The calculation of $G_{2D}$ at any temperature can be done as in
eq.\ref{eq:dIns}, by a convolution with the Fermi function.

\begin{figure}
%\vspace{0.45 cm}
\centering
\includegraphics[width=\columnwidth]{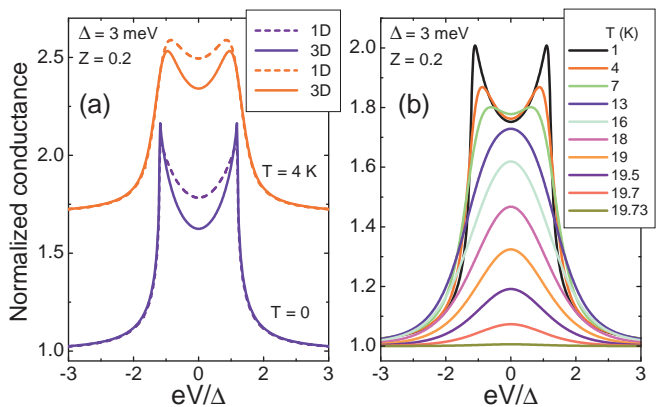}
\caption{(a) Normalized conductance curves calculated at $T=0$ and
$T=4$ within the 1D BTK model \cite{BTK} (dashed lines) and within
its 3D generalization \cite{kashiwaya96} (solid lines) using
$\Delta=3$ meV and $Z=0.2$. (b) Temperature dependence of the
conductance curves calculated within the 3D BTK model with $Z=0.2$
and assuming for the gap a BCS temperature dependence with
$\Delta(T=0)=3$ meV, $T_c=19.73$ K.}\label{fig:BTK_3D}
\end{figure}

When the system has rotational symmetry around the axis normal to
the interface (i.e. the gap is isotropic and the FS is spherical)
this approach can be considered as the 3D extension of the BTK
model. Figure \ref{fig:BTK_3D}(a) shows the comparison of two
normalized conductances at $T=0$ and $T=4$ K calculated with the
standard 1D BTK model and with its 3D version. The angular
integration leads to a remarkable depression of the AR signal when
$0<Z<10$. Obviously, when $Z=0$ (completely transparent junction) or
$Z>10$ (tunneling regime) the two approaches yield the same results.
In figure \ref{fig:BTK_3D}(b) a complete temperature dependency of
the normalized conductance calculated by using the 3D model and
assuming a BCS $\Delta(T)$ dependence is reported. It is trivial to
show that the 3D normalized conductance practically coincides with
the 1D one calculated for a properly enhanced $Z$ value. Probably
this fact explain why the standard 1D model is still largely used in
fitting the experimental data. Nevertheless, problems can arise when
comparing the $Z$ values obtained by the two different approaches,
particularly in the cases where the value of $Z$ has remarkable
consequences on the interpretation of the physical process occurring
at the interface, as, for example, in the study of
ferromagnet-superconductor PCAR junctions.

\subsubsection{Fermi velocity mismatch at the
interface}\label{sect:Zeff}
In a realistic system the Fermi velocities will be different on the
two sides of the contact. The mismatch of the Fermi velocities gives
rise to carrier reflections at the interface even when no barrier is
present. This effect was initially introduced in the original BTK
theory \cite{blonder83} by adopting an effective barrier parameter:
\begin{equation}\label{eq:Zeff}
Z_{eff}=\sqrt{Z^{2}+\frac{(1-r)^{2}}{4r}}
\end{equation}
where $r=v_{F1}/v_{F2}$ is the ratio of the Fermi velocities in the
superconducting and in the normal side. The normal-state resistance
at high voltage is given by $R_{N}=R_{S}(1+Z_{eff}^{2})$ where
$R_{S}$ is the Sharvin resistance \cite{blonder83}.

In the 3D version of the model \cite{kashiwaya96} the situation is
more complex. To account for the possibility of different effective
masses in N and S, the parameter $r$ of eq.\ref{eq:Zeff} is replaced
by $\lambda_0=\textbf{k}_S/\textbf{k}_N$. The ``refraction'' of
quasiparticles at the interface is due to the conservation of
transverse momentum, i.e. $sin(\theta_N)= \lambda_0 sin(\theta_S)$
where $\theta_N$ and $\theta_S$ are the incidence and transmission
angles, respectively. Under these conditions it is possible to show
\cite{kashiwaya96} that the normal transmission probability (eq.
\ref{eq:tau1}) becomes:
\begin{equation}\label{eq:tau2}
\tau_N(\theta_{N},\theta_{S})=\frac{4\lambda_{0}\cos\theta_{N}\cos\theta_{S}}{\left[\cos\theta_{N}+\lambda_{0}\cos\theta_{S}\right]^{2}+4Z^{2}}.
\end{equation}

By introducing this expression in the formula for the
superconducting transmission probability (eq.\ref{eq:sigma}) and
expressing $\theta_S$ as a function of $\theta_N$ by using the
``refraction'' relation $sin(\theta_N)= \lambda_0 sin(\theta_S)$,
one formally obtains the same expression for the normalized
conductance $G_{2D}(E)$ as in eq. \ref{eq:sigma2D} that now,
however, accounts for the mismatch in the Fermi velocities.
Incidentally, when $\lambda_0<1$, i.e. $\textbf{k}_N >
\textbf{k}_S$, a ``total reflection'' of electrons occurs at the
interface for injection angles $|\theta_N|>sin^{-1}\lambda_0$. In
this case the integral in $\theta_N$ has to be restricted to this
limit angle \cite{kashiwaya96}. It seems that the condition
$\lambda_0<1$ could easily apply in the case of a superconductor
with a small or very small FS and, thus, this problem could be
important in new unconventional superconductors. In the opposite
case, $\lambda_0<1$, $\theta_N$ can vary in the whole half-plane
while the range of $\theta_S$ is restricted. Anyway, whatever the
approach to the problem is, it turns out that the global effect of a
mismatch of Fermi velocities at the interface is simply described by
a sort of ``renormalization'' of the $Z$ values of the kind
described in eq. \ref{eq:Zeff}. As a consequence, apart from extreme
and hypothetical cases showing very large (or very small)
$\lambda_0$ values, the effect of the mismatch cannot be separated
from the standard experimental variability of $Z$ values, unless one
is able to determine the true $Z$ value at the interface.

\subsubsection{The broadening parameter}\label{sect:broadening}
Even if the BTK model allows a correct interpretation of some
experiments in low-temperature superconductors \cite{blonder83}, in
most cases it predicts much sharper gap features than those actually
observed in the low-temperature conductance curves. This means that
the AR structures in the experimental spectra are not only depressed
in amplitude but also \emph{spread} in energy. This effect can be
attributed to the reduction of the quasiparticle lifetime, resulting
from: i) the imaginary part of the quasiparticle self-energy. This
term is ``intrinsic'' but very small, as discussed in the tunnel
regime by Dynes et \emph{al.} \cite{dynes78}; ii) inelastic
quasiparticle scattering processes occurring near the N/S interface
(surface degradation, contamination etc. either at the N or the S
side) \cite{plecenik94}. This term is ``extrinsic'' and much larger
than the previous one. By properly solving the Bogoliubov-de Gennes
equations in the presence of an inelastic scattering term, it has
been shown \cite{plecenik94,srikanth92} that it is possible to
globally take these effects into account by including into the BTK
model a single broadening parameter $\Gamma$ in the form of an
imaginary part of the energy, i.e.  $E \rightarrow E+i\Gamma$.
$\Gamma$ can thus be considered as the sum of the ``intrinsic''
lifetime parameter $\Gamma_{i}=\hbar/\tau_i$ and the ``extrinsic''
one $\Gamma_{e}=\hbar/\tau_e$, being $\tau_{i,e}$ the corresponding
intrinsic and extrinsic lifetimes. There is actually a third
possible origin of broadening of the conductance curves that can be
accounted for by using $\Gamma$, i.e. a distribution of gap values
(in anisotropic superconductors). In this case, $\Gamma$ simulates
the effect of a convolution of the theoretical conductance with the
gap distribution (an example is presented in
Sect.\ref{sect:borocarbides}).

Introducing $\Gamma$ in the BCS quasiparticle density of states
leads to the modified expression \cite{dynes78,plecenik94}:
\begin{equation}\label{eq:N_Gamma}
N(E,\Gamma)=\Re\left[\frac{E+i\Gamma}{\sqrt{(E+i\Gamma)^{2}-\Delta^{2}}}\right].
\end{equation}

$\Gamma$ enters the BTK model or its generalizations through
$N_q(E)$ and $N_p(E)$ in eq. \ref{eq:gamma1}, thus modifying
$\sigma(E,\theta_N)$ and the conductance $G_{2D}(E)$ (eq.
\ref{eq:sigma2D}). Fig. \ref{fig:broad+phonons}(a) depicts the
normalized conductance $G_{2D}(E)$ calculated using $Z=0.25$ and
different values of the ratio $\Gamma/\Delta$. The broadening effect
of $\Gamma$ cannot be reproduced by any combination of parameters of
the standard BTK theory unless one convolutes the zero-temperature
conductance with the Fermi function at a fictitious temperature
higher than the actual one. This approach is sometimes implicitly
used indeed when the experimental smearing of the curves is treated
in terms of a Gaussian broadening. Such a procedure is not
theoretically founded and mixes the actual thermal smearing with the
other broadening effects, which are instead well distinct. Finally,
even if it is common (and reasonable) opinion that the best
conductance curves should allow a fit with $\Gamma/\Delta \lesssim
0.5$, large $\Gamma$ values might be sometimes necessary (for
example in the presence of a wide gap distribution). This does not
necessarily prevent the determination of the gap by means of a
fitting procedure, which is indeed possible even when
$\Gamma/\Delta>1$ (especially if $Z$ is sufficiently large).

\subsubsection{Energy dependence of the order
parameter}\label{sect:Delta(E)}
It is well known that the mean-field BCS definition of a
\emph{constant} superconducting order parameter $\Delta$ is only a
crude approximation of the physical reality. Actually, even in the
weak-coupling regime $\Delta$ is a function of the energy and shows
a small energy-dependent imaginary part. The signatures of this
energy dependence on the normalized tunneling (or AR) conductance
curves are extremely small but, when the intensity of the
electron-phonon coupling increases (strong-coupling regime) they
become visible. By solving the Eliashberg equations for the
strong-coupling regime starting from the electron-phonon spectral
function $\alpha^2F(E)$ and the Coulomb pseudopotential $\mu^*$
(direct solution) it is possible to obtain the full energy
dependence of the order parameter
$\Delta(E)=\Re\Delta(E)+i\Im\Delta(E)$. The imaginary part of
$\Delta(E)$ increases at the increase of the coupling and accounts
for the finite lifetime of Cooper pairs. By introducing the function
$\Delta(E)$ into the expression for the quasiparticle density of
states (eq. \ref{eq:N_Gamma} with $\Gamma=0$), small deviations from
the BCS DOS at the typical phonon energies are observed, due to the
electron-phonon interaction (EPI). It is well known that also the
inverse procedure works (but only approximately in multi-band
superconductors! \cite{dolgov03}) i.e. starting from the EPI
structures in the experimental tunneling conductance it is possible
to obtain $\alpha^2F(E)$ and $\mu^*$ by the inverse solution of the
Eliashberg equations.

\begin{figure}
%\vspace{0.45 cm}
\centering
\includegraphics[width=\columnwidth]{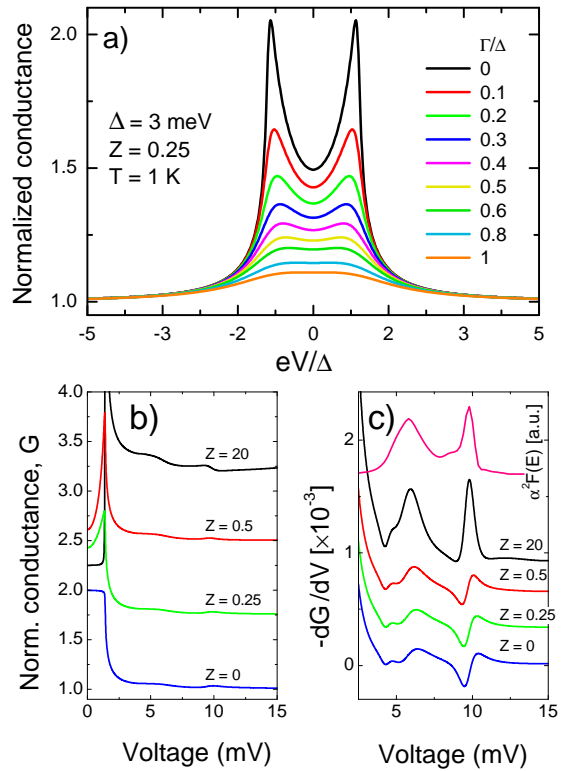}
\caption{(a) Conductance curves calculated at $T=1$K within the 3D
BTK model by using $\Delta=3.0$ meV, $Z=0.25$ and increasing values
of $\Gamma$. (b,c) Normalized conductance and its voltage derivative
($-dG/dV$) calculated for different $Z$ values within the 3D BTK
model, in the case of Pb, by using the energy-dependent order
parameter $\Delta(E)$ as obtained from the electron-phonon spectral
function $\alpha^2F(E)$ (top curve in panel (c), here shifted in
energy for ease of comparison).}\label{fig:broad+phonons}
\end{figure}

Since the BTK theory (and its modifications discussed so far)
coincides with the BCS theory for superconducting tunnel in the
limit of large $Z$ , it is easy to predict that the introduction of
$\Delta(E)$ into the BTK expressions will lead to EPI structures in
the normalized conductance for any $Z$ value in the ballistic
regime. This is indeed the case, as it can be explicitly
demonstrated \cite{ummarino10}. A simplified approach to the problem
was presented in Ref. \cite{yanson04c}, where simple asymptotic
expressions for the normalized conductance at $eV\gg \Delta$ in the
tunnel ($Z\rightarrow \infty$), ballistic and diffusive regime were
obtained by taking into account phonon self-energy effects on the
order parameter. Let us instead show here an example of the complete
procedure applied to a ``classic'' strong-coupling superconductor.
First, we calculated the $\Delta(E)$ function of lead starting from
its EPI spectral function (top curve in Fig.
\ref{fig:broad+phonons}(c)) and assuming $\mu^*=0.11$. $\Delta(E)$
was thus introduced in the expressions of $N_q(E)$ and $N_p(E)$
finally leading to the point-contact normalized conductance shown in
Fig. \ref{fig:broad+phonons}(b) for different $Z$ values. As
expected, the normalized conductance at $eV\lesssim\Delta_{Pb}$
coincides with the standard BTK one \cite{yanson04c}. At $eV \approx
\Delta_{Pb} + E_{ph}$ (where $E_{ph}$ represents the range of
energies where $\alpha^2F(E)_{Pb} \neq 0$) the EPI structures appear
for any $Z$ value but their amplitude increases with $Z$. Fig.
\ref{fig:broad+phonons}(c) shows the sign-changed first derivative
of the normalized conductance $-dG/dV = -d^2I_{NS}/dV^2$ vs. $V$
compared to the $\alpha^2F(E)_{Pb}$ (top red curve). Even if the EPI
structures shift to higher energies and their amplitude is depressed
at the decrease of $Z$, the use of DPCAR spectroscopy in very
high-quality single crystals to access quantitative information on
the $\alpha^2F(E)$ and its dependence on direction, temperature and
applied magnetic fields proved to be a feasible task
\cite{yanson04c}.

\subsubsection{Anisotropic order parameter}\label{sect:anisotropicDelta}
The assumption of an isotropic ($s$-wave) order parameter (OP) makes
the BTK model particularly simple, but this constraint must be
relaxed if one wants to describe systems in which the OP is instead
anisotropic, i.e. it depends on the wavevector $\mathbf{k}$ in the
reciprocal space. This happens for example in high-$T_c$ cuprates,
where at least one component of the OP has a $d$-wave symmetry
\cite{deutscher05}. Generally speaking, the anisotropy of the OP can
have two different origins: (i) the OP has a true $\textbf{k}$
dependence (at least along some planes of high symmetry) on the
single FS sheet where it opens; (ii) different isotropic OPs open on
different sheets of the FS of a multiband system. Strictly speaking,
in this case the OP is not anisotropic but appears so when it is
measured by techniques with null or poor resolution in the
$\textbf{k}$ space. Of course, more complex cases with multiple
anisotropic gaps can in principle occur, which could be probably
elucidated only by experimental techniques with full
$\textbf{k}$-space resolution (e.g. high-resolution ARPES). In this
section, we will show how to account for a single anisotropic OP
within the 2D BTK model. The more complex effect of multiple OPs on
different FS sheets and the influence of the shape of the FS itself
will be addressed in the next section.

\begin{figure}
%\vspace{0.45 cm}
\centering
\includegraphics[width=\columnwidth]{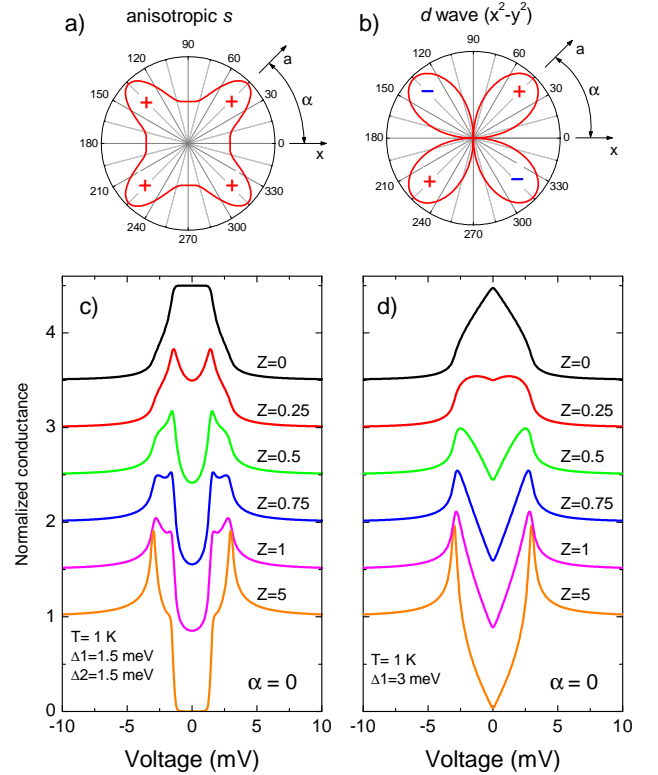}
\caption{(a,b) Polar plot of the magnitude of the OP in the case of
\emph{anisotropic} $s$-wave symmetry (a) and $d_{x^2-y^2}$ symmetry
(b). The sign of the OP is indicated by $+$ and $-$. The angle
$\alpha$ (here equal to $\pi/4$) between the crystallographic $a$
axis and the $x$ axis (normal to the interface) is also shown .
(c,d) Normalized conductance curves at $T=1$K calculated within the
generalized 2D BTK model \cite{kashiwaya96} with $\alpha=0$ for
different values of $Z$ and in the same OP symmetries as in (a,
b).}\label{fig:anisotropy}
\end{figure}

The problem of introducing the OP anisotropy into the expression of
the superconducting transmission probability $\sigma(E,\theta_N)$
was solved in Ref. \cite{kashiwaya96} in the most general case. Here
we will give a simplified ``operative'' description of the general
results. Let us suppose for simplicity that the OP has a
$\textbf{k}$ dependence only in the $k_xk_y$ plane and that $x$ is
the direction normal to the flat junction interface. Let the system
have a translational invariance along the $k_z$ axis so that the
problem reduces to a two-dimensional one, i.e. the FS is a cylinder.
We also suppose that the current injection occurs in the plane $xy$
($ab$-plane contact) and that $\lambda_0=1$, i.e. there is no
refraction of quasiparticles at the interface and both the
integration angles $\theta_N$ and $\theta_S$ span in the range
[$-\pi/2, \pi/2$]. Let the OP $\Delta$ be a function of the angle
$\theta_S$ with which electron-like quasiparticles (ELQ) are
injected in S. The specific expression of $\Delta(\theta_S)$ depends
on the kind of symmetry the OP shows in the $\textbf{k}$ space. To
take into account the possible rotation of the crystallographic $a$
axis with respect to the normal to the interface ($x$ axis) we also
introduce the angle $\alpha$ [see figs. \ref{fig:anisotropy}(a) and
\ref{fig:anisotropy}(b)]. Since ELQ and HLQ are injected in S with
angles $\theta_S$ and $-\theta_S$, respectively, they feel different
OPs, namely $\Delta_+=\Delta(\theta_S-\alpha)$ (for ELQ) and
$\Delta_-=\Delta(-\theta_S-\alpha)$ (for HLQ). Under these
conditions, the superconducting transmission probability becomes
\cite{kashiwaya96}:
\begin{equation}\label{eq:sigmad}
\sigma(E,\theta_N)=\tau_N\cdot\frac{1+\tau_N|\gamma_{+}(E)|^{2}+(\tau_N-1)|\gamma_{+}(E)\gamma_{-}(E)|^{2}}{|1+(\tau_N-1)\gamma_{+}(E)\gamma_{-}(E)\exp(i\varphi_{d})|^{2}}
\end{equation}
where
\[
\gamma_{\pm}(E)=\frac{E-\sqrt{E^{2}-|\Delta_{\pm}|^{2}}}{|\Delta_{\pm}|}
\]
and $\varphi_d=(\varphi_- - \varphi_+)$, $\varphi_\pm$ being the
phases of $\Delta_\pm$. When $\Delta_\pm$ are real quantities, then
their phase can only be either 0 or $\pi$ and the same holds for
$\varphi_d$. The choice of $\alpha$ determines the $\theta_S$
intervals in which the phase difference $\varphi_d$ is 0 or $\pi$.
If $\Delta_\pm$ do not show sign changes as a function of
$\theta_S$, then $\varphi_d=0$ independently of $\alpha$.  $\tau_N$
appearing in eq.\ref{eq:sigmad} has the same expression shown in eq.
\ref{eq:tau2}. Putting $\sigma(E,\theta_N)$ in eq. \ref{eq:sigma2D}
one finally obtains the total (integrated) normalized conductance at
$T=0$. The convolution with the Fermi function as in
eq.\ref{eq:dIns} will finally give the theoretical curves to be
compared with the experimental results at any $T$. Figure
\ref{fig:anisotropy}(c) shows the normalized conductance at $T=1$ K
for different $Z$ values and $\alpha=0$ in the case of
\textit{anisotropic} $s$-wave symmetry of the pair potential, where
$\Delta_+=\Delta_1 + \Delta_2 \cos^4[2(\theta_S-\alpha)]$,
$\Delta_-=\Delta_1 + \Delta_2 \cos^4[2(-\theta_S-\alpha)]$
($\Delta_1=1.5$ meV, $\Delta_2=1.5$ meV) and $\varphi_d=0$. Figure
\ref{fig:anisotropy}(d) shows the normalized conductance for the
same values of the parameters in the case of $d_{x^2-y^2}$-wave
symmetry of the OP, where $\Delta_+=\Delta_1
\cos2(\theta_S-\alpha)$, $\Delta_-=\Delta_1 \cos2(-\theta_S-\alpha)$
($\Delta_1=3$ meV). In both cases the shape of the normalized
conductance is quite different from the behavior shown in the
$s$-wave case for the same $Z$ values [figure \ref{fig:BTK_Z=0}
(a)]. In particular: i) the \emph{anisotropic} $s$ curves show a
four-peak (or two-peak and two-shoulder) structure similar to that
observed in MgB$_2$ ii) the $d_{x^2-y^2}$ curve with $Z=5$ presents
the well-known V-shaped conductance at low bias, while the one with
$Z=0$ shows a cusp at zero bias instead of the flat region typical
of the $s$-wave symmetry. In the $d$-wave symmetry,
$\alpha=\pm\pi/4$ gives $\varphi_d=\pi$ for any value of $\theta_S$
and the normalized tunneling conductance (for high Z) presents the
well-known zero-bias conductance peak \cite{deutscher05}.

\subsubsection{True shape of the Fermi surfaces and momentum dependence of the pair
potential}\label{sect:trueFS}
Taking into account in the calculations for the PCAR conductance the
true shape of the Fermi surfaces in N and in S, the possible
$\textbf{k}$ dependence of the pair potential and the possible
existence of multiple sheets of the FS -- where the OP can assume
different values -- is a rather complicated task, from both the
conceptual and the numerical point of view. Let us proceed step by
step following the approach reported in \cite{mazin99,brinkman02}.
We will neglect possible interference effects between bands, that
can lead to the formation of bound states at the surface as
discussed in Ref.\cite{golubov09}.

First of all the materials used in the N side are usually good
conductors (Au, Ag, Pt, Cu, Al) for which the approximation of a
spherical FS is reasonable. So here we restrict the analysis to the
shape of the FS in the superconducting material. In the most general
case the FS is divided into different sheets. Let us label them with
the subscript \emph{i} and call $\textbf{n}$ the unitary vector in
the direction of the total injected current, perpendicular to the
contact interface. As a consequence the components along the
direction $\textbf{n}$ of the Fermi velocities at wave vector
$\textbf{k}$ in the \emph{i}th FS sheet of the superconductor are
$\mathbf{\mathrm{v}_{\mathit{i\mathrm{\mathbf{k}}}}}$$\cdot\mathrm{\mathbf{n}}%
=v_{\mathit{i}\mathrm{\mathbf{k}},n}$
where
$\mathbf{\mathrm{v}_{\mathit{i\mathrm{\mathbf{k}}}}}=\frac{1}{\hbar}\left\{
\nabla_{\mathrm{\mathbf{k}}}\left[E_{i}(\mathrm{\mathbf{k}})\right]\right\}$.
Of course, due to the previous approximation, the corresponding
quantity in the normal metal is $v_{N,n}=
\mathrm{v}_{N}\cdot\mathrm{\mathbf{n}}$, being $\mathrm{v}_{N}$ the
(constant in magnitude) Fermi velocity in the normal material. The
\emph{i}th component of the total current flowing through a
perfectly transparent ($Z=0$) interface with no mismatch of the
Fermi velocities ($\lambda_{0}=1$) in a ballistic PCAR experiment on
a superconductor with isotropic OP is thus \cite{mazin99}:
\begin{equation}\label{eq:IaveFS}
I_{i}\varpropto\langle
N_{i\mathrm{\mathbf{k}}}v_{\mathit{i}\mathrm{\mathbf{k}},n}\rangle_{FS_{i}}=\oint_{FS_{i}}N_{i\mathrm{\mathbf{k}}}v_{\mathit{i}\mathrm{\mathbf{k}},n}dS_{F}=S_{i,n}
\end{equation}
where $N_{i\mathrm{\mathbf{k}}}(E_{F})=1/\left\{
4\pi^{3}|\nabla_{\mathrm{\mathbf{k}}}\left[E_{i}(\mathrm{\mathbf{k}})\right]|\right\}
_{E_{F}}$ is the density of states of the \emph{i}th band at the
Fermi energy and wave vector $\textbf{k}$ in S, $dS_F$ is the
elementary area on the FS in S and $\langle\rangle_{FS_{i}}$ is the
integral over the \emph{i}th FS sheet. The integral in eq.
\ref{eq:IaveFS} is limited to values
$v_{\mathit{i}\mathrm{\mathbf{k}},n}>0$. Obviously $S_{i,n}$ has the
meaning of area of the projection of the \emph{i}th FS sheet along
the $\textbf{n}$ direction, i.e. on the interface plane
perpendicular to $\textbf{n}$. It is the area of the \emph{i}th FS
sheet of the superconductor ``seen'' along the direction
$\textbf{n}$. Of course, under these restrictive conditions every
contribution to the total conductance from the \emph{i}th FS sheet
can be evaluated by using the same kind of integral, i.e. is
proportional to the projected area $S_{i,n}$. It means that the
total conductance ``seen'' along the direction $\textbf{n}$ is
$\langle\sigma(E)\rangle_{I\parallel\mathrm{\mathbf{n}}}=\sum_{i}\sigma_{i}(E)\langle
N_{i\mathrm{\mathbf{k}}}v_{\mathit{i}\mathrm{\mathbf{k}},n}\rangle_{FS_{i}}=\sum_{i}\sigma_{i}(E)S_{i,n}$
and the total normalized conductance is:
\[
\langle
G(E)\rangle_{I\parallel\mathrm{\mathbf{n}}}=\frac{\sum_{i}\sigma_{i}(E)\langle
N_{i\mathrm{\mathbf{k}}}v_{\mathit{i}\mathrm{\mathbf{k}},n}\rangle_{FS_{i}}}{\sum_{i}\langle
N_{i\mathrm{\mathbf{k}}}v_{\mathit{i}\mathrm{\mathbf{k}},n}\rangle_{FS_{i}}}=\frac{\sum_{i}\sigma_{i}(E)S_{i,n}}{\sum_{i}S_{i,n}}\]
where $\sigma_{i}(E)$ is the BTK superconducting transmission
probability (eq. \ref{eq:sigma}) of the \emph{i}th FS sheet. In the
case of different OPs $\Delta_{i}$ on the different sheets of the FS
the total normalized conductance will be dominated by the
contribution of the $\sigma_{i}(E)$ that corresponds the largest FS
projected area along the $\textbf{n}$ direction. As a consequence,
directional PCAR experiments at $Z=0$ and $\lambda_{0}=1$ can give
information on the distribution and values of the isotropic OPs on
the different FS sheets in a multiband, multigap superconductor. It
is quite obvious to expect a similar results also in the more
general case of an anisotropic OP and of $Z\neq0$ and
$\lambda_{0}\neq1$, but the calculation of the normalized
conductance is now much more complex.

\begin{figure}
\centering
\includegraphics[width=\columnwidth]{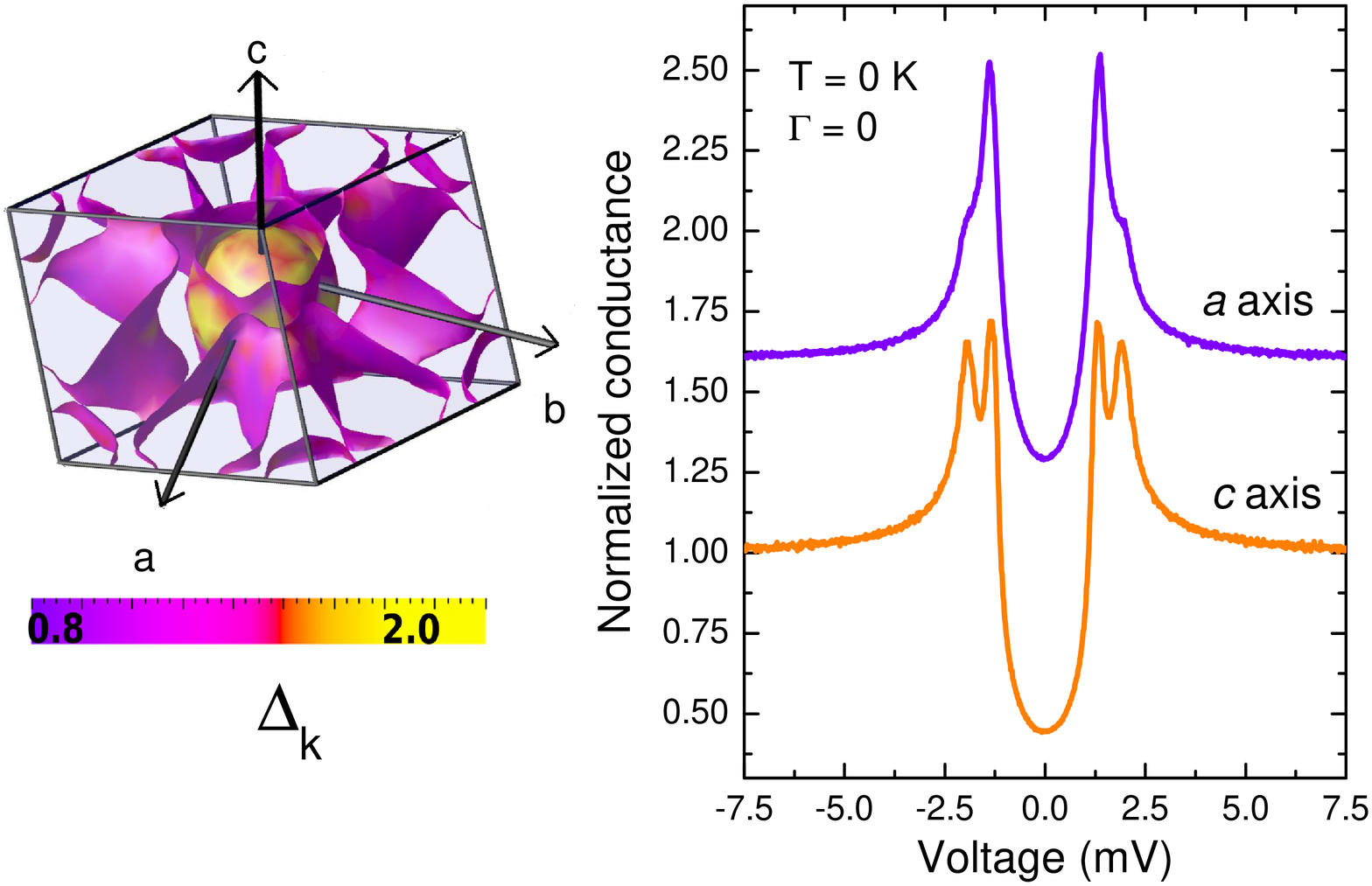}
\caption{Left: Distribution of the pair potential values over the
three different sheets of the FS of CaC$_6$ obtained by
first-principle calculations \cite{sanna07} (by courtesy of A. Sanna
and S. Massidda). Right: theoretical normalized conductance at $T=0$
for current injection along the \emph{a} axis ($Z_a=0.75$) and along
the \emph{c} one ($Z_c=1$). The values of $Z$ are taken from PCAR
experiments.}\label{fig:Sanna}
\end{figure}

First of all, if the OPs on the FS sheets are anisotropic, i.e.
$\Delta_{i}=\Delta_{i}(\mathrm{\mathbf{k}})=\Delta_{i\mathrm{\mathbf{k}}}$
(but still $Z=0$ and $\lambda_{0}=1$), then the superconducting
transmission probability becomes a function of $\mathrm{\mathbf{k}}$
and cannot be anymore extracted from the integral over the FS. The
total normalized conductance thus becomes:
\begin{equation}\label{eq:GaveFS}
\langle
G(E)\rangle_{I\parallel\mathrm{\mathbf{n}}}=\frac{\sum_{i}\langle\sigma_{i\mathrm{\mathbf{k}}}(E)N_{i\mathrm{\mathbf{k}}}v_{\mathit{i}\mathrm{\mathbf{k}},n}\rangle_{FS_{i}}}{\sum_{i}\langle
N_{i\mathrm{\mathbf{k}}}v_{\mathit{i}\mathrm{\mathbf{k}},n}\rangle_{FS_{i}}}
\end{equation}
where $\sigma_{i\mathrm{\mathbf{k}}}(E)$ is always expressed by eq.
\ref{eq:sigma} but using functions {\small $N^{q}_{i{\bf
k}}(E)=E/\sqrt{E^{2}-\Delta_{i{\bf k}}^{2}}$ }{\normalsize} and
{\small $N^{p}_{i{\bf k}}(E)=\Delta_{i{\bf
k}}/\sqrt{E^{2}-\Delta_{i{\bf k}}^{2}}$ }{\normalsize} that
substitute the standard ones in the definition of $\gamma(E)$ in eq.
\ref{eq:gamma1}. If the barrier has a finite transparency and there
is a N/S Fermi velocity mismatch, the normal transmission
probability of the barrier $\tau_{N}$ is no longer identically 1.
According to the standard 2D extension of the BTK model shown
before, $\tau_{N}$ (which here we call $\tau$ for simplicity of
notation) is given by eq. \ref{eq:tau2} that can be conveniently
rewritten as a function of the projections of the Fermi velocities
along the $\textbf{n}$ direction \cite{mazin99}:
\begin{equation}
\tau_{\mathit{i}\mathrm{\mathbf{k}},n}=\frac{4v_{\mathit{i}\mathrm{\mathbf{k}},n}v_{N,n}}{(v_{\mathit{i}\mathrm{\mathbf{k}},n}+v_{N,n})^{2}+4Z^{2}v_{N}^{2}}.
\end{equation}
By introducing this transmission probability inside the integrals
over the FS both at numerator and denominator of eq. \ref{eq:GaveFS}
and taking into account that
$N_{i\mathrm{\mathbf{k}}}v_{\mathit{i}\mathrm{\mathbf{k}},n}=v_{\mathit{i}\mathrm{\mathbf{k}},n}/v_{\mathit{i}\mathrm{\mathbf{k}}}$
we finally obtain the total normalized conductance at $T=0$ in the
most general case:
\begin{equation}\label{eq:sigmaFS}
\langle
G(E)\rangle_{I\parallel\mathrm{\mathbf{n}}}=\frac{\sum_{i}\langle\sigma_{i\mathrm{\mathbf{k}}n}(E)\frac{4v_{\mathit{i}\mathrm{\mathbf{k}},n}^{2}v_{N,n}}{v_{\mathit{i}\mathrm{\mathbf{k}}}[(v_{\mathit{i}\mathrm{\mathbf{k}},n}+v_{N,n})^{2}+4Z_{n}^{2}v_{N}^{2}]}\rangle_{FS_{i}}}{\sum_{i}\langle\frac{4v_{\mathit{i}\mathrm{\mathbf{k}},n}^{2}v_{N,n}}{v_{\mathit{i}\mathrm{\mathbf{k}}}[(v_{\mathit{i}\mathrm{\mathbf{k}},n}+v_{N,n})^{2}+4Z_{n}^{2}v_{N}^{2}]}\rangle_{FS_{i}}}
\end{equation}
where a subscript $n$ has been added to the expressions of
$\sigma_{i\mathrm{\mathbf{k}}}(E)$ and $Z$ just to include the
possibility to have different $Z$ values along the different
crystallographic directions, a thing that is often observed in DPCAR
experiments. In the case of large $Z$ (tunneling regime) the
weighting factor inside both the FS integrals of eq.
\ref{eq:sigmaFS} reduces to
$N_{i\mathrm{\mathbf{k}}}v_{\mathit{i}\mathrm{\mathbf{k}},n}^{2}$
and the calculations are simplified. As previously, the presence of
isotropic OPs on \emph{every} FS sheet allows extracting
$\sigma_{i}(E)$ from the integrals. This is the approach recently
followed by Brinkman \emph{et al.} in Ref.\cite{brinkman02}, where
the total normalized conductance of MgB$_2$ has been written as a
weighted sum of the partial conductances of the $\sigma$ and $\pi$
bands using the squares of the plasma frequencies along the
different crystallographic directions as weighting factors. Of
course, independently of the isotropic or anisotropic properties of
the OPs, if the current injection in a point-contact (or tunneling)
experiment was a fully directional process the gap should not be
seen along that directions where the FS has a null projected area.
Actually, as we have seen in the previous sections, this is not the
case, i.e. only a partial directionality is always present, which
depends on the $Z$ and $\lambda_0$ values. This explains why
\emph{c}-axis tunneling experiments on superconductors with a
quasi-2D FS (cylinder parallel to $k_z$) actually are able to
measure the gap averaged over the \emph{ab} plane. If the gap value
$\Delta_{i\mathrm{\mathbf{k}}}$ and the Fermi velocity
$v_{i\mathrm{\mathbf{k}}}$ are known at any $\mathrm{\mathbf{k}}$
point of the \emph{i}th FS sheet by first-principle calculations or
by high-resolution ARPES experiments, then eq. \ref{eq:sigmaFS}
allows the calculation of the PCAR normalized conductance at $T=0$
for a current injection along any crystallographic direction. An
example of the results of this procedure \cite{gonnelli08} is shown
in Fig. \ref{fig:Sanna}, which shows the distribution of the pair
potential values over the three different sheets of the FS of
CaC$_6$ obtained by first-principle calculations \cite{sanna07}
(left panel) and the theoretical AR normalized conductance at $T=0$
for current injection along the \emph{a} axis ($Z_a=0.75$) and along
the \emph{c} one ($Z_c=1$) (right panel). The theoretical curves of
Fig. 9, when properly broadened by $\Gamma$ values close to the
experimental ones, turned out to reproduce very well the
experimental DPCAR results in CaC$_6$ \cite{gonnelli08}, as it will
be shown in Sect.\ref{sect:CaC6}.

\section{Non-ideal effects in the contact}\label{sect:nonideality}
\subsection{Dips}\label{Dips_sect}\label{sect:dips} 
\begin{figure}
%\vspace{-0.2 cm}
\centering
\includegraphics[width=0.9\columnwidth]{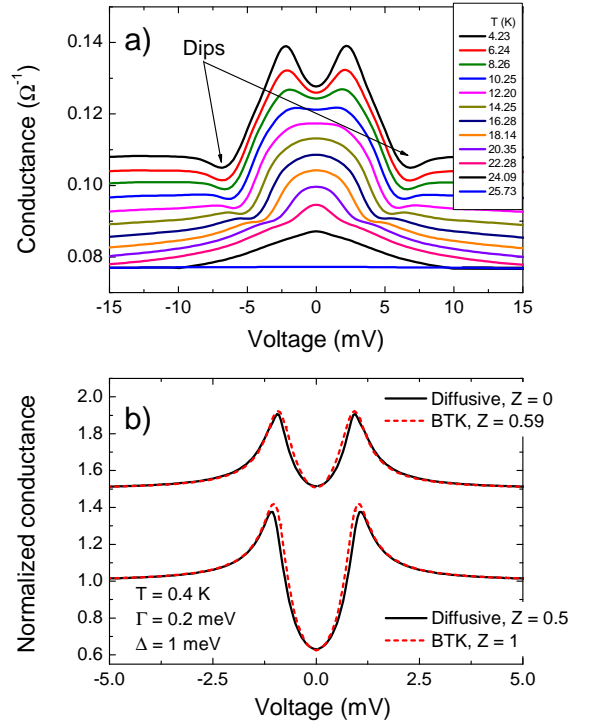}
\caption{(a) Temperature evolution of the conductance curve of a
Ag-paste contact on a Mg$_{0.85}$Al$_{0.15}$B$_2$ single crystal
(normal-state resistance $R_N=13\,\Omega$) featuring clear dips. All
the curves but the bottom ones are vertically offset for clarity.
(b) Upper curves (shifted for clarity): the theoretical conductance
curve obtained within the diffusive model with $Z=0$ (solid line) is
well reproduced by that obtained within the BTK model with the same
parameters but with $Z=0.59$ (dashed line). Lower curves:
conductance curve calculated in the diffusive model with $Z=0.5$
(solid line) and its BTK fit (dashed line) that requires $Z=1$. In
all cases, the other parameters of the models are always the same:
$T=0.4$ K, $\Delta=1$ meV and $\Gamma=0.2$
meV.}\label{fig:non_ideal}
\end{figure}

The PCAR
differential conductance often shows unexpected sharp dips at
voltage values larger than the superconducting gap, but sometimes
very close to it, as shown in Fig. \ref{fig:non_ideal}(a). These
dips are related to the superconducting properties of the S
electrode since they never show up in NN junctions, but the BTK
theory is unable to reproduce them. On increasing temperature, they
generally shift to lower energies and generally affect the shape of
the gap structures, as shown in Fig.\ref{fig:non_ideal}(a). For
example, they can make a broad maximum centered at zero bias look as
a sharp zero-bias conductance peak.

It is commonly accepted that these dips indicate a non-ideal
conduction through the contact. The detailed mechanism leading to
their emergence was studied by Sheet et al. \cite{sheet04} who
measured the evolution of the PCAR spectra of various N/S contacts
(made with the needle-anvil technique) on progressively reducing the
diameter of the point contact by withdrawing the tip in small steps,
and found indeed that the dips are related to the regime of
conduction through the junction, but also to the bias current. As a
matter of fact, according to Wexler's formula, the point-contact
resistance is generally the sum of a Sharvin and a Maxwell
contribution, whose relative weight depends on $a$. If $a$ is large,
the dominant term is the Maxwell one, which contains the bulk
resistivity of the two electrodes (eq. \ref{eq:R_diff_hetero}). As
long as the current flowing through the contact is small, the
resistivity of the superconductor is zero; however, when the current
reaches the critical value ($I_c$) in the S side, a normal-state
region can be created in S close to the junction, as discussed in
Sect.\ref{sect:andreev}. If this happens, the resistivity of the
superconductor starts playing a role and enters
eq.\ref{eq:R_diff_hetero}, giving a sharp increase in the voltage
across the junction and a dip in the differential conductance. The
same mechanism can be described as being due to the sudden
disappearance of the excess current. Numerical simulations of the
conductance, obtained by summing the $I-V$ curves of a ballistic
contact (given by BTK) to those of a typical bulk superconductor,
give indeed results in good agreement with observations
\cite{sheet04}.

An alternative explanation of the dips as being due to proximity
effect was given in Ref.\cite{strijkers01}. The idea is that, if a
proximity layer with depressed order parameter $\Delta_{prox}$ is
present at the interface, Andreev reflection is limited to energies
$eV < \Delta_{prox}$, while quasiparticles can enter the S side only
when $eV>\Delta_{bulk}$. This gives rise to dips in the conductance
curves, at energies between $\Delta_{prox}$ and $\Delta_{bulk}$,
which also necessarily shift to lower energies on increasing
temperature because of the temperature dependence of the gaps. Ref.
\cite{strijkers01} also provides a model for the fit of the
conductance curves that requires $\Delta_{bulk}$, $\Delta_{prox}$
and $Z$ as adjustable parameters and can be generalized to include a
broadening term $\Gamma$.

Very often, when analyzing conductance spectra with dips, a BTK fit
is done ignoring the dips. However, even if this procedure
introduces only a small error in the determination of the gap when
the dips are small, it has been shown \cite{sheet04} that a
considerable overestimation of the gap can occur when they become
large.

\subsection{Diffusivity in the contact}
In Sect.\ref{sect:diffusive} we mainly discussed the effects of a
diffusive contacts in the case of a N/N junction. In N/S junctions,
the diffusivity in the contact has been theoretically addressed by
Mazin \emph{et al.} \cite{mazin01,mazin01b} and turns out to affect
only the $Z$ parameter. For instance, the conductance of a diffusive
junction with a given barrier parameter $Z$ can be fitted with a
ballistic (BTK) model with an effective $Z^*>Z$. This is shown in
figure \ref{fig:non_ideal}(b) where the conductances obtained within
the diffusive model (solid lines) are compared with those calculated
with the standard BTK model (dashed lines). All the curves are
calculated for $\Gamma=0.2$ meV, $\Delta=1$ meV and $T=0.4$ K. The
upper curves show that the conductance in the diffusive model with
$Z=0$ is well reproduced by the BTK model with $Z=0.59$.
Analogously, the lower curves indicate that when $Z=0.5$ is
introduced in the diffusive model, the obtained conductance
corresponds reasonably to that obtained within the BTK model, but
with $Z=1$. This conclusion is also, and even more, true at higher
temperatures and for higher values of the lifetime broadening, i.e.
when the curves are more smeared out.

\subsection{Inelastic scattering in the vicinity of the contact}
The inelastic scattering due to some layer with different
composition at the N/S interface has been clearly singled out
experimentally in ref. \cite{chalsani07} where ballistic
Andreev-reflection measurements were performed in Cu-Pb junctions
with and without a very thin ($\approx 2$ nm) Pt layer in between.
The PCAR curves of the Cu/Pt/Pb junctions were shown to be more
broadened than those of the Cu-Pb contacts, and were well fitted by
the BTK model by systematically using larger $\Gamma$ values --
though giving a good determination of the gap amplitude (note that,
already in the original paper by Plecenik et al. \cite{plecenik94},
$\Gamma$ was introduced in the BTK model to take into account
exactly these effects).

Something similar is likely to happen in the ``soft'' point
contacts, whose normalized conductance curves show a reduced
amplitude and a larger broadening than those obtained with the
conventional needle-anvil technique. To identify the scattering
layer in this case, we carefully measured the temperature dependence
of the resistivity of the particular Ag paint used for the contacts.
We found a residual resistivity at low temperature of $0.34$
m$\Omega$ cm (about $10^5$ times higher than that of pure Ag), and
an enormously increased slope of $\rho(T)$ at higher temperature.
The former indicates a huge contribution of intergrain connectivity
to the resistivity, and the latter a drastic reduction of the
inelastic mean free path on the grain surface, which could well give
rise to the observed broadening of the conductance curves. It must
be said, however, that a contribution from a layer at the surface
\emph{of the sample} cannot be completely ruled out, and is instead
proved by the fact that a similar broadening has been observed also
in some PCAR spectra taken with the needle-anvil technique. This
will be further discussed in the experimental survey (see sections
\ref{sect:MgB2gaps},\ref{sect:FeAs}).

\subsection{Spreading resistance}
For spectroscopic measurements to be reliable, electrons must not
lose a significant energy while traveling through the electrodes. If
at least one of the electrodes is highly resistive, a so-called
spreading resistance $R_{sp}$ must be considered in series with the
contact resistance, and this results in a shift of the conductance
peaks to higher energies, leading to an overestimation of the gap
\cite{woods04,baltz09}. Actually, a spreading resistance $R_{sp}$ is
always present but usually plays a role only in measurements
performed in thin films, while in bulk or highly conductive samples
it is much smaller than the contact (junction) resistance and can
thus be neglected. In the case of ``soft'' point contacts, one can
wonder whether the Ag paste between the Au wire and the sample
surface can give a significant contribution to $R_{sp}$. Actually,
the resistance of the Ag-paste spot (approximately modeled as a
cylinder with a diameter of 50 $\mu$m) is as small as 0.086 $\Omega$
even if a (largely overestimated) thickness of 50 $\mu$m is assumed.
This value is clearly negligible when compared to the contact
resistance that is usually in the range $5-100 \Omega$ (depending on
the material under study).

\section{Point-contact spectroscopy in multiband
superconductors}\label{sect:Experiments}
\subsection{Two-band model for superconductivity}\label{sect:2bands}
The first theoretical study of multiband superconductivity dates
back to the late Fifties when Suhl, Matthias and Walker
\cite{suhl59} generalized the BCS theory to the simple case of a
superconductor with two overlapping bands. The corresponding BCS
Hamiltonian contains two intraband terms of the kind
$\sum_{\textbf{k}\textbf{k}^{\prime}}V_{ii}c^{\dagger}_{i,\textbf{k}\uparrow}c^{\dagger}_{i,\textbf{-k}\downarrow}c_{i,\textbf{-k}^{\prime}\downarrow}c_{i,\textbf{k}^{\prime}\uparrow}$
and two interband terms of the kind
$\sum_{\textbf{k}\textbf{k}^{\prime}}V_{ij}c^{\dagger}_{i,\textbf{k}\uparrow}c^{\dagger}_{i,\textbf{-k}\downarrow}c_{j,\textbf{-k}^{\prime}\downarrow}c_{j,\textbf{k}^{\prime}\uparrow}$
(where $ij=1,2$ is the band index). $V_{ij}$ is the (constant in the
BCS approach) averaged pairing potential which results from boson
emission and absorption by an $i$-$j$ process, minus the
corresponding shielded Coulomb interaction. In the absence of
interband coupling ($V_{ij}=0$), the two bands would be completely
independent, each featuring its own BCS gap and critical
temperature. In the opposite case (only interband coupling,
$V_{ii}=0$) the critical temperature is the same, but there are
still two gaps unless the partial density of states is the same in
the two bands ($N_1=N_2$). In general, through interband coupling
the band with the higher superconducting $T_c$ raises the critical
temperature of the weaker, or even induces superconductivity in a
nonsuperconducting band. The critical temperature is defined as
$k_{B}T_{c}=1.14k_{B}\theta_{D}e^{-1/\lambda_{eff}}$ where
$\lambda_{eff}$ is the effective coupling constant and is simply the
maximum eigenvalue of the matrix $\Lambda_{ij}=V_{ij}N_{j}$ where
$N_{j}$ is the density of states at the Fermi energy (per spin) in
the $j$th band.

\begin{figure}
\centering
\includegraphics[width=0.9\columnwidth]{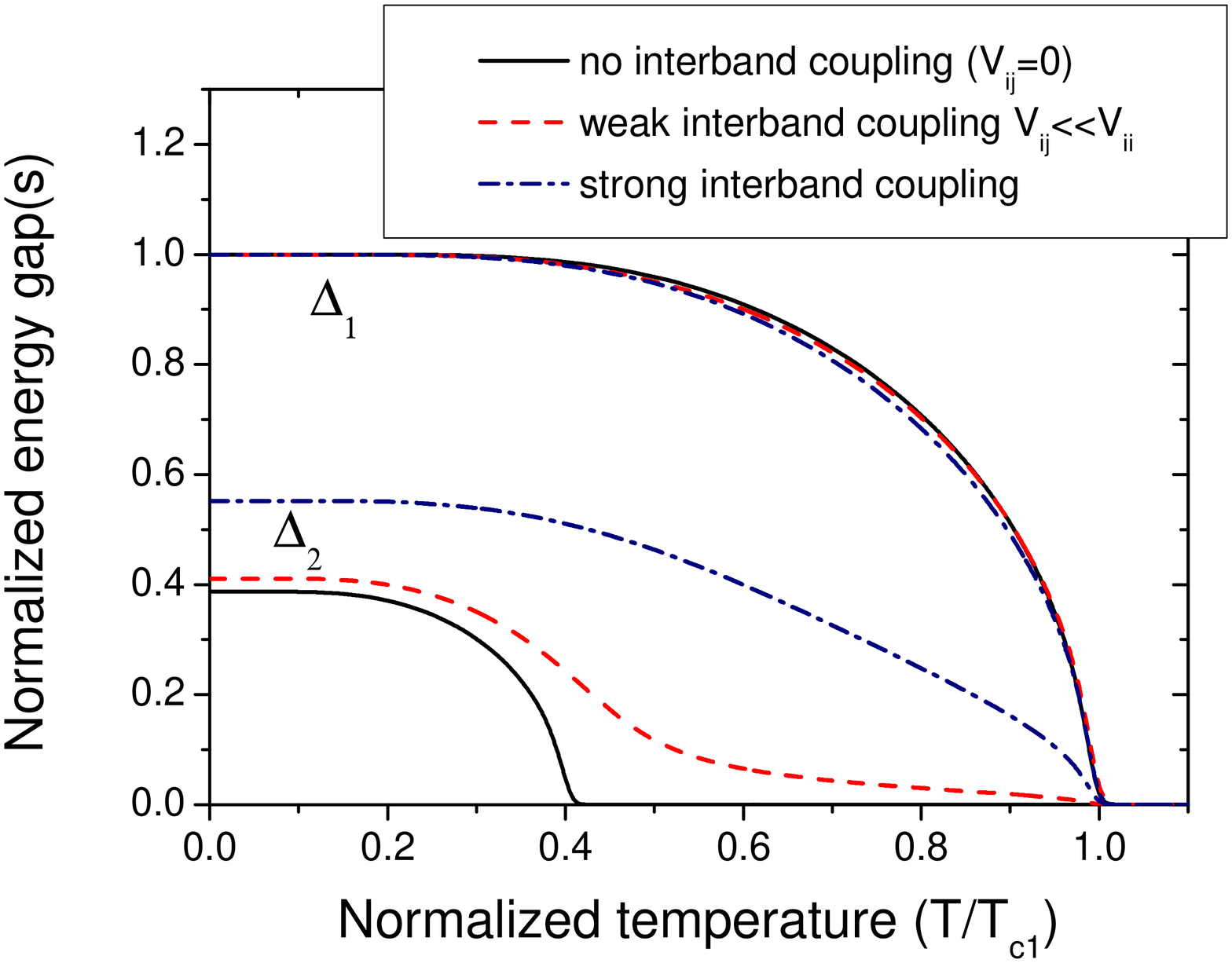}
\caption{Temperature dependence of the gaps $\Delta_1$ and
$\Delta_2$ in a two-band BCS model, calculated in the cases of: no
intraband coupling (solid lines); weak intraband coupling (dotted
lines); strong interband coupling (dash-dot lines). The intraband
coupling constants are arbitrary; here we used those for
MgB$_2$.}\label{fig:Suhl_Matthias}
\end{figure}
Figure \ref{fig:Suhl_Matthias} shows the temperature dependence of
the (normalized)  gaps as a function of the normalized temperature
in the BCS two-band model in different cases: i) bands completely
decoupled ($V_{ij}=0$, solid lines). The $T_c$'s of the bands depend
on the relevant intraband coupling;  ii)  weakly coupled bands
(dashed lines). While $\Delta_{1}$ follows the same standard BCS
temperature dependence (but with $2\Delta/k_{B}T_{c}> 3.53$),
$\Delta_2$ features a high-temperature tail and closes at the same
$T_c$ as $\Delta_1$; iii) strongly coupled bands (dash-dot lines).
The small gap still deviates from a BCS-like behavior but smoothly
decreases on heating, to finally close rather quickly at $T_c$. The
gap ratios 2$\Delta/k_BT_c$ for the two gaps are greater and smaller
than the single-band BCS value 3.53, respectively. As we will show
in the following experimental survey, PCAR measurements in multiband
superconductors have provided examples of all these three cases.

\subsection{Magnesium diboride} \label{sect:MgB2}
After the publication of the theory for two-band superconductivity,
some of its consequences on various measurable quantities were
calculated and possible marks of multiband superconductivity were
found in conventional materials like Nb \cite{hafstrom70,carlson70}.
In 1980 a clearer experimental evidence of multiband
superconductivity was found in Nb-doped SrTiO$_3$ \cite{binnig80} by
means of tunnel spectroscopy. Despite the fundamental importance of
the result, the very low transition temperature of this compound (a
few hundred mK) made its experimental investigation rather demanding
and prevented its study from becoming very popular. The situation
changed completely in 2001 when superconductivity below 39 K was
discovered in MgB$_{2}$, which remains up to now the most known and
the most studied example of multiband superconductor. MgB$_2$ has a
layered structure with graphite-like, honeycomb B layers
intercalated by Mg planes with hexagonal close-packed structure
\cite{buzea01}. Its electronic structure includes four $\sigma$
bands originating from $sp^2$-hybrid B orbitals, and two $\pi$ bands
due to the overlapping of the residual $p_{z}$ orbitals. The Fermi
surface is made up of nearly-2D cylinders around the $\Gamma-A$ line
(due to the $\sigma$ bands) and a 3D tubular network related to the
$\pi$ bands \cite{kortus01}. Superconductivity develops in the
$\sigma$ bands below $T_c=39$ K mainly because of their coupling to
the $E_{2g}$ phonon modes \cite{kong01}, and is induced in the $\pi$
bands through interband coupling.

\begin{figure}[hb]
\begin{center}
\includegraphics[width=0.8\columnwidth]{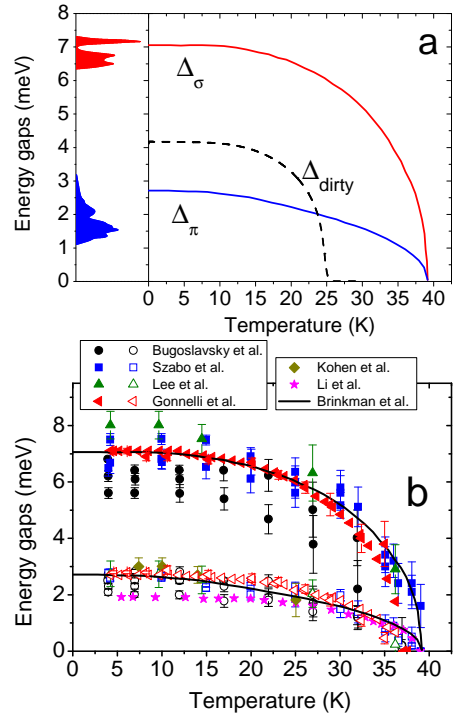}
\end{center}
\caption{(a): Theoretical predictions for the gaps in MgB$_2$ within
the Eliashberg two-band model \cite{brinkman02}. The dotted line
refer to the ``dirty'' limit (i.e. strong interband scattering).
Similar results are obtained in the BCS approach. On the left, the
gap distribution given by a fully-anisotropic Eliasherg calculation
are shown \cite{choi02}. (b) Experimental results of PCAR in
MgB$_2$, from Ref. \cite{bugoslavsky02} (circles), \cite{szabo01}
(squares), \cite{lee02} (up triangles), \cite{gonnelli02c} (left
triangles), \cite{kohen01} (diamonds), \cite{li02} (stars). Lines
indicate the predictions of the two-band model.}
\label{fig:MgB2gaps}
\end{figure}

The key role of this two-band-system structure was soon witnessed by
the failure of all the conventional single-band theories in
describing the phenomenology of MgB$_{2}$ \cite{liu01,buzea01}. An
effective two-band model was thus proposed, in which the four bands
were grouped into two band systems ($\sigma$ and $\pi$). The
anisotropic effective coupling constant for superconductivity
$\lambda_{eff}=1.01$ actually indicates an intermediate coupling
regime which is best described by the Eliashberg theory
\cite{eliashberg60,choi03,nicol05}. The calculation of the gaps
within a two-band Eliashberg model \cite{brinkman02}  gave
$\Delta_{\sigma}=7.1$ meV and $\Delta_{\pi}=2.7$ meV (see
Fig.\ref{fig:MgB2gaps}). Similar values can be obtained within a BCS
approach \cite{liu01}.

An interesting feature of multiband superconductivity in MgB$_2$ is
the role played by impurity scattering in the intraband and
interband channels. According to Anderson's theorem
\cite{anderson59}, it can be shown \cite{golubov97} that, at least
for small impurity concentrations, the intraband non-magnetic
scattering has no effect on $T_{c}$ and the gaps. The interband
scattering on the other side, has a \emph{pair-breaking effect} and
thus decreases the critical temperature $T_{c}$. According to the
two-band model, in the limit of very strong interband scattering
(dirty limit) a complete isotropization is asymptotically achieved,
and the two gaps assume the same value so that one single gap is
actually observed (dotted line in Fig.\ref{fig:MgB2gaps}a). This is
often referred to as ``gap merging''. According to Eliashberg
calculations in Ref.\cite{brinkman02}, at low temperature
$\Delta_{dirty}=4.1$ meV with a corresponding reduced $T_c=25$ K.
For the sake of completeness, Fig.\ref{fig:MgB2gaps}a also shows the
results of a fully-anisotropic Eliashberg calculation, based on the
actual momentum dependence of the electron-phonon coupling
calculated ab-initio \cite{choi02}. This approach gives two distinct
and non-overlapping distributions of gap values with average $6.8$
meV and $1.8$ meV. The differences from the two-band model arise
from details in the calculations that are not worth discussing here.
In any case, all calculations show that the gap values on the two
band systems are sufficiently different to be distinguishable also
experimentally.

According to the discussion of Sect.\ref{sect:trueFS}, the shape of
the FS (and in particular of the quasi-2D $\sigma$-band sheets)
suggests a dependence of the PCAR or tunneling spectra on the
direction of (main) current injection. Brinkman \emph{et al.}
\cite{brinkman02} calculated the conductance curves of an ideal
MgB$_2$-I-N junction with various barrier transparencies within the
Eliashberg theory. They expressed the normalized conductance $G$ of
the junction as the (weighted) sum of the BTK contributions of the
two band systems: \makebox{$G=w_{\pi}G_{\pi}+(1-w_{\pi})G_{\sigma}$}
\cite{brinkman02} \footnote{In this calculation, interference
effects between bands were not taken into account. In a recent paper
\cite{golubov09}, it has instead been shown that such effects can in
principle give rise to observable features in the Andreev
conductance spectra not only in iron pnictides, where the order
parameter changes sign on different bands, but also in MgB$_2$ where
the order parameter has the same sign on both $\sigma$ and $\pi$
bands.}. As expected, the weight $w_{\sigma}=1-w_{\pi}$ depends on
the direction of current injection. For $I\parallel c$ (and parallel
to the axis of the nearly cylindrical $\sigma$-band sheets)
$w_{\sigma}$ is no more than 1\% so that only the small gap
$\Delta_{\pi}$ should give detectable structures in the conductance
curve. For $I\parallel ab$, $w_{\sigma}$ is maximum and equal to 33
\%, so that four peaks corresponding to the small and the large gaps
$\Delta_{\pi}$ and $\Delta_{\sigma}$ are found in the conductance
curves \cite{brinkman02}. Note that the theoretical values of
$w_{\sigma}$ are referred to ideal tunnelling current injection;
slight differences are expected in PCAR experiments where the angle
of effective current injection as defined in Sect.\ref{sect:2DBTK}
can be considerably larger.

\subsubsection{Determination of the gaps in MgB$_2$}\label{sect:MgB2gaps}
% misure in campioni polcristallini: dipendenza degli spettri dalla pressione e effetti di superficie
The earliest PCAR investigations carried out in MgB$_2$ polycrystals
gave evidence of a single isotropic ($s$-wave) gap. Schmidt \emph{et
al.} \cite{schmidt01} obtained $\Delta=4.3\div 4.6$ meV, while Kohen
\emph{et al.} \cite{kohen01} measured a gap $\Delta=3.8\div 4.0$ meV
in higher-resistance contacts, while in a lower-resistance junction
a smaller gap ($\Delta=3$ meV) was found, with reduced $T_{c}'=29$
K. Laube \emph{et al.}\cite{Laube01} obtained an accumulation of gap
values around 1.7 meV and 7 meV but never observed both of them in
the same spectrum. Plecenik \emph{et al.}\cite{plecenik02} studied
the Andreev reflection curves of MgB$_{2}$/N junctions obtained in
different ways, whose fit with the modified BTK model (with
$\Gamma=0.8$ meV) gave a gap $\Delta=4.2$ meV. A discontinuity in
the temperature evolution of the gap suggested the existence of
parallel contacts in clean and dirty regions of the sample, with a
gap $\Delta_{S}=\Delta_{\pi}=2.6$ meV closing at $T_c\simeq 38$ K
and a gap $\Delta_{dirty}=4.0$ meV closing at $T_c=22$ K,
respectively. The absence of $\Delta_{\sigma}$ was probably due to a
preferred $c$-axis current injection. The presence of a degraded
layer on the sample surface suggested in \cite{kohen01} and
\cite{plecenik02} was confirmed by PCAR measurements performed with
electrochemically sharpened tips of different hardness
\cite{gonnelli02a}, which showed indeed a decrease in the height of
the conductance peaks (from 1.8 down to 1.25) and an increase in
$\Gamma$ from zero up to 1.2 meV on decreasing the pressure in the
contact region from about 0.6 GPa down to 0.1 GPa. Spectra taken
with the ``soft'' pressure-less technique had a height of only 1.15,
and their fit with a single-gap BTK model gave $\Gamma\simeq 3$ and
$\Delta_{dirty}= 4-5$ meV with a reduced $T_c$. A reduced $T_c$ was
found also in Ref.\cite{bugoslavsky02}, together with an increase in
$\Gamma$ and $Z$ on decreasing the barrier transparency. All these
results indicate an extrinsic contribution to $\Gamma$ from
inelastic carrier scattering in the barrier, not easily accountable
for in the theoretical model, and possibly due to a degraded or
reconstructed layer covering the sample which can be broken by a tip
but remains intact when the pressure is small or absent
\cite{gonnelli02a}. Indeed, it was shown experimentally
\cite{chalsani07} that this effect can be simply accounted for by
increasing the broadening parameter(s) in the modified BTK model.

% PCAR con 2 gap: il problema del fit

With the improvements in the sample quality, spectra with multiple
gap features were readily obtained in films and polycrystals
\cite{bugoslavsky02,szabo01}, that allowed a fit by the two-band BTK
model. In principle, the fitting function contains seven parameters:
the two gap amplitudes $\Delta_{\sigma}$ and $\Delta_{\pi}$, the
broadening parameters $\Gamma_{\sigma}$ and $\Gamma_{\pi}$, two
barrier parameters $Z_{\sigma}$ and $Z_{\pi}$, plus the weight
$w_{\pi}$ (so that $w_{\sigma}=1-w_{\pi}$) for a total of 7
parameters. Some authors decided to use only one $Z$ for both bands
\cite{szabo01,bugoslavsky02} but, owing to the different Fermi
velocities in the two bands, keeping $Z_{\sigma}$ and $Z_{\pi}$ as
independent parameters is more general. Some authors also take
$\Gamma_{\sigma}=\Gamma_{\pi}$  or even replace them with a
convolution of the $T=0$ conductance with a Gaussian of width
$\omega$ \cite{bugoslavsky02}. Others (including us) prefer instead
to calculate the conductance at the correct temperature, and add
$\Gamma_{\sigma}$ and $\Gamma_{\pi}$ as imaginary parts of the
energy in the BTK equation \cite{plecenik94} to account for all the
sources of broadening discussed in Sect.\ref{sect:broadening}.
Despite the number of free parameters, reliable values of the gaps
can be obtained. This is certainly true for $\Delta_{\pi}$ which is
quite strictly determined by the energy position of the relevant
conductance peaks. The same holds for $\Delta_{\sigma}$ when the
relevant peaks are observable -- that means, for $I \parallel ab$
\cite{brinkman02}. When the structures related to the large gap are
only smooth shoulders (as in $c$-axis contacts or in $ab$-plane
contacts at higher temperature), the uncertainty on
$\Delta_{\sigma}$ increases. The evaluation of this uncertainty is
not straightforward, because of the complex expression for the
conductance in the two-band BTK model and the number of parameters.
Indeed, an automated fitting procedure is destined to fail, and one
has to manually search for the parameters that allow minimizing the
chi-square or the sum of squared residuals (SSR). Once the ``best''
fit is found, a range of parameters that give ``acceptable'' fits
must be determined. This can be done by fixing a level of confidence
for the chi-square or allowing a percent increase in the SSR. Then,
the fit has to be repeated many times by changing \emph{all} the
free parameters so as to find the maximum variation of the gaps
compatible with the fixed limits. Several fits made independently by
different people normally ensure a good estimate of this range.
Fortunately, some physical constraints limit the range of
variability of some parameters. For example, $w_{\pi}$, $Z_{\sigma}$
and $Z_{\pi}$ should not depend on either the temperature and the
magnetic field; the intrinsic (lifetime) part of $\Gamma_{\sigma}$
and $\Gamma_{\pi}$ can increase with temperature, but their usually
much larger extrinsic part, related to the interface properties,
should probably not.

Fig.\ref{fig:MgB2gaps}(b) reports the experimental results of
various PCAR experiments in MgB$_2$. In all cases apart from
Ref.\cite{li02} a two-band fit was used. All the data sets
approximately agree with each other, apart from the early data by
Bugoslavsky in thin films which show a reduced $T_c$. The error bars
are indicated only for some data sets and clearly increase on
approaching $T_c$ because of the thermal smearing of the gap
features. Because of the same effect, one may wonder whether the two
gaps really close at the same temperature, since at high temperature
the spectra show only a broad maximum and the two-band fit could be
questioned. A conclusive answer to this issue and in favor of the
two-gap model in MgB$_{2}$ was found already in 2001 by Szab\'{o}
\emph{et al.} \cite{szabo01}, who performed PCAR measurements in
polycrystalline samples (squares in Fig.\ref{fig:MgB2gaps}(b))
obtaining gap values in very good agreement with theoretical
predictions. They found that the application of magnetic fields to
the junctions resulted in a much faster suppression of the
$\pi$-band features with respect to the $\sigma$-band ones. At high
temperature or in $c$-axis contacts where no $\sigma$-band features
are apparent, the disappearance of the dominant $\pi$-band
structures allows unveiling the underlying $\sigma$-band
contribution, with the emergence of two much well resolved maxima
related to $\Delta_{\sigma}$ even at $T=30$K.

%PCAR in cristalli singoli: direzionalità

The synthesis of single crystals large enough to be used for PCAR
allowed a step forward in the experimental investigation of
multiband superconductivity in MgB$_2$, and in particular a study of
the anisotropy of the spectra \cite{brinkman02}) by controlling the
direction of (main) current injection. The soft-PCAR technique
allowed us to make the contacts either on the flat surface of the
crystals ($c$-axis contacts in the following, according to the
nominal direction of current injection) or on their thin (50-100
$\mu$m) side ($ab$-plane contacts), which is very difficult by using
a tip.

\begin{figure}[ht]
\begin{center}
%\vspace{-6mm}
\includegraphics[width=0.8\columnwidth]{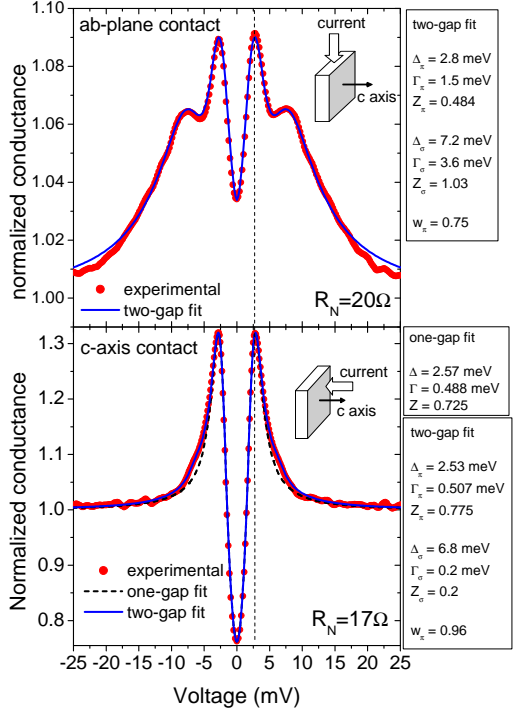}
\end{center}
%\vspace{-8mm}
\caption{``Soft'' PCAR spectra in MgB$_2$ single crystals for
different directions of current injection. The experimental,
normalized conductance curves for a $ab$-plane contact (upper panel)
and a $c$-axis contact (lower panel) are shown (symbols) and
compared to the two-band BTK fit (solid lines). In the lower panel,
the single-band BTK fit is also reported (dotted line). The
normal-state resistance of the contacts, $R_N$, is indicated in the
labels. }\label{fig:directionalityMgB2}
\end{figure}

Figure \ref{fig:directionalityMgB2} shows two examples of
conductance spectra measured in $ab$-plane and $c$-axis contacts
whose normal-state resistance is indicated in the labels. Note that
for all contacts with $R_N > 10\,\Omega$ the rather large mean free
path of these samples ($\ell=80$ nm) ensures the fulfillment of the
conditions for ballistic conduction (see Sect.\ref{sect:ballistic})
\emph{even if a single contact is hypothesized}. Clearly, if several
parallel contacts are present, they must be necessarily ballistic
\cite{gonnelli02c}. The spectra are normalized, i.e. divided by the
differential conductance at $T_c^A$ (being $T_c^A$ the critical
temperature of the junction). The experimental curves in
Fig.\ref{fig:directionalityMgB2} clearly show the predicted
anisotropy \cite{brinkman02}, but the non-perfect directionality of
PCAR prevents the weight of the $\pi$-band conductance from assuming
the theoretical extremal values ($w_{\pi}=0.66$ for $ab$-plane
tunneling, $w_{\pi}=0.99$ for $c$-axis tunneling). This is
particularly clear in $c$-axis contacts, where the single-gap BTK
fit (dashed line) does not work well and a two-band fit (solid line)
is instead necessary (with $w_{\pi}<0.99$). The values of the
fitting parameters are indicated in the labels. The temperature
dependence of the gaps obtained in different contacts on single
crystals \cite{gonnelli02c} is shown in Fig.\ref{fig:MgB2gaps}(b)
(left triangles).

\subsubsection{PCAR in magnetic field}
\label{sect:magneticfield} 
\begin{figure}[hb]
\begin{center}
\includegraphics[width=\columnwidth]{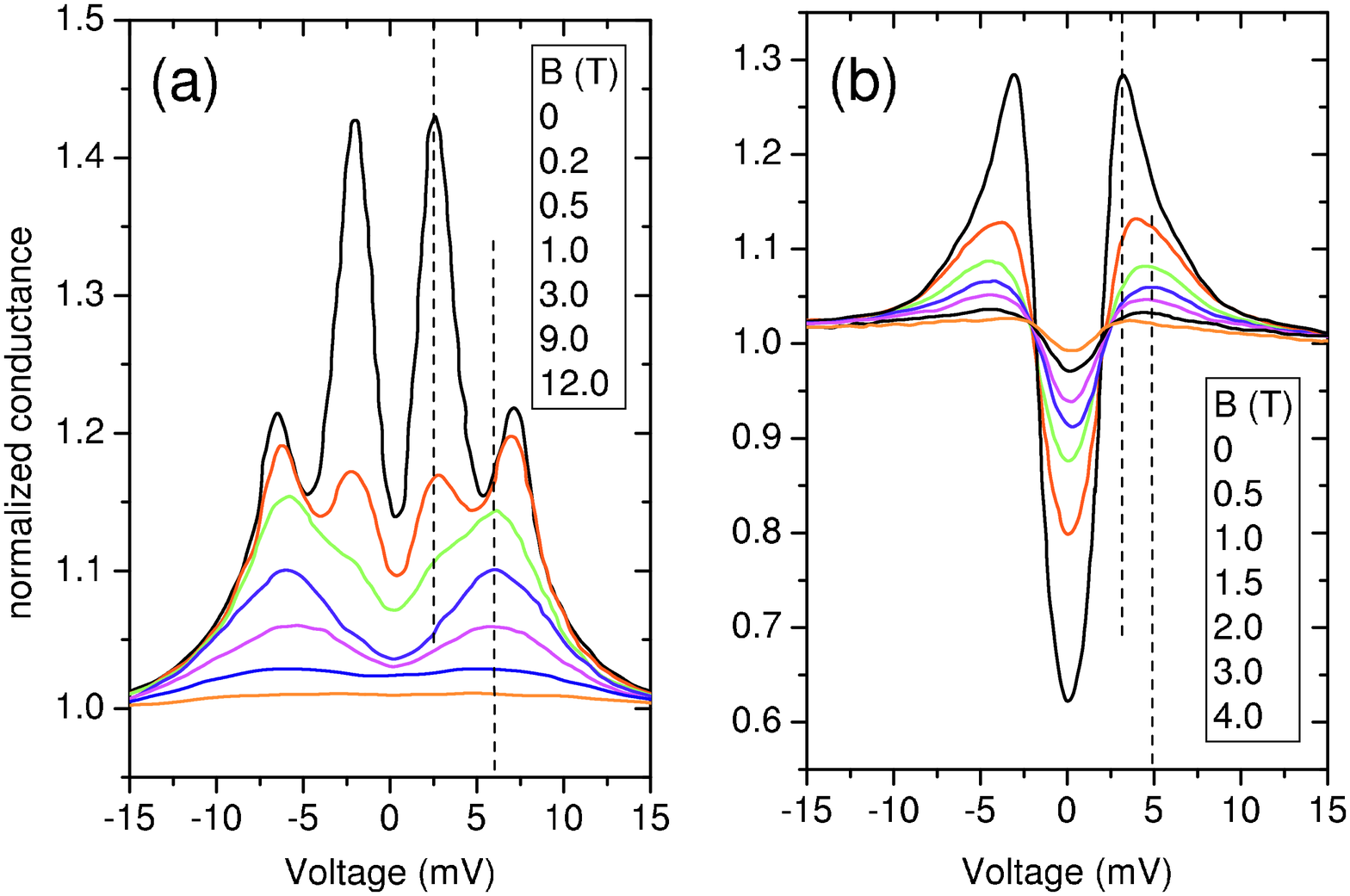}
\end{center}
\vspace{-5mm} \caption{Magnetic field dependence of low-temperature
normalized conductance curves measured in MgB$_2$ polycrystals. The
curves refer to contacts with large $ab$-plane contribution (a) and
dominant $c$-axis contribution (b). Data taken from Refs.
\cite{szabo01} (a) and \cite{szabo03b} (b).}\label{fig:szabo_field}
\end{figure}

As mentioned above, the first PCAR
measurements in MgB$_2$ in the presence of a magnetic field were
carried out by Szab\'{o} \emph{et al.} \cite{szabo01} in
polycrystals. Fig. \ref{fig:szabo_field} shows the magnetic field
dependence of the low-temperature PCAR spectra for contacts  with a
large $ab$-plane contribution (a) and a dominant $c$-axis
contribution (b). In the first case, the peaks related to
$\Delta_{\pi}$ are fast depressed by weak fields and become barely
detectable at $B=1$T, at which the large-gap maxima are still
clearly visible. In the second case, where no clear peaks related to
$\Delta_{\sigma}$ are observed in zero field \cite{szabo03b}, the
suppression of the $\pi$-gap at 1-1.5 T causes an apparent outward
shift of the conductance peaks (from about 3 meV to 5 meV in
Fig.\ref{fig:szabo_field}) that then start to shrink, because of the
suppression of the $\sigma$-band gap. Actually, the use of
polycrystals made it impossible to control the direction of both the
probe current and the magnetic field. This is not irrelevant because
of the anisotropy of the critical fields in MgB$_2$
\cite{sologubenko02,welp02,angst02}. Indeed, PCAR measurements in
single crystals \cite{gonnelli02c,daghero03,gonnelli04a} showed
that: i) A field of about 1 T ``completely'' suppresses the small
gap irrespective of the field direction. This does not mean that the
$\pi$ band becomes nonsuperconducting, but simply that above 1 T its
contribution to the Andreev signal becomes experimentally
undetectable and the conductance curves can be fitted to a function
like $G=w_{\pi}1 + (1-w_{\pi})G_{\sigma}$ (where $G_{\pi}=1$).
Incidentally, this also confirms that the $\pi$ band is rather
isotropic; ii) the direction of the field instead affects the
behavior of $\Delta_{\sigma}$, which is reasonable due to the
almost-2D character of this band. When $\mathbf{B}
\parallel ab$, the $\Delta_{\sigma}$ peaks in the conductance remain clearly distinguishable
up to 9 T, with only some signs of gap closing. Instead, when
$\mathbf{B} \parallel c$, they merge together at $\mathbf{B} \geq 4$
T giving rise to a broad maximum \cite{gonnelli04a}; iii) In any
case, at least at 4.2 K, $\Delta_{\sigma}$ is very little affected
by a field of 1 T, either parallel or perpendicular to the ab plane
\cite{gonnelli02c}; iv) in $c$-axis contacts, the suppression of the
$\pi$-band contribution to the conductance at about 1T is
accompanied by an outward shift of the conductance peaks and by an
abrupt decrease in the amplitude of the spectrum.

A quantitative study of the effect of the field on the gaps requires
a fit of the experimental curves. Here the main problem is: can the
BTK model or its generalized version be used to fit the conductance
curves when a magnetic field is present? In conventional
superconductors, Naidyuk \emph{et al.} \cite{naidyuk96} showed that
the pair-breaking effect of the field can be mimicked, within a
generalized BTK model, by the broadening parameter $\Gamma$. In
other words, the total broadening parameter $\Gamma$ can be
considered as the sum of an intrinsic (field-independent) $\Gamma_i$
(due to self-energy and inelastic scattering effects, see
Sect.\ref{sect:broadening}) and an extrinsic $\Gamma_f(B)$ due to
the magnetic field. This approach assumes that the pair-breaking
effect of the field can be completely represented by the broadening
$\Gamma_f(B)$ while its effect on the DOS is negligible in a
first-order approximation.

Unfortunately, it is not really so. At $T=0$, the theoretical
effects on the DOS curves of both the magnetic field and the actual
shape of the FS are very relevant \cite{dahm02,dahm04a,koshelev03}
so that the BTK model (which in its simplest formulation assumes a
spherical FS) even if modified to account for lifetime effects and
even if a high $Z$ is used (tunnel regime), fails in fitting them.
One could expect something similar to occur in the AR regime.
However, i) the inadequacy of the BTK model is dramatic only at
$T=0$, while at low but finite temperatures (T=4 K) the thermally
broadened theoretical DOS curves become much more similar to those
given by the standard BTK-lifetime model; ii) as we will show in the
following, the model seems to fit much better the experimental
curves than the calculated ones, giving results in good agreement
with theory. As a matter of fact, the attempts to use the BTK model
to obtain the field dependence of the gaps from the conductance
curves not only have given good results, but have been able to
extract quantitative information about the \emph{diffusivities} in
the two bands \cite{gonnelli04a,bugoslavsky04,naidyuk05b}. This
information is crucial since, as shown in
Refs.\cite{koshelev03,dahm03,dahm04a,gurevich03}, varying the ratio
of electron diffusivities in the two bands will change the resulting
macroscopic superconducting properties. Incidentally, the
possibility to obtain information also on the \emph{interband
coupling} by means of PCAR measurements in the presence of a
supercurrent parallel to the interface has been recently proposed
\cite{Lukic07}.

\begin{figure}[ht]
\begin{center}
%\vspace{-6mm}
\includegraphics[width=0.9\columnwidth]{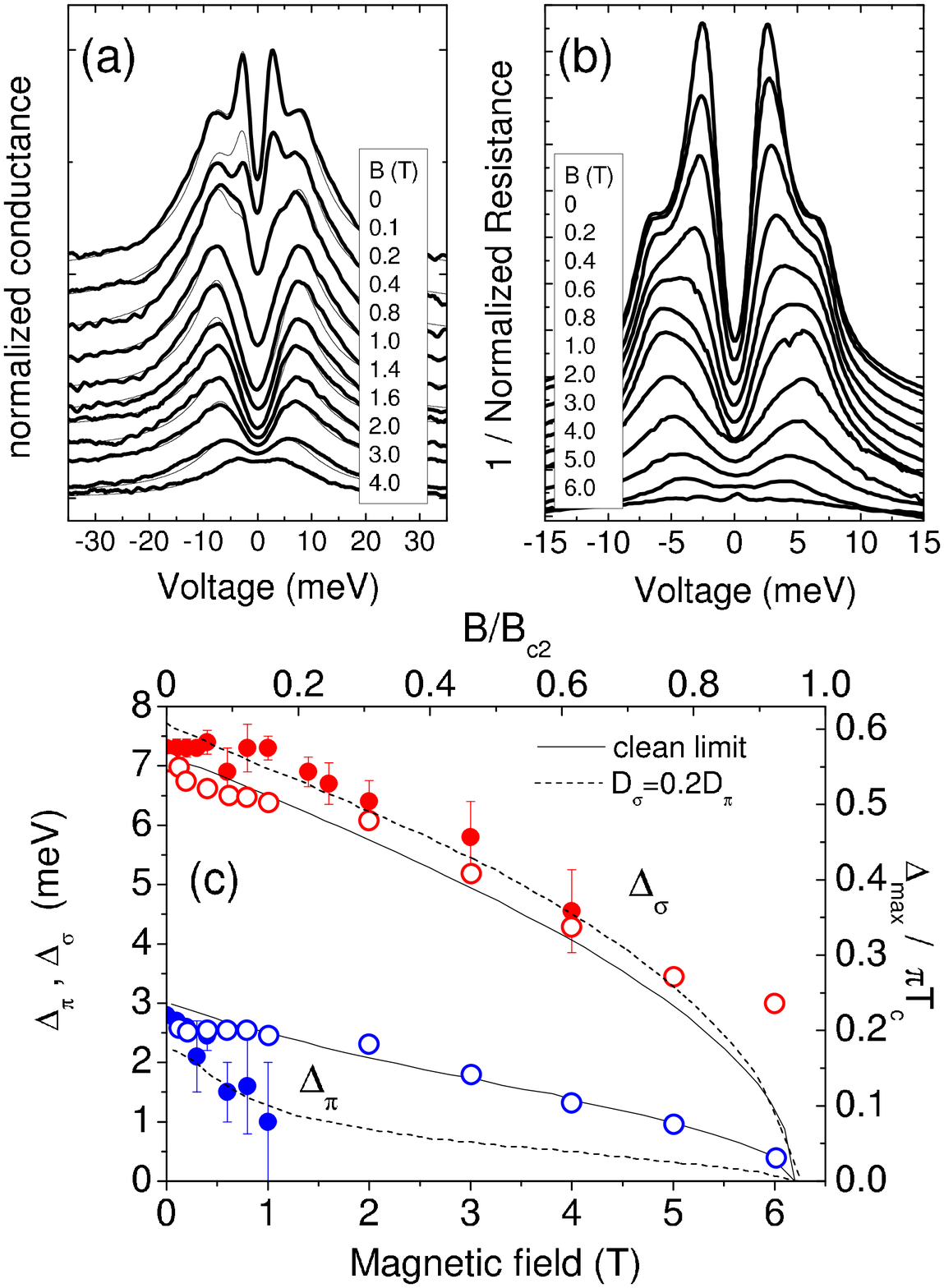}
\end{center}
\caption{(a,b) Two examples of magnetic-field dependence of
$ab$-plane point-contact spectra in MgB$_2$ single crystals. Thick
lines are experimental spectra, thin lines in (a) are the relevant
two-band BTK fit. In (a) the ``soft'' PCAR technique was used, with
a Ag-paste contact (from Ref.\cite{gonnelli04c}). In (b) the contact
was made by pressing a Cu tip against the sample edge (from
Ref.\cite{naidyuk05b}). (c) Magnetic field dependence of the gaps
extracted from the fit of the conductance curves in (a) (solid
circles) and (b) (open circles). The values of the gaps are compared
to theoretical predictions in the clean limit (solid lines)
\cite{dahm04a} and in the dirty limit \cite{koshelev03} in the
particular case $D_{\sigma}=0.2 D_{\pi}$. The scale for the dashed
lines is on the top and left axes; the curves for the clean limit
have been re-scaled vertically and horizontally using the actual
$T_c$ of the samples and the relevant critical
field.}\label{fig:conductance_fieldMgB2}
\end{figure}

Figure \ref{fig:conductance_fieldMgB2} shows the PCAR spectra of
$ab$-plane contacts on MgB$_2$ single crystals, in magnetic fields
parallel to the $c$ axis. The curves of panel (a) were obtained in a
``soft'' point contact \cite{gonnelli04c}, while the curves of panel
(b) were measured by Naidyuk \emph{et al.} with a Cu tip
\cite{naidyuk05b}. In both panels, the experimental curves (thick
lines) are compared to the two-band generalized BTK fit (thin
lines). In panel (a), the fit was carried out with the two-band BTK
model up to about 1 T, while above this field the $\pi$-band
features became undetectable so that we took $\sigma_{\pi}=1$, in
agreement with our previous findings in single crystals
\cite{daghero03,gonnelli04a} and with those by Szab\'{o} \emph{et
al}. In panel (b), instead, the fit was performed with both the
$\sigma$ and the $\pi$ contributions up to the highest field. A
similar result was also obtained by Bugoslavsky \emph{et al.} in
epitaxial thin films of MgB$_2$, where the $\pi$-band gap was found
to survive up to 5 T \cite{bugoslavsky04}.

The values of the gaps extracted from the fit are shown in
Fig.\ref{fig:conductance_fieldMgB2}(c) as solid and open symbols.
The data are compared to the predictions by Dahm \emph{et al.}
\cite{dahm04a} in the clean limit (solid lines) as well as to those
by Koshelev and Golubov \cite{koshelev03} in the dirty limit (dashed
lines) in the particular case where the diffusivities in the two
bands are $D_{\sigma}=0.2 D_{\pi}$, which is the case that allows
best fitting the data in panel (a). The fit of the experimental PCAR
curves seems to work very well and the resulting values of
$\Delta_{\sigma}$ are perfectly compatible with the expected field
dependence. In all cases \cite{bugoslavsky04,gonnelli04c} the
broadening parameters of the BTK model, $\Gamma_{\sigma}$ and
$\Gamma_{\pi}$ increase linearly with field, giving further support
to the distinction between intrinsic lifetime broadening and
field-induced broadening (proportional to B).

According to Fig.\ref{fig:conductance_fieldMgB2}(c), a fast
suppression of the $\pi$-band gap features in weak fields
\cite{szabo01,szabo03a,szabo03b,gonnelli02c,daghero03} indicates
that the $\pi$-band diffusivity of the samples under study is a few
times greater than the $\sigma$-band one. Since the theoretical
field dependence of the gaps in clean limit \cite{dahm04a} is
identical to that predicted by the dirty-limit model in the case
$D_{\sigma}=D_{\pi}$ (equal diffusivities) \cite{koshelev03}, the
difference between our results (a) and those by Naidyuk (b) or
Bugoslavsky \cite{bugoslavsky04} are simply due to sample-dependent
variations in the diffusivity ratio $D_{\sigma}/D_{\pi}$, that
occurs also in different crystals from the same batch
\cite{gonnelli04c}.

\begin{figure}[hb]
\begin{center}
%\vspace{-6mm}
\includegraphics[width=0.9\columnwidth]{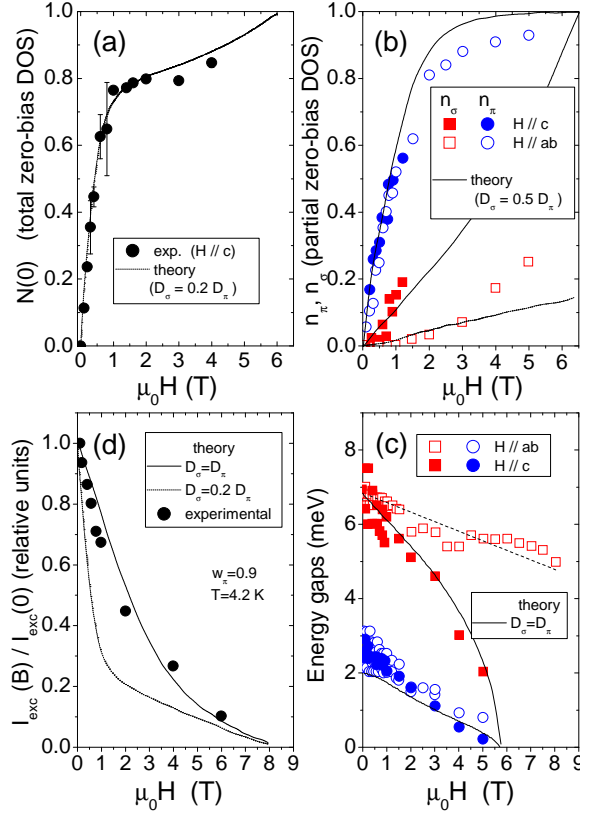}
\end{center}
\caption{(a,b) Magnetic field dependence of the zero-bias density of
states evaluated from PCAR. (a) The total ZBD obtained by simulating
the zero-temperature tunneling conductance curves (from
Ref.\cite{gonnelli04c}) (symbols) compared to the theoretical
prediction in the dirty limit for $D_{\sigma}=0.2 D_{\pi}$. The
perfect agreement confirms the result given by the field-dependence
of the gaps in fig.\ref{fig:conductance_fieldMgB2}(c). (b) The
partial ZBD $n_{\pi}$ and $n_{\sigma}$ obtained from the fit of the
conductance curves with the model by Bugoslavsky \emph{et al.}
\cite{bugoslavsky05}, compared to the theoretical predictions in the
case $D_{\sigma}=0.5 D_{\pi}$ (solid lines). The dashed line is an
estimate of the expected behavior of $N_{\sigma}$ for the case
$\mathrm{H} \parallel ab$. The results agree with the gap
measurements in the same samples (shown in (c)) with the persistence
of the small gap up to 5 T. (d) Magnetic field dependence of the
excess current $I_{exc}$ (from Ref.\cite{naidyuk05b}). Symbols:
experimental values of $I_{exc}$ from integration of the conductance
curves in Fig.\ref{fig:conductance_fieldMgB2}(b). Lines:
$I_{exc}(B)$ calculated from eq. \ref{eq:naidyuk} based on the
theoretical predictions of
ref.\cite{koshelev03}.}\label{fig:DOS_field_MgB2}
\end{figure}

A check of internal consistency of the results described above was
achieved by studying the partial averaged zero-bias density of
states (ZBD) in the two bands, $N_{\sigma}(0)$ and $N_{\pi}(0)$,
whose magnetic-field dependence is predicted to depend again on the
diffusivity ratio \cite{koshelev03}. In Ref.\cite{gonnelli04c} we
used the gaps $\Delta_{\sigma}$ and $\Delta_{\pi}$, the weight
$w_{\pi}$ and the field-induced broadening parameters
$\Gamma_{f}^{\sigma,\pi}$ obtained from the fit of the PCAR spectra
in Fig.\ref{fig:conductance_fieldMgB2}(a) to calculate the
\emph{zero-temperature ideal tunneling conductance} by setting
$Z_{\sigma,\pi}=20$, $T=0$ and $\Gamma_i^{\sigma,\pi}=0$. The values
of the total ZBD as a function of the magnetic field are shown in
Fig.\ref{fig:DOS_field_MgB2}(a) (symbols), and compared to the
theoretical total ZBD $N=w_{\pi}N_{\pi}(0)+(1-w_{\pi})N_{\sigma}(0)$
were $N_{\pi}(0)$ and $N_{\sigma}(0)$ are those calculated in
\cite{koshelev03} suitably scaled to the actual critical field
$B_{c2\parallel c}=6.25$ T. It is clear that the ZBD follows the
theoretical predictions for the case $D_{\sigma}=0.2 D_{\pi}$, in
perfect agreement with the conclusions drawn from the field
dependence of the gaps.

Bugoslavsky \emph{et al.} \cite{bugoslavsky05} were able to directly
extract the partial ZBD from the fit of their PCAR spectra in
epitaxial MgB$_2$ films with a suitable model developed and tested
in conventional type-II superconductors \cite{miyoshi05}. The model
is based on the existence of parallel contacts and on the fact that
a fraction of them (increasing with field) occurs in normal-state
regions of the sample (vortex cores) \cite{bugoslavsky05}. Hence,
what PCAR measures is an effective average over the vortex lattice
and the conductance of the point contact should be considered as
being the sum of a ``normal channel'' and of a ``superconducting
channel'' contributions. In MgB$_2$, the normalized conductance in
magnetic field thus becomes
\begin{eqnarray}
G(V)&=&w_{\pi}[n_{\pi}+ (1-n_{\pi})G_{\pi}]+ \nonumber
\\
& &+(1-w_{\pi})[n_{\sigma}+
(1-n_{\sigma})G_{\sigma}]\label{eq:bugoslavsky}
\end{eqnarray}
where $n_{\pi}$ and $n_{\sigma}$ represent the fraction of
normal-state core excitation and were thus identified
\cite{bugoslavsky05} with the partial ZBD $N_{\sigma}$ and
$N_{\pi}$. To reduce the number of free fitting parameters, the
authors assumed $Z_{\sigma}=Z_{\pi}$ and used the convolution with a
Gaussian of width $\omega$ to account for both the thermal smearing
and inelastic interface scattering. $Z$ and $w_{\pi}$ were fixed to
their zero-field values. The values of $n_{\sigma}$ and $n_{\pi}$
obtained from the fit of different series of conductance curves, in
magnetic fields either parallel or perpendicular to the $c$ axis,
are reported in Fig. \ref{fig:DOS_field_MgB2}(b). The data in
$B\parallel c$ approximately agree with the theoretical curves
(solid lines) for $D_{\sigma}=0.5D_{\pi}$. This is consistent with
the field dependence of the gaps shown in (c), that shows the
persistence of the small gap up to 5 T \cite{bugoslavsky04}. Note
that the field dependence of $\Delta_{\pi}$ and $n_{\pi}$ does not
depend on the field direction (indicating that the $\pi$ band
diffusivity is isotropic) while the $\sigma$-band quantities show a
marked anisotropy.

Another interesting way to determine the diffusivity ratio
$D_{\sigma}/D_{\pi}$ from a PCAR experiment is described in
Ref.\cite{naidyuk05b}, where the excess current $I_{exc}$ --
obtained by integration of the reduced $dI/dV$ after subtraction of
the background -- is directly plotted as a function of the magnetic
field. The result is reported in Fig.\ref{fig:DOS_field_MgB2}(d) for
the curves in Fig.\ref{fig:conductance_fieldMgB2}(b). Let us just
recall here for convenience that the excess current is approximately
$I_{exc}\simeq \Delta/eR_{N}$; when a magnetic field is applied, a
fraction $N(0, B)$ of the contact becomes nonsuperconducting (vortex
cores) and does not contribute any longer to $I_{exc}$. Taking into
account the presence of two bands (whose partial ZBD behave
differently in field) Naidyuk \emph{et al.} arrived to the
expression
\begin{eqnarray}
I_{exc}(B) &\propto &w_{\pi}\Delta_{\pi}(B)[1-N_{\pi}(0,B)] + \nonumber \\
& & + (1-w_{\pi})\Delta_{\sigma}(B)[1-N_{\sigma}(0,B)]
\label{eq:naidyuk}
\end{eqnarray}
that was used to fit the experimental data. In this function, the
gaps vales $\Delta_{\pi}(B)$ and $\Delta_{\sigma}(B)$ as well as the
ZBD $N_{\pi}(0,B)$ and $N_{\sigma}(0,B)$ are taken from the
theoretical curves of Ref.\cite{koshelev03} suitably scaled to the
actual critical field. The zero-field values of the gaps and the
weight were obtained from the fit of the zero-field Andreev spectra.
Fig.\ref{fig:DOS_field_MgB2}(d) shows that the experimental values
of $I_{exc}$ are in good agreement with the theoretical predictions
in the case $D_{\sigma}=D_{\pi}$, which further confirms the
conclusion drawn from the magnetic-field dependence of the gaps in
Fig.\ref{fig:conductance_fieldMgB2}(c). It is worth mentioning that
both in Ref.\cite{gonnelli04c} and \cite{naidyuk05b}, different
contacts resulted in different values of the diffusivity ratio and
the curves shown here are just an example.

% separazione delle conduttanze
The low-temperature magnetic-field dependence of the gaps in single
crystals shown in Fig.\ref{fig:conductance_fieldMgB2}(c) indicates
that, in these samples, a field of 1 T makes the PCAR spectra look
\emph{as if} the small gap was completely suppressed, but does not
seriously affect the $\sigma$-band gap. A fit of the spectra with a
function like $G(B=1T)=w_{\pi}1+(1-w_{\pi})G_{\sigma}(B=1T)$ is thus
possible and gives a gap $\Delta_{\sigma}$ which coincides with the
zero-field one -- but has a smaller uncertainty because of the
reduced number of fitting parameters.  If now one subtracts the
experimental normalized spectrum $G_{exp}(B=1T)$ from the zero-field
one, $G_{exp}(B=0)$, one obtains a curve that only contains the
\emph{zero-field} $\pi$-band contribution and can thus be fitted by
a 3-parameters function like
$G(B=0)-G(B=1T)=w_{\pi}[G_{\pi}(B=0)-1]$ from which the small gap
$\Delta_{\pi}$ can be obtained with a small uncertainty. In this
fitting process, all the parameters must be adjusted to ensure a
consistency between the different fits. The final result, as shown
in Ref.\cite{gonnelli02c}, is a rather strict determination of the
gaps, that turn out to be $\Delta_{\pi}=2.80 \pm 0.05 $ meV and
$\Delta_{\sigma}=7.1 \pm 0.1$ meV, in excellent agreement with
theoretical predictions of Ref.\cite{brinkman02}.

\begin{figure}[h]
\begin{center}
%\vspace{-6mm}
\includegraphics[width=0.9\columnwidth]{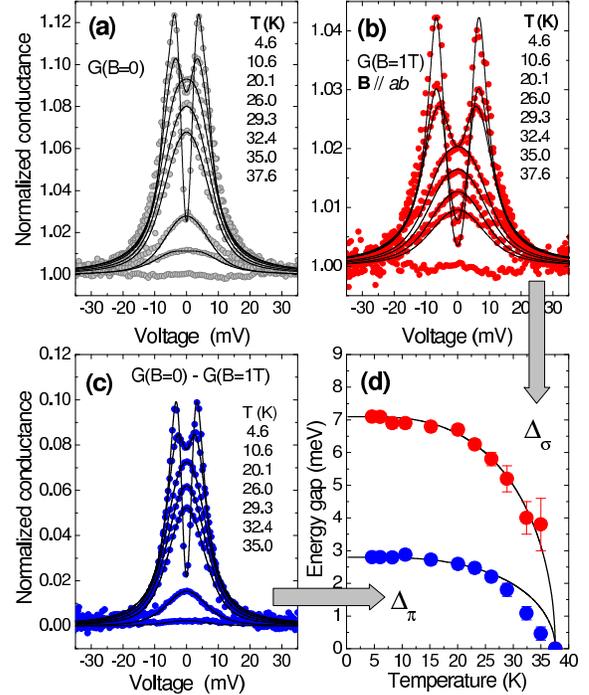}
\end{center}
\vspace{-1.5cm}\caption{(a) Temperature dependence of the zero-field
conductance curves $G(B=0)$ of a $c$-axis contact on a MgB$_2$
single crystal (symbols) and the relevant two-gap BTK fit. (b) Same
as in (a), but in a field of 1 T applied parallel to the $ab$ plane.
The $G(B=1T)$ curves are compared to the relevant fit with a
single-gap BTK model containing only the $\sigma$-band contribution
to the Andreev conductance. (c) Temperature dependence of the
difference $G(B=0)-G(B=1T)$, compared to the relevant single-gap BTK
fit that contains only the $\pi$-band Andreev reflection
conductance. (d) Temperature dependence of the gaps
$\Delta_{\sigma}$ and $\Delta_{\pi}$ extracted from the single-gap
BTK fits of the partial $\sigma$ and $\pi$-band conductances (panels
(b) and (c), respectively). Data taken from Refs. \cite{daghero03}
and \cite{gonnelli03a}.}\label{fig:field_ab}
\end{figure}

The process works well at 4.2 K, and can be extended to higher
temperatures with some caution, because of temperature-dependent
anisotropy of the critical fields in MgB$_2$
\cite{sologubenko02,welp02,angst02}. In particular, the field must
be parallel to the $ab$ plane \cite{daghero03,gonnelli04a}; in this
case the single-band BTK fit of the PCAR spectra in a field of 1 T
(see Fig.\ref{fig:field_ab}(b)) gives values of $\Delta_{\sigma}$
that agree very well with those determined by the two-band fit of
the zero-field curves (Fig.\ref{fig:field_ab}(a)) apart from a much
smaller uncertainty. The fit of the difference $G_{exp}(B=0) -
G_{exp}(B=1T)$ (Fig.\ref{fig:field_ab}(c)) gives very good results
for the small gap $\Delta_{\pi}$ as well. The resulting temperature
dependence of the gaps \cite{gonnelli02c} is shown in
Fig.\ref{fig:field_ab}(d).

In Refs. \cite{gonnelli04a} we also determined by means of PCAR the
temperature dependence of the critical field $B_{c2 \parallel c}$,
and $B_{c2 \parallel ab}$, identified with the field that suppresses
superconductivity and restores the normal-state conductance. For
$T>0.8 T_c$, the critical fields measured by PCAR fall on the curves
given by bulk measurements like thermal conductivity
\cite{sologubenko02}, torque magnetometry \cite{angst02} and
specific heat \cite{welp02}. At lower temperature, they depart from
these curves and tend to the (larger) values given by resistivity
measurements \cite{welp02,sologubenko02}. Since the latter rather
determine the surface critical field $B_{c3}$ \cite{welp02}, some
surface effects clearly play a role in PCAR (especially at low
temperature).  Even if PCS is a surface-sensitive technique, this
effect cannot simply be a surface nucleation of superconductivity at
a field $B_0$ ($B_{c2} < B_0 < B_{c3}$) because the
magnetoresistivity of the electrode is always much lower than that
of the MgB$_2$ crystals.

\subsubsection{Determination of the electron-phonon spectral function by
PCS}\label{sect:MgB2_phonons} Point contact spectroscopy in the
normal and superconducting state of MgB$_2$ was also used to obtain
the electron-phonon spectral function and elucidate the role of the
in-plane B stretching mode $E_{2g}$ in determining the
superconducting properties of this compound. In ref.
\cite{yanson03}, Yanson \emph{et al.} investigated the point-contact
spectra in $c$-axis oriented films, in the superconducting ($T<
T_c$) and in the normal ($T>T_c$) state. They directly measured the
differential resistance ($dV/dI$) as well as the second derivative
of the I-V curves, i.e. $d^2V/dI^2$, which is proportional to the
electron-phonon spectral function $\alpha_{PC}^2F(\omega)$ (see
Sect.\ref{sect:ballistic}). The authors observed clear structures in
the superconducting state, with a signature of the $E_{2g}$ mode of
the same amplitude as other phonon peaks. Owing to the preferential
current injection along the $c$ axis, this is in agreement with the
calculated $\alpha^2F_{\pi}(\omega)$ \cite{golubov02}. However,
these structures were found to disappear at $T>T_c$ where only much
smaller nonlinearities persisted. This indicates a superconducting
origin of the structures, i.e. due to the energy-dependence of the
order parameter (``elastic'' or self-energy term) rather than to the
actual inelastic e-ph scattering. This point was also addressed
theoretically in Ref.\cite{yanson04c} where a simple asymptotic
formula for the order parameter self-energy effects in the
superconducting point contact was derived. Later, the same authors
performed PCS measurements in single crystals in an inverse
needle-anvil configuration (superconducting crystal as the needle)
\cite{naidyuk03}. The smaller critical field of crystals with
respect to films allowed at least a partial suppression of
superconductivity at 4.2K by means of fields as high as 9 T. In
low-temperature, zero-field spectra with a major contribution from
$ab$-plane current injection, the second derivative showed a very
broad maximum around 60-70 meV that was almost insensitive to
magnetic field and was identified with the signature of the $E_{2g}$
phonon mode, largely smeared by the e-ph coupling, as observed by
X-ray inelastic scattering \cite{shukla03}. The $d^2V/dI^2$ spectra
of contacts with a predominant $c$-axis tunneling contribution
showed instead much weaker structures such as shallow maxima at 30
and 50 mV that were claimed to reflect bulk (isotropic) phonons, and
were put in connection with  the first two maxima in the phonon
density of states or $\alpha^2F(\omega)$ \cite{golubov02}. Further
details on this subject can be found in refs. \cite{yanson04a} and
\cite{yanson04b}.

\subsubsection{Effect of chemical doping on the gaps of MgB$_2$}
\begin{figure}[t]
\begin{center}
%\vspace{-6mm}
\includegraphics[width=0.8\columnwidth]{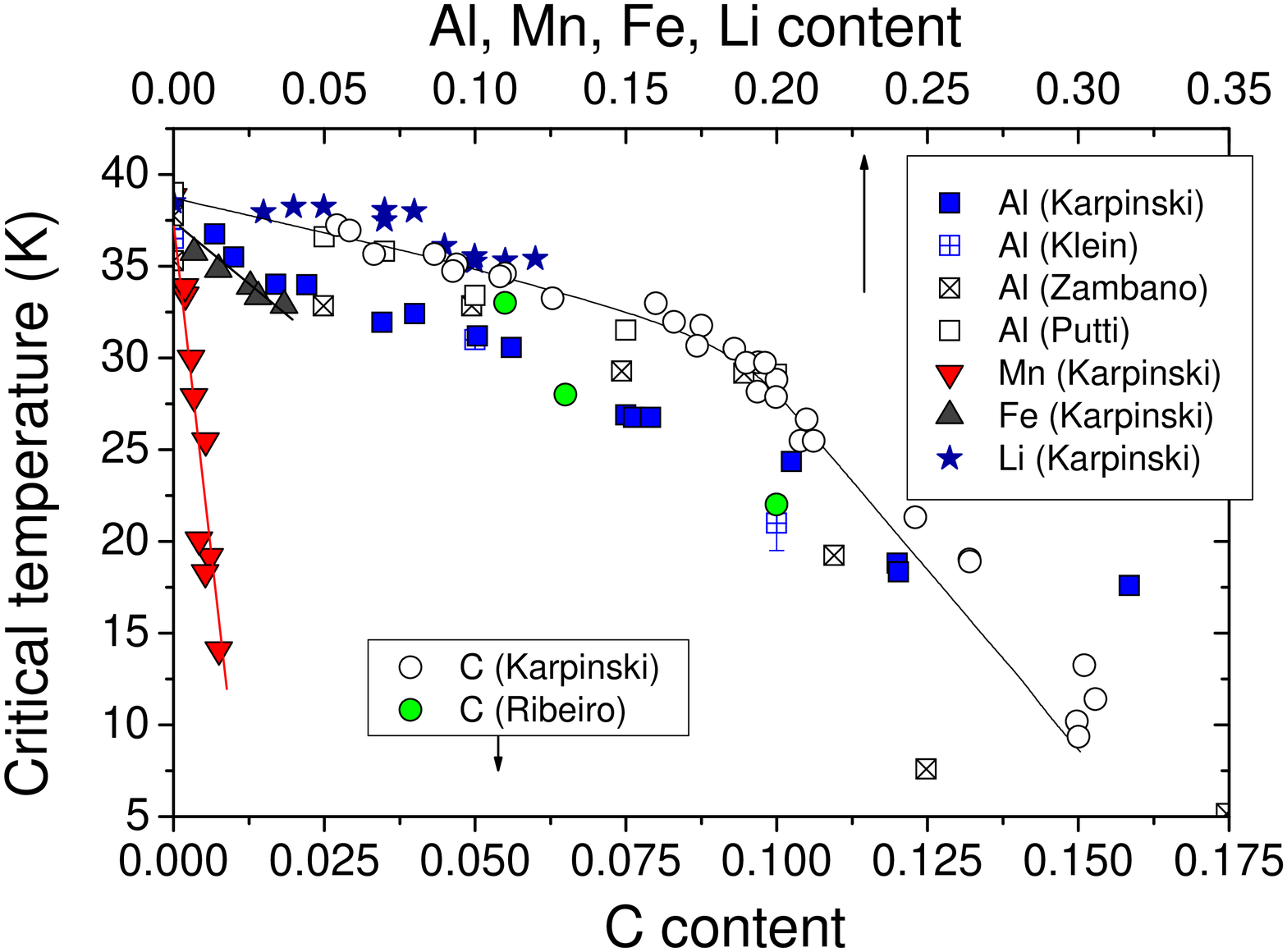}
\end{center}
\caption{Bulk critical temperature vs. doping content in various
series of substituted MgB$_2$. The doping content for substitutions
in the Mg site (general formula Mg$_{1-x}$M$_x$B$_2$) must be read
on the top axis. The content of carbon in Mg(B$_{1-x}$C$_x$)$_2$ is
on the bottom axis. The two scales differ by a factor 2 to show that
all the curves for non-magnetic doping look very similar if plotted
as a function of the number of substitutional atoms per formula
unit. The data are taken from Refs. \cite{ribeiro03} (C, Ribeiro),
\cite{klein06} (Al, Klein), \cite{zambano05} (Al, Zambano),
\cite{putti05} (Al, Putti), \cite{karpinski07} (Al, Mn, Fe, Li, C,
Karpinski).} \label{fig:Tc_doping}
\end{figure}

Chemical substitutions in MgB$_2$ were tried very soon after the
discovery of superconductivity in this compound \cite{cava03}. The
huge experimental work carried out in substituted MgB$_2$ has
allowed a deeper understanding of the pure compound, but has also
unveiled a surprisingly rich an complex physics. Even in the
simplest effective two-band model \cite{liu01,brinkman02} the
quantities needed to describe MgB$_2$ are manifold: \emph{four}
Eliasberg functions $\alpha^2 F_{i,j}(\omega)$ (where $i,j=\sigma,
\pi$) \cite{dolgov03}, \emph{four} quasiparticle scattering rates
(intraband, $\Gamma_{ii}$, and interband, $\Gamma_{ij}$) \emph{two}
densities of states (DOSs), and a prefactor $\mu_{0}$ to the Coulomb
pseudopotential (which is a 2x2 matrix whose elements only contain
the densities of states $N_{\sigma}$ and $N_{\pi}$). Not all these
parameters are independent (for example $\Gamma_{\pi \sigma}=
(N_{\pi}/N_{\pi})\Gamma_{\sigma \pi}$, being $N_{\sigma,\pi}$ the
zero-bias DOS in the two bands) and some of them (like the DOS, the
phonon spectrum) can be either calculated from first principles or
determined experimentally. As we will see in the following, chemical
doping in MgB$_2$ always gives a decrease in $T_c$ (see
Fig.\ref{fig:Tc_doping}) and in $\Delta_{\sigma}$, that can be due
either to a variation in the DOS (and in the phonon frequencies,
which however play a minor role) or to an increase in interband
scattering. Fortunately, the effects of the latter are rather easily
distinguishable from those of other sources of $T_{c}$ reduction,
since an increase in $\Gamma_{\sigma \pi}$ suppresses $T_{c}$ and
$\Delta_{\sigma}$ but increases $\Delta_{\pi}$. As a consequence,
some indications about the effects of doping on the DOS and on the
interband scattering can be extracted from the analysis of the
doping dependence of the gaps, measured by PCAR, within the two-band
Eliashberg theory. Indications about the relative role of
\emph{intraband} scattering rates in the two bands can instead be
obtained from the magnetic-field dependence of the ZBD or the excess
current, as shown in Sect.\ref{sect:magneticfield}. A detailed
review about PCAR measurements in doped MgB$_2$ is reported in
Ref.\cite{gonnelli07}; in the following we will thus give only a
general discussion of the main findings.

In general, a problem with PCAR in doped MgB$_2$ is that the
structures related to the large gap $\Delta_{\sigma}$ become less
and less clear on increasing the dopant content. The comparison of
the two-band fit to the single-band one can clarify whether two gaps
are still present or not, but sometimes the magnetic-field
dependence of the conductance curves can be more conclusive in this
sense: an outward shift of the conductance peaks is a strong
indication in favor of two gaps. Another problem is the
determination of the actual doping content, especially in the case
of light atomic species (Li, C), which makes the trend of the gaps
vs. the doping content rather uncertain. Moreover, the doping
content is intrinsically inhomogeneous on the scale probed by PCAR
so that different contacts on the same sample can even show
different gaps and different $T_c^A$. For these reasons, we always
prefer to report the values of the gaps as a function of the local
critical temperature $T_c^A$. This representation is also the most
suited to compare the results for different substitutions. For
example, one can learn that there is an interesting universal
scaling law of $\Delta_{\sigma}$ with $T_c^A$, at least for $T_c^A >
20 $ K, independently of the main mechanism of $T_c$ reduction.

%%%%%%%%%%%%%% Carbonio %%%%%%%%%%%%%%%%%%%%%%%%%%%%%%%%

Carbon is the only chemical substitution in the site of boron that
the structure of MgB$_2$ accepts. PCAR experiments in
Mg(B$_{1-x}$C$_x$)$_2$ were carried out in nearly single-phase
polycrystals with $0.09\leq x \leq 0.13$ \cite{ribeiro03,avdeev03}
and in single crystals grown at high pressure and high temperature
\cite{kazakov05} with $0.047\leq x \leq 0.132$. The critical
temperature \emph{decreases} on C doping, as shown in
Fig.\ref{fig:Tc_doping}, although the difficulty in determining the
actual C content gives rise to some minor differences in the actual
$T_c$ vs. $x$ curve for crystals (black circles) and polycrystals
(green circles).

\begin{figure}[ht]
\begin{center}
%\vspace{-6mm}
\includegraphics[width=\columnwidth]{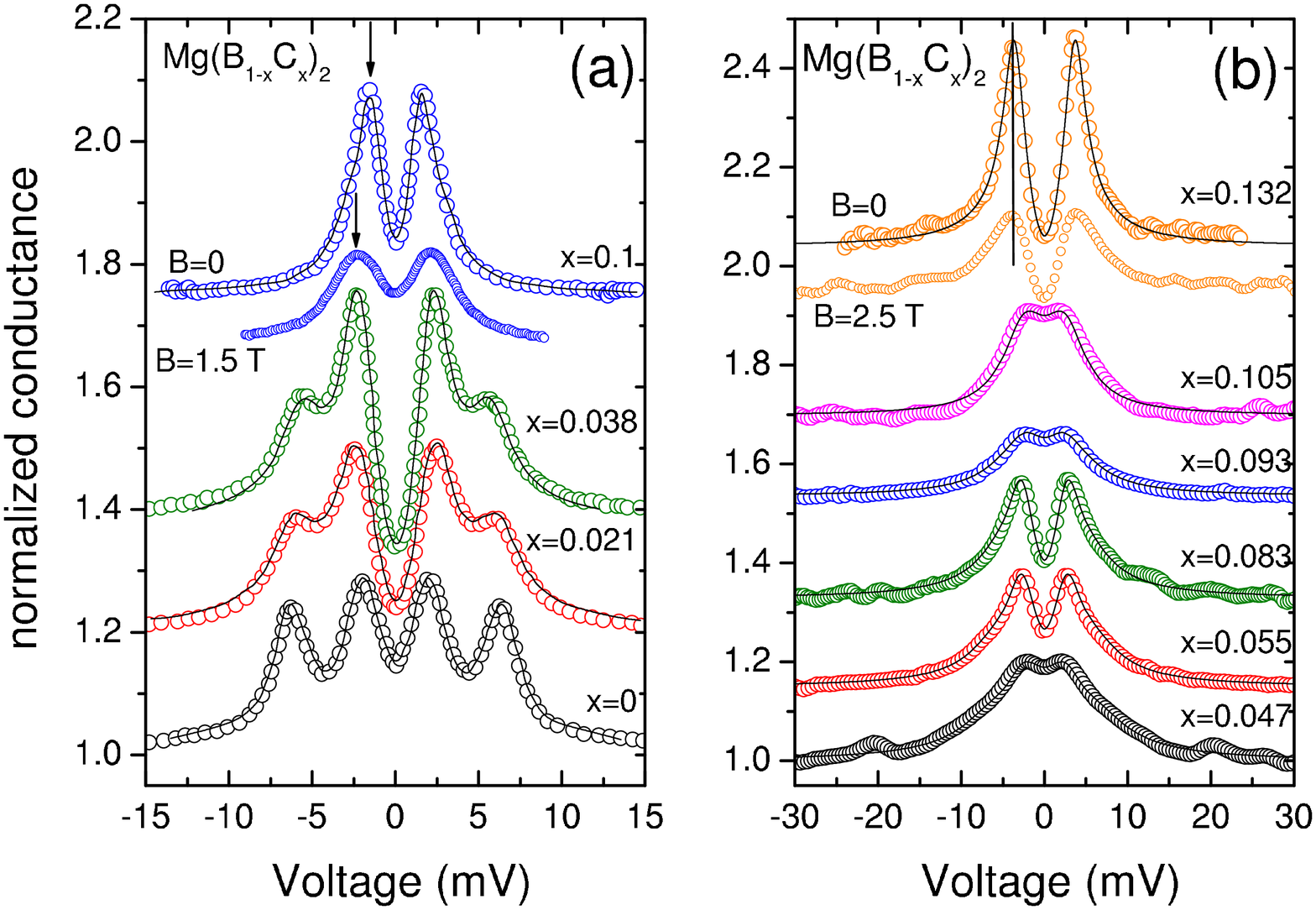}
\end{center}
\caption{Doping dependence of PCAR spectra in C-doped MgB$_2$
polycrystals (a) and single crystals (b). The point contacts were
made with a metallic tip in (a) and with a Ag-paste spot in (b).
Symbols represent experimental curves, solid lines the relevant
two-band BTK fit, apart from the top curve in (b) that represents
instead the single-band BTK fit. Data are taken from Ref.
\cite{holanova04a} (a) and \cite{gonnelli05} (b).
}\label{fig:Cdoping_conductance}
\end{figure}

Fig. \ref{fig:Cdoping_conductance} shows some examples of normalized
PCAR curves and the relevant fit at different C contents, in
polycrystals and wires \cite{holanova04a} and in single crystals
\cite{gonnelli05}. Already at $x=0.1$, the experimental curves seem
to be fittable by a single-band BTK model with lifetime broadening.
Actually, at this C content the second gap is still retained
\cite{samuely03a} as it can be easily shown by applying a magnetic
field: as in pure MgB$_2$, the small gap is fast suppressed and the
large-gap features emerge clearly \cite{holanova04a,gonnelli05}. As
shown in Fig.\ref{fig:Cdoping_conductance}(b), in single crystals
with the highest doping content ($x=0.132$) the application of the
magnetic field makes the conductance peaks decrease in amplitude but
no shift in energy is observed as long as the applied field is much
smaller than the critical field (this is certainly true for
$\mathrm{H}\parallel ab$ \cite{gonnelli05} because C doping
increases the critical field). The absence of a shift indicates
that, if two gaps are present, they have very similar amplitude --
indeed, the two-band BTK fit of all the conductance curves at
$x=0.132$ requires two gap values that are experimentally
indistinguishable. The single-band BTK fit shown in
Fig.\ref{fig:Cdoping_conductance}(b) gives a gap $\Delta=2.8 \pm
0.2$ meV. One could thus conclude that the ``gap merging'' is
obtained in heavily C-doped MgB$_2$, also relying on the fact that:
i) the ratio $2\Delta/k_B T_c= 3.8$ is close to the BCS value; ii)
the conductance curves recorded in this sample at different
temperatures all admit a very good single-band BTK fit; iii) the
temperature dependence of the gap $\Delta$ extracted from the fit is
perfectly BCS (within the experimental uncertainty) and the critical
temperature of the junction, $T_{c}^{A}=19$ K is in perfect
agreement with the bulk $T_{c}$ measured by DC zero-field-cooling
magnetization.

\begin{figure}[ht]
\begin{center}
%\vspace{-6mm}
\includegraphics[width=0.8\columnwidth]{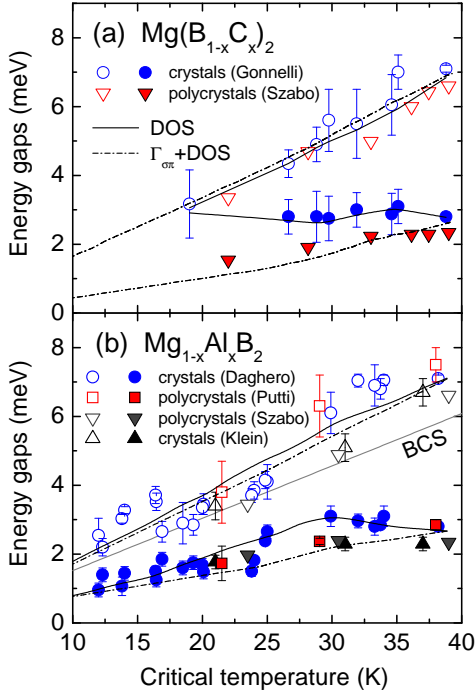}
\end{center}
\caption{(a) Energy gaps in Mg(B$_{1-x}$C$_x$)$_2$ as a function of
the critical temperature, measured in single crystals
\cite{gonnelli05} (circles) and polycrystals and wires
\cite{holanova04a,holanova04b,szabo07} (triangles). Each point of
the single-crystal series is actually the average of different gap
values measured in different contacts, whose spread is indicated by
the error bar. (b) Energy gaps measured by PCAR in
Mg$_{1-x}$Al$_x$B$_2$ as a function of the critical temperature.
Circles are taken from \cite{daghero08}, squares from
\cite{putti05}, down triangles from \cite{szabo07} and up triangles
from \cite{klein06}. In both (a) and (b), dash-dot lines represent
the gap values calculated within the two-band Eliashberg theory by
using the DOS and the phonon frequencies from ab-initio
calculations. Solid lines indicate the fit of the gaps vs. $T_c$
obtained by adding to the model an adjustable interband scattering
rate $\Gamma_{\sigma \pi}$. }\label{fig:gaps_C_Al_doping}
\end{figure}

However, the situation is not so simple. Fig.
\ref{fig:gaps_C_Al_doping}(a) reports the values of the gaps
obtained in C-doped crystals \cite{gonnelli05} and polycrystals
\cite{holanova04a,holanova04b,szabo07} as a function of the critical
temperature. It is clear that the two data sets agree rather well as
far as $\Delta_{\sigma}$ is concerned, but disagree on the values
and trend of $\Delta_{\pi}$. In particular, in wires and bulk
polycrystalline samples there is no tendency to the gap merging
observed instead in single crystals. To analyze the data within the
two-band Eliashberg model one can use the $\sigma$- and $\pi$-band
DOS and the phonon frequencies calculated for C-doped MgB$_2$
\cite{ummarino05a} thus leaving $\mu_0$ and $\Gamma_{\sigma \pi}$ as
the only adjustable parameters. The overall trend of the gaps in
C-doped polycrystals is reproduced by keeping $\Gamma_{\sigma
\pi}=0$ as in pure MgB$_2$, adjusting $\mu_0$ to reproduce the
experimental $T_{c}$, and calculating the corresponding gaps (dashed
lines in Fig.\ref{fig:gaps_C_Al_doping}(a)). This means that the
decrease in $T_c$, $\Delta_{\sigma}$ and $\Delta_{\pi}$ in these
samples can be completely explained by band filling \cite{kortus05}.
In single crystals, instead, the trend of $\Delta_{\pi}$ can only be
reproduced by increasing the interband scattering rate (solid lines
in Fig.\ref{fig:gaps_C_Al_doping}(a)). This contrasts with the
theoretical prediction \cite{erwin03} that substitutions in the B
plane (for example by carbon) preserving the different parity of
$\sigma$ and $\pi$ bands have little or no effect on $\Gamma_{\sigma
\pi}$. The key to this puzzle could be the presence of microscopic
defects in C-doped single crystals, acting as scattering centers,
suggested by the doing-induced increase in flux pinning and in the
normalized resistance \cite{kazakov05}. The nature of these defects
and the reason why they should be able to create interband
scattering is however not completely clear \cite{gonnelli05}. It is
worth mentioning that the analysis of the zero-bias DOS as a
function of the magnetic field in C-doped polycrystals, carried out
in ref.\cite{szabo07} by using the fitting model developed by
Bugoslavsky \cite{bugoslavsky05}, clearly proves that the ratio
$D_{\sigma}/D_{\pi}$ increases from 0.2 (pure MgB$_2$) towards 1 on
increasing the doping content, indicating that C doping
(surprisingly) increases the intraband $\pi$ scattering more than
the $\sigma$-band one.

%%%%%%%%%%%%%%%%%% Alluminio %%%%%%%%%%%%%%%%%%%%%%%%%%
Doping in the Mg site has been obtained with different chemical
species: Al, Li, Mn, Fe. The first two are heterovalent and result
in electron and hole doping, respectively. According to theoretical
predictions, Al should give the maximum increase in interband
scattering (for 2\% of Al a value of $\Gamma_{\sigma \pi}=1.1$ meV
is predicted, which already has measurable effects on the critical
temperature and on the gaps \cite{erwin03}). On the contrary, Li
should have little or no effect on $\Gamma_{\sigma \pi}$
\cite{erwin03}. PCAR measurements in Mg$_{1-x}$Al$_x$B$_2$
polycrystals \cite{putti05,szabo07} and crystals \cite{klein06} up
to $x=0.2$, carried out with either the conventional or the ``soft''
technique, showed an almost linear decrease of $\Delta_{\sigma}$ and
$\Delta_{\pi}$ as a function of the Al content, in agreement with
the findings of specific-heat measurements \cite{putti05}. No clear
tendency of the gaps to merge was observed; an extension to higher
doping ($x=0.32$) was later obtained by us in single crystals
\cite{daghero08}. All these results are reported, as a function of
the critical temperature, in Fig.\ref{fig:gaps_C_Al_doping}(b). As
in the case of carbon doping, all the data sets agree on the
behavior of $\Delta_{\sigma}$, which is also directly related to the
suppression of the critical field by Al doping \cite{putti05}.
Interestingly, this relationship implies the validity of a
clean-limit description of the system, as it follows from the
analysis of the critical field \cite{putti05,szabo07} but also from
the analysis of the zero-bias DOS as a function of the magnetic
field (carried out in ref.\cite{szabo07} by using Bugoslavsky's
fitting model \cite{bugoslavsky05}) which shows that the diffusivity
ratio of pure MgB$_2$ is preserved on Al doping at least up to
$x=0.20$. The persistence of two gaps even at the highest doping
content is not clear in the spectra; the single-band and two-band
BTK fit are also very similar to each other although a statistical
test (the Fisher F test) clearly indicates that the latter is
preferable for any level of confidence. In any case, the
magnetic-field dependence of the conductance curves shows the
outward shift of the conductance maxima on increasing the field
\cite{szabo07,daghero08} that always indicates the presence of two
gaps of different amplitude -- even though, as explained in
Ref.\cite{daghero08}, the suppression of $H_{c2}$ by Al doping
prevents the separation of the partial $\sigma$ and $\pi$ band
contributions to the conductance, as we instead did in pure MgB$_2$.

The dependence of $\Delta_{\sigma}$ on the critical temperature
shown in Fig.\ref{fig:gaps_C_Al_doping}(b) can be superimposed to
that observed in C-doped MgB$_2$ \cite{szabo07,daghero08} shown in
Fig.\ref{fig:gaps_C_Al_doping}(a). The trend of the small gap
$\Delta_{\pi}$ is similar, but not identical, to that  observed in
C-doped MgB$_2$ polycrystals (see
Fig.\ref{fig:gaps_C_Al_doping}(a)). Actually, a small tendency to an
increase in $\Delta_{\pi}$ at low doping content (with a maximum
around $T_c^A = 30$ K) is very clear, outside the experimental
uncertainty, in our data on single crystals (solid circles). Cooley
\emph{et al.} \cite{cooley05} noticed the same trend only in samples
produced via a long reaction at high temperature so as to reduce the
strain and the inhomogeneity in the Al content. This would indicate
that the enhancement in $\Delta_{\pi}$ is intrinsic to Al doping but
is often masked by other effects, and could explain why it is barely
detectable in PCAR results by the Slovak group
\cite{szabo07,klein06} (triangles) as well as in the results of
specific-heat measurements in some polycrystals samples
\cite{putti05,putti03}. In any case, this trend cannot be reproduced
within the two-band Eliashberg model if only the proper variation in
the DOS \cite{profeta03} and in the phonon frequencies due to Al
doping \cite{daghero09a} are taken into account (dashed lines in
Fig.\ref{fig:gaps_C_Al_doping}(b)). Although the effect on the DOS
is certainly dominant \cite{kortus05}, an increase in interband
scattering at low Al contents in quantitative agreement with
expectations \cite{erwin03} is also necessary to catch the
experimental trend of $\Delta_{\pi}$ in our single crystals.
However, in order to fit the data, $\Gamma_{\sigma \pi}$ must again
decrease for $x> 0.1$. The reason for this is not completely clear
but might be related to extrinsic effects like inhomogeneity and
lattice stress (not taken into account in the Eliashberg model) that
start playing a role for $x>0.1$, as suggested in
Ref.\cite{zambano05,cooley05} and by the increase in the width of
the superconducting transition for $x>0.1$ \cite{daghero08}.

\begin{figure}[ht]
\begin{center}
%\vspace{-6mm}
\includegraphics[width=0.8\columnwidth]{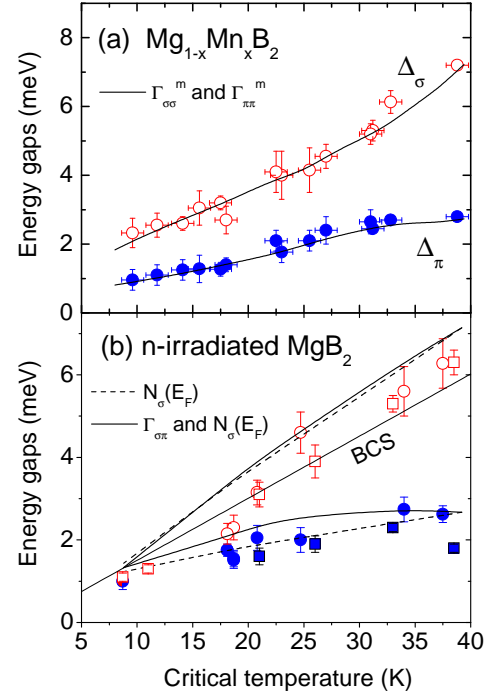}
\end{center}
\caption{(a) Energy gaps measured by PCAR in Mg$_{1-x}$Mn$_x$B$_2$
single crystals \cite{gonnelli06a} as a function of $T_c^A$. Lines
indicate the gap trend calculated within the Eliashberg theory by
assuming that Mn doping only gives rise to an increase in the
$\sigma $ intraband magnetic scattering ($\Gamma_{\sigma
\sigma}^m$), with a smaller contribution from the $\pi-\pi$ channel
($\Gamma_{\pi \pi}^m$). (b) Energy gaps measured by PCAR (circles
\cite{daghero06c}) and specific heat (squares \cite{putti06}) in
neutron-irradiated Mg$^{11}$B$_2$ polycrystals. Lines indicate the
trend of the gaps calculated within the two-band Eliashberg theory
by assuming that: i) the only effect of irradiation is a decrease in
the $\sigma$-band DOS at the Fermi level $N_{\sigma}(0)$ (dash
lines), and ii) this effect is accompanied by an increase in
interband scattering $\Gamma_{\sigma \pi}$ (solid lines). The
straight line represents the BCS gap vs. $T_c$
curve.}\label{fig:gaps_Mn_irradiated}
\end{figure}

%%%%%%%%%%%%%%% Manganese %%%%%%%%%%%%%%%%%%%%%%%%%%

Mn doping is peculiar for two reasons: i) Mn is homovalent with Mg;
ii) its magnetic moment gives rise to spin-flip, pair-breaking
scattering that is considerably larger than the non-magnetic one:
indeed, $T_c$ is very fast suppressed by small Mn contents (see
Fig.\ref{fig:Tc_doping}). One can thus expect the DOS to be
unaffected and the interband non-magnetic scattering to play little
role in this compound. PCAR measurements with the soft technique
were carried out in single crystals \cite{gonnelli06a} of
Mg$_{1-x}$Mn$_x$B$_2$ with $x$ up to 0.015 \cite{rogacki06}. The
trend of the gaps as a function the $T_c^A$ is shown in
Fig.\ref{fig:gaps_Mn_irradiated}(a). For $x>0.004$ (i.e. for $T_c^A
\lesssim 33$ K) the persistence of two gaps was not evident in the
conductance curves and had to be proved by using a magnetic field.
Unfortunately, Mn doping also suppresses $H_{c2}$ so that, when $T_c
< 17$ K, this procedure becomes unreliable and the conclusion that
two gaps persist down to the lowest $T_c^A$ can be based on: i) the
better quality of the two-band fit \cite{gonnelli06a}, and ii) the
fact that the presence of a single gap in the low-$T_c^A$ region
would imply a sudden change in the slope of the $\Delta_{\sigma}$
and $\Delta_{\pi}$ vs. $T_c^A$ curves \cite{gonnelli06a} that is not
justified by any observed discontinuity in the physical properties
of the compound \cite{rogacki06}. The trend of $\Delta_{\sigma}$ and
$\Delta_{\pi}$ vs $T_c^A$ in Mn-doped MgB$_2$ is surprisingly
similar to that observed in Al-doped samples, apart from the
low-doping enhancement of $\Delta_{\pi}$. By the way,
$\Delta_{\sigma}$ follows the universal scaling law with $T_c$.
Unlike in previous cases, the analysis of the data within the
two-band Eliashberg model can here give precise information on the
magnetic scattering rates, either interband ($\Gamma_{ij}^M$) or
intraband ($\Gamma_{ii}^M$). The gap trend can indeed be reproduced
very well by using the same phonon spectra, DOS values, and Coulomb
pseudopotential as in pure MgB$_2$, neglecting all non-magnetic
scattering rates, and taking $\Gamma_{\sigma \sigma}^M$,
$\Gamma_{\pi \pi}^M$ and $\Gamma_{\sigma \pi}^M$ as the only
adjustable parameters \cite{gonnelli06a}. The fit of the gaps vs.
$T_c$ indicates a dominant intraband spin-flip scattering in the
$\sigma$ band, $\Gamma_{\sigma \sigma}^M$, with possible smaller
contributions from either the $\pi$-intraband $\Gamma_{\pi \pi}^M$
or the interband $\Gamma_{\sigma \pi}^M$ channels. A large
$\sigma-\sigma$ scattering was predicted theoretically as being due
to the hybridization of the $\sigma$ bands of MgB$_2$ with the $d$
orbitals of Mn \cite{joseph07}. The dominance of this term on the
$\pi-\pi$ or $\sigma-\pi$ channels was instead demonstrated by
first-principle calculations of the electronic structure of MgB$_2$
near a Mn impurity \cite{gonnelli06a}.

\subsubsection{Effects of irradiation on the gaps of
MgB$_2$}
The effects of intentional introduction of disorder in MgB$_2$ by
means of neutron irradiation have been recently discussed in a
review \cite{ferrando07}. Here we will just quickly mention the
results of PCAR measurements in neutron-irradiated Mg$^{11}$B$_2$
polycrystals \cite{daghero06c}. As explained in \cite{tarantini06},
the use of isotopically enriched $^{11}$B, was necessary to ensure a
homogeneous distribution of defects in the bulk and avoid
self-shielding effects. Neutron flux densities up to 1.6 $\times$
10$^{13}$ cm$^{-2}$s$^{-1}$) were used, which suppressed the bulk
$T_c$ down to 8.7 K.  The defect distribution is very homogeneous,
as shown by X-ray diffraction and by the small width (0.9 K at most)
of the superconducting transition \cite{tarantini06}. The PCAR
measurements were performed with the ``soft'' technique
\cite{daghero06c}. The severe shortening of the electronic mean free
path \cite{tarantini06} made fulfilling the conditions for ballistic
conduction be more and more difficult. In the most irradiated
sample, even the contact with the highest normal-state resistance
(40 $\Omega$) turned out to be in the diffusive regime and showed
the typical dips at $V > V_{peak}$ \cite{sheet04}, as well as a
moderate heating, which was shown to be negligible as long as the
voltage drop at the junction was of the order of $V_{peak}$
\cite{daghero06c}. The trend of the gaps as a function of $T_c^A$ is
shown in Fig.\ref{fig:gaps_Mn_irradiated}(b) (circles).  In the
region of $T_c^A$ around 18-19 K, the results of the two-band BTK
fit are shown even though a single-band fit (with
$\Delta\approx\Delta_\pi$) is possible as well. The
$\Delta_\sigma(T_c^A)$ and $\Delta_\pi(T_c^A)$ curves clearly
indicate a transition from two-band to single-band superconductivity
below 20 K, in excellent agreement with the findings of
specific-heat measurements \cite{putti06} in the same samples
(squares).  The initial small increase in $\Delta_\pi$ suggests that
neutron irradiation increases interband scattering. However, this is
not the only effect since a decrease in the $\sigma$-band DOS
(indeed observed experimentally \cite{gerashenko02}) is necessary as
well to approximately reproduce the overall trend of the gaps within
the two-band Eliashberg model (solid line in
Fig.\ref{fig:gaps_Mn_irradiated}b). Actually, the DOS decrease is
dominant and can, alone, qualitatively explain the experimental data
(dashed lines) and the inclusion of interband scattering only
improves the agreement in the high-$T_c$ region. The fit in the
low-$T_c$ region (below 20 K) is poor but cannot be improved since
here both gaps are smaller than the BCS value and this is forbidden
within the 2-band Eliashberg theory (although often observed in
disordered superconductors \cite{putti08}).

\subsection{Point-contact spectroscopy in novel Fe-based
superconductors}\label{sect:FeAs}

At the beginning of 2008, a new class of Fe-based superconductors
with unexpectedly high critical temperatures -- with a record $T_c$
(up to now) of 57 K -- was discovered . These materials are the
first real term of comparison for cuprates and thus provide a unique
opportunity to test the generality of the theories for high-$T_c$
superconductivity and to identify more clearly the conditions for
its occurrence. Many compounds of this class have been (and are
being) discovered and studied; in the following we will only refer
to the most widely studied families of iron-arsenide
superconductors. The so called ``1111'' family includes the
compounds REFeAsO (RE= rare earth) that become superconducting upon
doping in the O site, with max $T_c$=55 K as well as the recently
discovered oxygen-free Ca$_{1-x}$RE$_x$FeAsF which shows the record
$T_c$ = 57 K. The ``122'' family has general formula AFe$_2$As$_2$
(A=Ba, Sr, Ca, Eu) and, upon doping in the A site, develops $T_c$ up
to 38 K. The state of the research on Iron-Pnictide up to May 2009
is (partially) summarized in ref.\cite{physicaCFeAs}.

Like cuprates, these materials are layered, with alternating RE-O
and Fe-As layers, the latter apparently playing the key role for the
occurrence of superconductivity. Bandstructure calculations
\cite{singh08} and ARPES measurements \cite{ding08} showed that the
Fermi surface is quasi-2D, and is generally made up of two or three
hole-like sheets around the $\Gamma$ point of the first Brillouin
zone, and two electron-like cylinders at the M point. This
immediately suggests, in analogy with MgB$_2$, the possibility of
multigap superconductivity and a dependence of the tunneling or PCAR
spectra on the direction of current injection. Experimental
indications of multigap superconductivity came very soon from
measurements of the critical field and NQR in LaFeAsOF
\cite{hunte08,kohama08,kawasaki08}, from direct ARPES measurements
in Ba$_{0.6}$K$_{0.4}$FeAs \cite{ding08}, from NMR in
PrFeAsO$_{0.89}$F$_{0.11}$ \cite{matano08} and so on
\cite{physicaCFeAs}.

The main issue that PCAR spectroscopy has been asked to address is
the determination of the number, the amplitude and the symmetry of
the order parameter(s). This information is crucial for the
development of theoretical models and to test the existing ones.

The first PCAR measurements, carried out with the conventional
needle-anvil technique, seemed to support a nodal symmetry of the
order parameter, because of the systematic observation of a
zero-bias conductance peak (ZBCP). This happened in
LaFeAsO$_{0.9}$F$_{0.1}$, where Shan \emph{et al.} \cite{shan08}
only observed spectra either featureless or with a large ZBCP and
much smaller additional features. By fitting the conductance curves
with the 2D BTK model \cite{kashiwaya96} (see Sects.\ref{sect:2DBTK}
and \ref{sect:anisotropicDelta}), they obtained a $d$-wave gap
$\Delta=3.9 \pm 0.7$ meV, corresponding to
$2\Delta/k_{B}T_{c}=4.11$. Also Wang \emph{et al.} \cite{wang08}
often observed a ZBCP in polycrystalline SmFeAsO$_{0.9}$F$_{0.1}$
and interpreted it as a signature of nodal gap. All their spectra
admitted a fit with the $d$-wave 2D BTK model, including the few
spectra with no ZBCP.

The interpretation of the ZBCP as an indication of nodal symmetry
was however soon questioned. Yates \emph{et al.} \cite{yates08a}
performed PCAR in oxygen-deficient NdFeAsO$_{0.85}$ with $T_c=45.5$
K and observed that a ZBCP (always vanishing at $T_c^A$) develops on
increasing the pressure applied by the tip (and thus on decreasing
$Z$ and $R_N$). This made the authors warn that the ZBCP may be an
artifact, not related to the gap symmetry. A temperature dependence
of the ZBCP incompatible with a $d$-wave symmetry was found by the
same authors also in TbFeAsO$_{0.9}$F$_{0.1}$ with $T_c=50$ K, where
the presence or absence of the ZBCP was found to even depend on the
sample region probed by PCAR \cite{yates09}. Samuely \emph{et al.}
\cite{samuely09a} observed, in NdFeAsO$_{0.9}$F$_{0.1}$, a large
predominance of low-temperature spectra \emph{without} ZBCP.
However, the ZBCP was found to emerge at a temperature $T^*\propto
1/Z$, to grow in amplitude until it overwhelms the gap structures
(as in \cite{yates08a}), and finally to disappear at $T_c^A$. A ZBCP
very robust against the magnetic field was found in low-resistance
junctions by Chen \emph{et al.} \cite{chen08b}  who performed PCAR
measurements in polycrystalline SmFeAsO$_{0.85}$F$_{0.15}$ with bulk
$T_c=42$ K. In contrast with ref. \cite{shan08}, Chen \emph{et al.}
also found that the ZBCP is very little affected by magnetic fields
up to 9T -- which excludes its relationship with a $d$-wave symmetry
of the gap.

The actual nature of this peak is not completely clear. Kondo
scattering coming from the magnetic impurities in or near the
barrier \cite{shan08}, or to the magnetic moment of Nd or Tb
\cite{samuely09a,yates08a}, can be excluded. The same holds for
intergrain Josephson coupling \cite{chen08a}. Owing to the decrease
in $R_N$ that accompanies the increase in tip pressure, one could
hypothesize that the ZBCP is due to the contact not being in the
ballistic regime so that critical current effects \cite{sheet04} or
heating in the contact region occurs. It is true that, as shown in
Ref.\cite{samuely09a}, the ZBCP coexists with clear gap features at
finite energy \cite{yates08a,wang08,samuely09a}, but this does not
necessarily exclude the possibility of parallel ballistic and
thermal contacts. What instead seems to exclude this picture is that
soft point-contact measurements carried out in
LaFeAsO$_{1-x}$F$_{x}$ \cite{gonnelli09a,gonnelli09b} and
SmFeAsO$_{0.8}$F$_{0.2}$ \cite{gonnelli09b,daghero09b} never gave
evidence of ZBCP irrespective of the contact resistance, which
ranged between a few $\Omega$ and more than 250 $\Omega$
\cite{gonnelli09b}. Therefore the ZBCP is probably related to the
pressure rather than to the contact resistance, and this points
towards its relationship with local lattice deformations.

\begin{figure}[hb]
\begin{center}
%\vspace{-6mm}
\includegraphics[width=0.9\columnwidth]{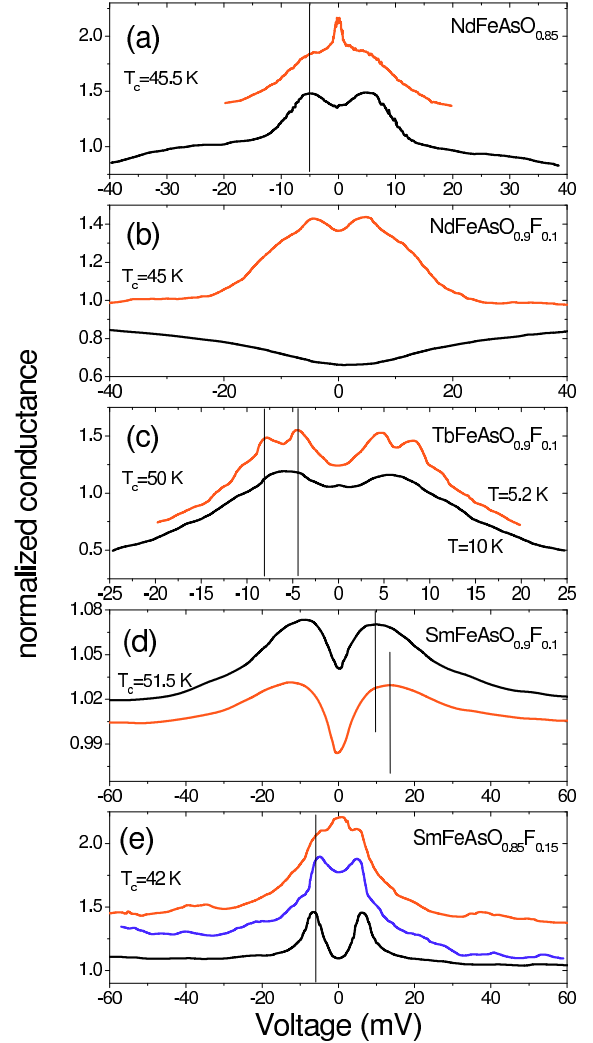}
\end{center}
\caption{A few examples of normalized low-temperature PCAR spectra
measured in various Fe-As compounds of the 1111 family (indicated in
the labels) by means of the conventional needle-anvil technique. The
data are taken from Refs. \cite{yates08a} (a), \cite{samuely09a}
(b), \cite{yates09} (c), \cite{wang08} (d), \cite{chen08b} (e). Some
spectra are vertically offset for clarity. In (c), the bottom curve
is measured in a different region of the sample.
}\label{fig:FeAs_spectra}
\end{figure}
Apart from the zero-bias anomaly, the spectra measured by Yates
\emph{et al.} in NdFeAsO$_{0.85}$ show conductance maxima in the
low-pressure ZBCP-free spectra (bottom curve in
Fig.\ref{fig:FeAs_spectra}(a)) which evolve smoothly into shoulders
in the high-pressure ones (top curve in
Fig.\ref{fig:FeAs_spectra}(a)). The position of these features is
robust against pressure and, if taken as an indicator of the gap
amplitude, it gives $\Delta=7$ meV corresponding to $2\Delta/k_B T_c
=3.57$. A BTK fit (in $s$ wave-symmetry) of the bottom spectrum in
Fig.\ref{fig:FeAs_spectra}(a) gives $\Delta\simeq 6 $ meV. In
F-doped Nd-1111 (NdFeAsO$_{0.9}$F$_{0.1}$) \cite{samuely09a}, the
low-temperature spectra are sometimes featureless (bottom curve in
Fig.\ref{fig:FeAs_spectra}(b)) but more often display conductance
peaks at $V=\pm 5-7 $ mV \emph{and} shoulders at $V=10 $ mV (top
curve in Fig.\ref{fig:FeAs_spectra}(b)), which makes them strikingly
resemble the PCAR spectra in MgB$_2$. Indeed, they were rather well
fitted by a two-band BTK model where $G=\alpha G_1+ (1-\alpha)G_2$
being G the normalized junction conductance, and $\alpha$ the
(unknown) weight of the contribution of band $1$. The best-fitting
values of the gaps are $\Delta_1 \simeq 4-6$ meV and $\Delta_2
\simeq 9-13$ meV that, being $T_c=45$ K, imply gap ratios of 2.6 and
5.7, respectively. Multigap features were clearly observed in
Tb-1111 \cite{yates09}, in the form of peaks at about $4.5$ and $8$
meV (see top curve in Fig.\ref{fig:FeAs_spectra}(c)). The two-band
BTK fit of the spectra gave, at low temperature, $\Delta_1 =5-6$ meV
and $\Delta_2=8-9$ meV. The PCAR measurements carried out by Wang
\emph{et al.} in polycrystalline SmFeAsO$_{0.9}$F$_{0.1}$ gave two
distinct families of spectra, with conductance maxima in different
positions. Two of them, with no ZBCP, are shown in
Fig.\ref{fig:FeAs_spectra}(d). Their fit with a single-band $d$-wave
model gave $\Delta_1=10.5 \pm 0.5$ meV and $\Delta_2=3.7 \pm 0.4$
meV \cite{wang08}.  Similar gap values were given by the two-band
$d$-wave fit of the only spectrum with clear two-gap structures (and
a ZBCP). However, the spectra shown in Fig.\ref{fig:FeAs_spectra}(d)
also admit a two-band BTK fit in $s$-wave. Although the very small
amplitude of the signal requires huge values of the broadening
parameters, the two $s$-wave gaps $\Delta_1\simeq 5$ meV and
$\Delta_2=14 -18$ meV can be obtained, which are much higher than
those given by the single-band $d$-wave fit.

The high-resistance spectra measured by Chen \emph{et al.} in
Sm-1111, showing no ZBCP up to $T_c^A$, can be fitted in the whole
temperature range with a $s$-wave BTK model. The raw conductance
curves reported in Refs.\cite{chen08b,chen09} feature a
characteristic left-right asymmetry -- also observed in Nd-1111 by
Samuely \emph{et al.} \cite{samuely09a} and by us in La-1111
\cite{gonnelli09a} and Sm-1111 \cite{daghero09b} -- probably
intrinsic to these materials, as it was in the case of cuprates
\cite{deutscher05}. The normalized spectra (of which three examples
are reported in Fig.\ref{fig:FeAs_spectra}(e)) have very high
Andreev signal, always show clear conductance peaks at $V\simeq \pm
7$ meV related to a superconducting gap, and often also display
additional structures at higher energy that were seen by the authors
as extrinsic features and thus disregarded in the fitting procedure.
The single-band BTK fit done in \cite{chen08b} clearly reproduces
only the two peaks at about 7 meV and, in the best cases, the
zero-bias minimum between them. The resulting gap is very robust
against the contact resistance and turns out to be almost perfectly
BCS, i.e. $\Delta=6.67 \pm 0.15$ meV at low $T$, corresponding to
$2\Delta/k_B T_c=3.68$.

The existence of a single BCS gap in iron pnictides is actually
surprising since, at the present state of knowledge, there seems to
be no possible weak-coupling mechanism able to justify the high
$T_c$ of these compounds. The electron-phonon coupling is very weak
\cite{boeri08} while the coupling mechanisms mediated by spin
fluctuations \cite{mazin08} proposed for these materials require a
strong interband coupling, and are rather unlikely to give the same
BCS gap on all the sheets of the Fermi surface. On the other hand,
many results, not only from PCAR, speak in favor of a multiband
picture. The same spectra measured by Chen \emph{et al.} may also
give indications of multiple gaps, if the additional shoulders at
$V>V_{peak}$ are interpreted as the hallmarks of a larger
superconducting gap. This interpretation is questioned by Chen
\emph{et al.} because the position and the amplitude of the
additional features are contact-dependent \cite{chen09}. This
argument holds in conventional superconductors but, for example, it
does not in MgB$_2$ (where the observability of the
$\Delta_{\sigma}$ peaks actually depends on the current direction)
or in cuprates (where a large spatial inhomogeneity in the gap
values has been observed by STM \cite{lang02}).

\begin{figure}[ht]
\begin{center}
%\vspace{-6mm}
\includegraphics[width=0.9\columnwidth]{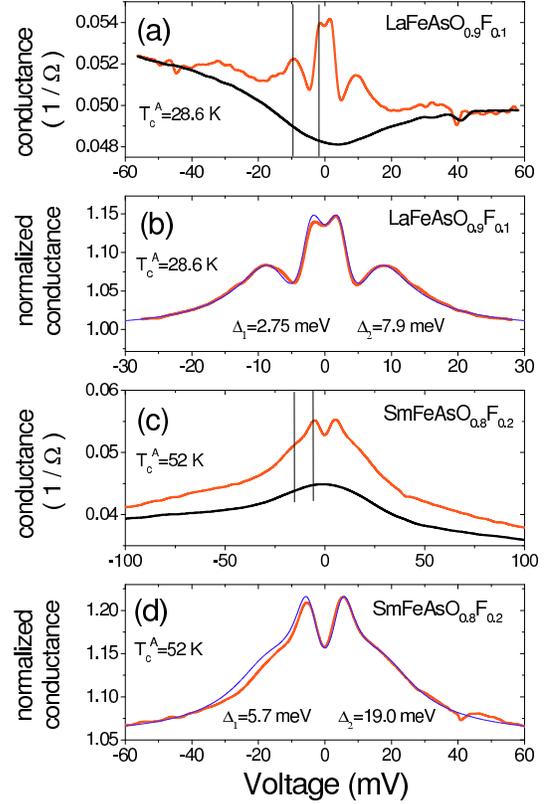}
\end{center}
\caption{(a) The raw conductance curve in a soft point contact on
La-1111 measured at 4.3 K (red curve) and in the normal state at
$T_c$ (black curve). (b) The normalized low-temperature curve (red
thick line) and the relevant 2-band generalized BTK fit (thin blue
line). (c,d)  Same as in (a) and (b) but for Sm-1111. Data are taken
from Refs. \cite{gonnelli09a} and
\cite{daghero09b}.}\label{fig:La_Sm_nostro}
\end{figure}

The differential conductance curves measured in ``soft'' point
contacts on SmFeAsO$_{1-x}$F$_{x}$ polycrystals synthesized at high
pressure with $x=0.20$ ($T_c=52$ K) and $x=0.09$ ($T_c = 42$ K)
\cite{daghero09b} generally look very similar to those by Chen
\emph{et al.} and always present two clear conductance peaks and
additional shoulders that we chose not to disregard in the fit.
Similar structures, even more marked, were observed by soft PCAR in
LaFeAsO$_{1-x}$F$_x$ polycrystals with bulk $T_c=27$ K. The
normalization of the spectra prior to fitting is complicated by the
fact that: i) the normal-state spectrum at low $T$ is not accessible
due to the high $H_{c2}$ \cite{hunte08} ii) the normal state at
$T_c$ is not flat but shows a zero-bias hump in Sm-1111 and a
pseudogaplike feature in La-1111, both progressively washed out on
increasing temperature until, around the N\'{e}el temperature of the
parent compound (about 140-150 K), the conductance becomes flat (but
retains its right-left asymmetry). This suggests that the normal
state conductance may change with temperature also below $T_c$. In
the case of Sm-1111, we chose to divide all curves at $T\leq T_c$ by
the normal state at $T_c$. In La-1111, we tried different
normalizations \cite{gonnelli09a} and showed that their choice has a
small effect on the small gap (less than 2\%), while it can change
the larger one by about 10\% though preserving its trend as a
function of temperature, magnetic field and critical temperature.
Fig. \ref{fig:La_Sm_nostro} reports an example of raw conductance
curves at low temperature (4.3 K) and at $T_c^A$ in La-1111 (a) and
Sm-1111 (c). The low-temperature normalized curves are shown,
together with the relevant two-band BTK fit, in panels (b) and (d).
The resulting values of the nodeless gaps are reported in the
labels. At least in Sm-1111, the normalized curves still feature an
asymmetry that might be due to the normalization (i.e. the asymmetry
of the normal state might depend on temperature) or might be an
intrinsic feature of these compounds, as it was for cuprates.
Certainly, it increases the uncertainty on the gaps ($\Delta_2$ in
particular) since the model can only fit either side of the
experimental spectrum (in this case, the right-hand one).

\begin{figure}[ht]
\begin{center}
%\vspace{-6mm}
\includegraphics[width=0.8\columnwidth]{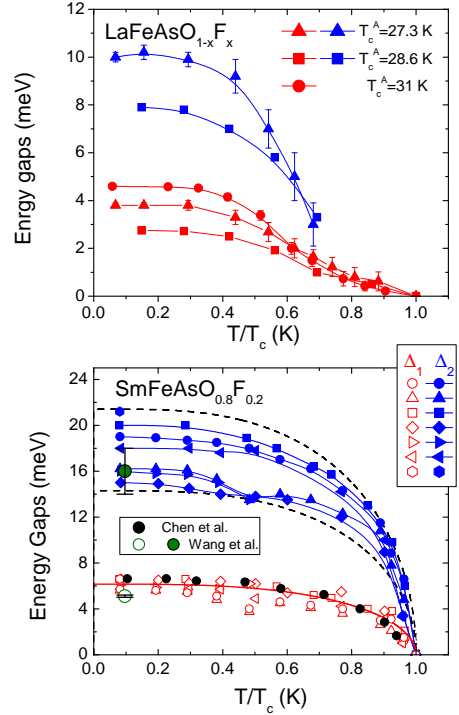}
\end{center}
\caption{Upper panel: Energy gaps in La-1111 as a function of
normalized temperature. The three sets of curves refer to different
contacts with different $T_c^A$ (indicated in the legend)
\cite{gonnelli09a}. Lower panel: The gaps in Sm-1111 as a function
of the normalized temperature \cite{daghero09b}. The data taken from
Ref.\cite{chen08a} (black circles) are also included, as well as the
result of the two-band $s$-wave fit of the low-temperature curves by
Wang \emph{et al.} \cite{wang08} shown in Fig.
\ref{fig:FeAs_spectra}(c) (big green circles).
}\label{fig:La_Sm_gaps}
\end{figure}

The temperature dependence of the gaps in La-1111 and Sm-1111
\cite{gonnelli09a,gonnelli09b,daghero09b} is reported in
Fig.\ref{fig:La_Sm_gaps}(a) and (b). In La-1111 (a), three curves
are shown, obtained in contacts with different $T_c^A$ and thus
possibly made in regions with different local doping. Note that
$T_c^A=31$K corresponds to the very beginning of the resistive
transition. At low temperature, the large gap $\Delta_2$ decreases
on increasing $T_c^A$ and apparently disappears at $T_c^A=31$ K,
while the small gap $\Delta_1$ increases. This finding could be
compatible with recent predictions of a doping dependence of the
gaps \cite{mazin09,benfatto08} although any conclusion in this sense
is definitely premature. The anomalous $T$ dependence of the gaps in
La-1111, where $\Delta_2$ seems to close at a $T^*<T_c^A$ above
which $\Delta_1$ shows a ``tail'', could be due to the shortening of
the mean free path on increasing temperature, so that the junction
ceases to be ballistic at a voltage that decreases with temperature.
This would mean that the superconducting features are progressively
weakened starting from the high-bias ones. An indication in this
sense is given by the high-temperature curves that are more peaked
at zero bias than expected, but if this was the case an apparent
decrease in $T_c^A$ with respect to $T_c$ could be expected, which
is instead not observed.

Fig.\ref{fig:La_Sm_gaps}(b) shows the gaps values extracted from the
fit of various soft-PCAR curves in Sm-1111
\cite{gonnelli09b,daghero09b}. Here the situation is clearer:
$\Delta_1$ almost coincides with that determined by Chen \emph{et
al.} \cite{chen08a}  (although in our case $T_c^A=51-53$ K while in
\cite{chen08a} $T_c=42$ K); its values are well reproducible and
follow a BCS-like temperature dependence up to $T_c^A$. The trend of
$\Delta_2$ vs. $T$ in each set of data is compatible with a BCS-like
curve, but the absolute values are scattered within a region bounded
by two BCS-like curves with $2\Delta_2/k_B T_c=7-9$. This spread
could both due to the residual asymmetry of the normalized
conductance curves and to the uncertainty introduced by the
normalization. The possibility has also been explored theoretically
that \emph{two} large gaps exist, of similar amplitude and thus
virtually indistinguishable by PCAR \cite{benfatto08,ummarino09}.
The gaps obtained by fitting Wang's curves in
Fig.\ref{fig:FeAs_spectra}(c) (green circles) with a two-band
$s$-wave 2D BTK model turn out to be in agreement with the other
results.

In summary, the interpretation of PCAR spectra in 1111 compounds
within a multigap picture gives reasonable results, with a certain
degree of universality for the different compounds: i) the
low-temperature small gap is a little smaller than BCS in amplitude:
$2\Delta_1/k_BT_c$ ranges between 2.23 and 3.44 in La-1111
\cite{gonnelli09a}, between 2.54 and 2.95 in Sm-1111
\cite{daghero09b} and is around 2.1 in Tb-1111 \cite{yates09}; ii)
the large gap is larger or much larger than BCS with
$2\Delta_2/k_BT_c$ equal to about 4 in Tb-1111, ranging from 6.42 to
8.68 in La-1111 (when $\Delta_2$ is detectable) and from 6.7 to
almost 9 in Sm-1111. Such high values are confirmed, among others,
by ARPES measurements in Nd-1111 \cite{kondo08} and by infrared
ellipsometry in Sm-1111 \cite{dubroka09}. They indicate that a
non-conventional paring mechanism is taking place -- and indeed they
can be obtained within the Eliashberg theory \cite{ummarino09} by
supposing a spin-fluctuation-mediated pairing mechanism related to
the nesting of the Fermi surface \cite{mazin08,benfatto08} and which
should give rise to the so-called $s\pm$ superconductivity, with
nodeless order parameters of opposite sign on the electron-like and
hole-like sheets of the Fermi surface. Unfortunately, PCAR in
polycrystals cannot give indications about this expected $\pi$-phase
change. A definite answer to the single gap vs. multigap debate
could come from ARPES, while phase-sensitive techniques are needed
to establish whether the symmetry is really $s\pm$, but probably
these developments will need the growth of large enough single
crystals. It must be said that preliminary measurements in
$ab$-plane contacts on Sm-1111 single crystals \cite{karpinski09}
have perfectly confirmed the results mentioned here
\cite{daghero09b}, supporting the existence of two nodeless gaps.
\begin{figure}[ht]
\begin{center}
%\vspace{-6mm}
\includegraphics[width=0.9\columnwidth]{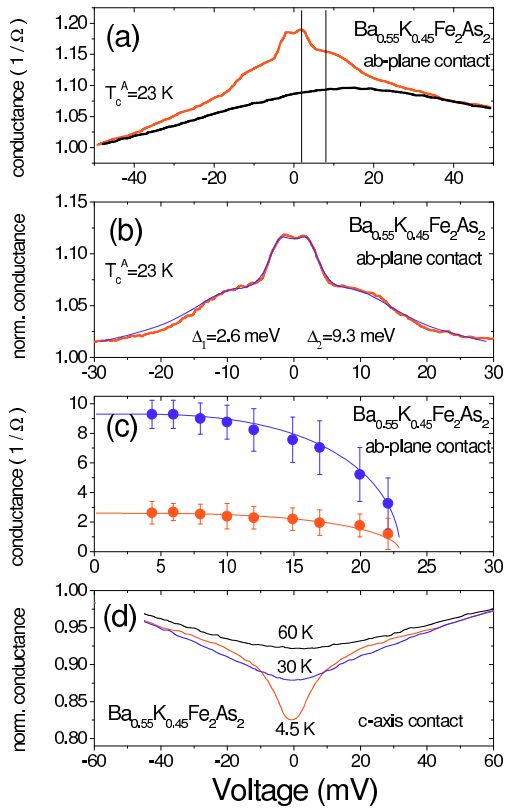}
\end{center}
\caption{(a) An example of spectra at low temperature (red line) and
at $T_c^A$ (black line) measured in $ab$-plane contacts in
hole-doped Ba-122. (b) The normalized curve (red line) and he
relevant two-gap BTK model (blue line). (c) Temperature dependence
of the gaps. (d) Typical spectra measured in $c$ axis contacts, at
low temperature (4.5 K), just above $T_c$ (30 K) and well above
$T_c$ (60 K). Al the data are adapted from Refs.\cite{szabo09} and
\cite{samuely09b}. }\label{fig:122}
\end{figure}

The 122 compounds have been soon grown in the form of large
crystals, with markedly layered structure and easily cleavable.
ARPES measurements \cite{ding08} in (Ba,K)Fe$_2$As$_2$ gave
unambiguous evidences of multiple nodeless gaps ($\Delta_2 \simeq
12$ meV on the two small holelike and electron-like FS sheets, and
$\Delta_1 \simeq 6$ meV on the large hole-like FS). Directional PCAR
measurements with a Pt tip were performed in hole-doped
Ba$_{0.55}$K$_{0.45}$Fe$_2$As$_2$ single crystals by the Slovak
group \cite{szabo09,samuely09b}. The majority of $ab$-plane point
contacts showed a broadened Andreev-reflection feature at zero bias
(but no ZBCP) and a pseudogaplike feature in the normal state that
persists well above $T_c^A$ and is very similar to that observed in
La-1111 \cite{gonnelli09a}. As in La-1111, it is not clear whether
the contemporaneous observation of superconducting signatures and a
pseudogaplike feature means that they spatially coexist or, instead,
they belong to spatially separated superconducting and
antiferromagnetic phases \cite{park09}.

In some contacts, clear double-gap structures were observed, i.e.
symmetric peaks at $\pm 2-4$ meV and additional shoulders at higher
energy (about 10 meV) as in Fig.\ref{fig:122}(a). In these contacts,
the normal-state spectrum at $T_c^A$ showed a broad hump at zero
bias (black curve in Fig.\ref{fig:122}(a)). The conductance curves
were then normalized to the normal-state spectrum and fitted with
success to a two-band BTK model, as shown in Fig.\ref{fig:122}(b).
The low-temperature gaps obtained in several contacts are
$\Delta_1=2.5 - 4$ meV and $\Delta_2=9-10$ meV, corresponding to gap
ratios 2$\Delta_1 /k_BT_c =2.5 -4$ and 2$\Delta_1 /k_BT_c=9-10$.
Both $\Delta_1$ and $\Delta_2$ were found to follow rather well a
BCS-like trend and to close at the same $T_c^A$ (see
Fig.\ref{fig:122}(c)). The values of these gaps differ from those
observed by ARPES in $\rm{Ba_{0.6}K_{0.4}Fe_2As_2}$ \cite{ding08}
$\Delta_1 \simeq 6$ meV and $\Delta_2 \simeq 12$ meV but this
difference is partly justified by the different $T_c$ of the samples
($T_c=37$ K in Ref.\cite{ding08}, $T_c^A=23$ K in
Ref.\cite{szabo09}).

No trace of gap features were instead observed in $c$-axis contacts
(Fig.\ref{fig:122}(d)), but only a V-shaped conductance
progressively flattening on increasing temperature, with no apparent
signature of the $T_c$ crossing. The filling effect cannot be simply
explained by the thermal broadening of the spectra and continues up
to about $70-80$ K, the temperature at which the magnetic transition
in the system takes place \cite{ni08}.

This marked anisotropy of the spectra is interesting and it is
tempting to associate it with an anisotropy of the FS. However, the
122 compounds seem to have a nearly-3D FS
\cite{vilmercati09,mazin09} so that the complete absence of gap
signature along the $c$ direction is difficult to explain; moreover,
as the authors discuss, the inability to observe the gaps in $c$
axis contacts might be as well related to surface contamination or
reconstruction.

PCAR measurements were also carried out in the electron-doped system
$\rm{Ba(Fe_{0.93}Co_{0.07})_2As_2}$ with bulk $T_c=23$K
\cite{samuely09b}. No multigap features were ever observed in this
material; the spectra allowed instead a fit to a standard BTK model,
although with large broadening, that gave a single isotropic gap of
about 5-6 meV. However, ARPES measurements in the same system
$\rm{Ba(Fe_{0.925}Co_{0.075})_2As_2}$ ($T_c=25.5$ K) gave evidence
of two nodeless gaps $\Delta_1=4.5$ meV and $\Delta_2=6.7$ meV
\cite{terashima09}. Evidences of multiple gaps have also been
observed in recent (preliminary) soft PCAR measurements in the same
material.

\subsection{PCS and PCAR in borocarbides}\label{sect:borocarbides}
Since their discovery in 1994, borocarbides with formula
$RT_2\rm{B_2C_2}$ ($R$=rare earth, $T$=transition metal, usually Ni)
have been the subject of an intensive study. Their crystal structure
resembles that of high-$T_c$ cuprates but their electronic
properties are rather those of 3D metals.  Point contact
measurements have been performed in these materials both to
determine the electron-boson spectral function (which allows
clarifying the nature of the superconducting coupling) and to
investigate the symmetry of the superconducting state and/or the
possibility of  multiband superconductivity.

%YNiBC
Among the non-magnetic quaternary borocarbides, $\rm{YNi_2B_2C}$ is
probably the most widely studied. Initially claimed to be a $s$-wave
superconductor, it has later been found to display a large
anisotropy in the superconducting state but basically isotropic
normal-state properties, which suggests an anisotropic order
parameter. The proposed symmetries were $s+g$ (or anisotropic
$s$-wave), with point nodes in the $ab$ plane (and precisely along
the [100] and [010] directions) and the $d$-wave symmetry, with
nodes both in the $ab$ plane (in the [100] and [010] direction) and
along the $c$ axis (where instead the $s+g$ gap has a finite value).
However, already in 1998, well before the discovery of MgB$_2$,
Shulga \emph{et al.} \cite{shulga98} proposed a two-band model to
explain the temperature dependence of the critical fields measured
in YNi$_2$B$_2$C and LuNi$_2$B$_2$C. As we will see in the
following, the debate about the structure and the amplitude of the
OP in these compounds is far from being settled, but the multiband
scenario is definitely the most supported by the many experimental
results.

\begin{figure}[hb]
\begin{center}
%\vspace{-6mm}
\includegraphics[width=0.8\columnwidth]{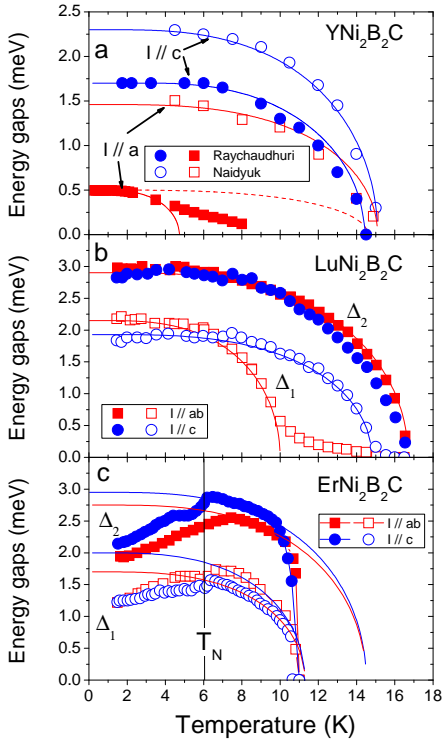}
\end{center}
\caption{(a) Temperature dependence of the gaps in $\rm{YNi_2B_2C}$
single crystals along different directions, measured by  Naidyuk
\emph{et al.} \cite{naidyuk07a} (open symbols) and by Raychaudhuri
\emph{et al.} \cite{raychaudhuri04} (solid symbols). The
measurements with $I\parallel a$ down to 300 mK are unpublished (by
courtesy of P. Raychaudhuri). Solid lines are BCS-like temperature
dependencies. (b) Temperature dependence of the gaps in
$\rm{LuNi_2B_2C}$ single crystals determined from the two-band fit
of the PCAR conductance curves by Naidyuk \emph{et al.}
\cite{naidyuk07b}. (c) Gaps in $\rm{ErNi_2B_2C}$ given by the
two-band fit of the PCAR spectra along the $ab$ plane (squares) and
along the $c$ axis (circles) taken from
Ref.\cite{bobrov08}.}\label{fig:borocarbides}
\end{figure}

Low-$T$ PCS measurements in the normal state of YNi$_2$B$_2$C,
obtained by suppressing superconductivity with a magnetic field
\cite{yanson97,naidyuk07a} identified a well-resolved maximum in the
second derivative $d^2V/dI^2$ at 12 meV that corresponds to a soft
phonon mode contributing to about 90\% of the total electron-phonon
coupling. Other phonon peaks at 20, 24 and 32 meV were not observed.
This indicates a superconducting coupling mainly mediated by soft
phonons, which usually makes an unconventional gap symmetry rather
unlikely. Gap measurements in the superconducting state were carried
out by PCAR in single crystals
\cite{raychaudhuri04,mukhopadhyay05,naidyuk07a} and $c$-axis
oriented films \cite{bashlakov05}. In single crystals with
$T_c=14.6$ K Raychaudhuri \emph{et al.} performed directional PCAR
measurements by injecting the current either along the $a$ or the
$c$ axis. Their conductance curves always admitted a single-band BTK
fit that however gave clearly different gaps, i.e. a small
$\Delta_{I\parallel a}=0.37 - 0.49$ meV and a much larger
$\Delta_{I\parallel c}=1.8-2.2$ meV. The ratio $\Gamma/\Delta$ was
found to be larger for $I\parallel a$, possibly indicating (see
Sect.\ref{sect:broadening}) a greater angular variation of the gap
for $ab$-plane contacts, as expected for a $s+g$ symmetry
\cite{raychaudhuri04}. The temperature dependence of the large gap,
which closes at $T_c^A=T_c=14.6$ K, was found to slightly deviate
from a BCS-like curve (solid circles in
Fig.\ref{fig:borocarbides}(a)), which in principle is compatible
with a gap with nodes. However, the small gap $\Delta_{I\parallel
a}$ was found to be fast suppressed on heating, falling below a
BCS-like curve with $T_c=14.6$ K (dashed line in
Fig.\ref{fig:borocarbides}(a)) and to become undetectable above 8 K.
The recent measurements of $\Delta_{I\parallel a}$ down to 300 mK
carried out by the same group and shown in
Fig.\ref{fig:borocarbides}(a) (solid squares) indicate that, for
$T\leq 3.5$ K, this gap follows very well a BCS-like curve with
$T_c=4.75$. As remarked by the authors \cite{raychaudhuri04}, this
temperature dependence is inconsistent with a $s+g$ symmetry or with
any gap function of the form $\Delta(k)=\Delta_0 f(k)$, and is
indeed more compatible with a picture in which the two gaps open on
different, weakly coupled bands \cite{suhl59} (see
Fig.\ref{fig:Suhl_Matthias}). The magnetic-field dependence gives
similar results \cite{mukhopadhyay05}, i.e. the small gap
$\Delta_{I\parallel ab}$ is fast suppressed by the magnetic field
and ``disappears'' well below $H_{c2}^{ab}$. This is very similar to
what happens in MgB$_2$ when the diffusivities in the two bands are
different \cite{koshelev03}. Also the zero-bias DOS, calculated by
using $N(E)=\Re[(E+i\Gamma)/\sqrt{(E+i\Gamma)^2-\Delta^2}]$ and
taking $E=0$, increases with field \cite{mukhopadhyay05} in a way
similar to that predicted by the two-band model for dirty
superconductors \cite{koshelev03}. If this is the case, the fact
that the small gap almost only contributes to the conductance for
$I\parallel a$ suggests that it opens on the nearly-cylindrical,
fast-electron Fermi surface sheets that have the maximum
cross-section (and dominate the conductance) for $I\parallel a$ but
play almost no role for $I\parallel c$ \cite{mukhopadhyay05}. The
multiband picture is also strongly supported by the effect of
nonmagnetic (Pt) doping on the critical temperature and the upper
critical field \cite{mukhopadhyay09}. The decrease of both $T_c$ and
$H_{c2}$ (the latter for either orientations of the field,
$\mathbf{H}\parallel a$ and $\mathbf {H}\parallel c$) and their
subsequent saturation on increasing doping was indeed shown to be
explainable, within a two-band picture, as being due to an increase
in interband scattering as in doped MgB$_2$ \cite{mukhopadhyay09}.
Going back to PCAR measurements, further support to the multiband
picture also came from the magnetic-field dependence of the excess
current $I_{exc}$ obtained by integrating the normalized PCAR
spectra in YNi$_2$B$_2$C $c$-oriented films \cite{bashlakov05} and
single crystals \cite{naidyuk07a,naidyuk07b}. As in MgB$_2$ (see
Fig.\ref{fig:DOS_field_MgB2}(d)) $I_{exc}(B)$ shows a positive
(although small) curvature. In these measurements, however, the
small gap measured by Raychaudhuri \emph{et al.} was never seen. The
conductance curves were found to display only one peak (at either
positive or negative bias) and were thus fitted to a single-band BTK
model giving a distribution of gap values. In single crystals with
$T_c=15.4$ K Naidyuk \emph{et al.} \cite{naidyuk07a} found  for the
$a$ axis $\Delta_{\rm{[100]}}=1.5 - 1.7 $ meV, for the $c$ axis
$\Delta_{\rm{[001]}}=1.8-2.5$ meV, and for the [110] direction
$\Delta_{\rm{[110]}}=1.0 - 2.5$ meV. A representative temperature
dependence of the gaps for $I\parallel a$ and $I\parallel b$ is
shown in Fig.\ref{fig:borocarbides}(a) as open symbols.

%LuNiBC

In LuNi$_2$B$_2$C single crystals, PCAR measurements carried out by
Bobrov \emph{et al.} \cite{bobrov05} gave spectra with a single
conductance peak that were however shown to be poorly fitted by a
single-band BTK model, and a little better, but still
unsatisfactorily, by a two-band one. Two different continuous
distributions of gap values were thus used to reproduce the shape of
the low-temperature spectra, with $\Delta_{I\parallel ab}$ ranging
between 1 meV to 3.35 meV (with maxima at 2 meV and 3.1 meV) and
$\Delta_{I\parallel c}$ ranging from 0.7 to 4 meV (with maxima at
1.8-2 meV and 2.5 meV). In \cite{bobrov06,naidyuk07b}, a more
conventional two-gap approach was used to fit the data in
\emph{either} direction, although the spectra always featured a
single conductance peak -- possibly because the two gaps are too
close to each other to be clearly distinguishable. As shown in
Fig.\ref{fig:borocarbides}(b), a qualitatively similar (but
quantitatively different) gap trend as a function of temperature was
observed for $I\parallel ab$ (squares) and $I\parallel c$ (circles).
The results seem to indicate the existence of two bands with a weak,
anisotropic interband coupling \cite{suhl59}. In a more recent PCAR
study in LuNi$_2$B$_2$C single crystals \cite{lu08} the spectra were
fitted to a single-band BTK model and gave almost equal gaps in the
[001] ($c$ axis, $\Delta_{[001]}\simeq 2.4$ meV) and [110]
directions ($\Delta_{[110]}\simeq 2.6$ meV), which clearly excludes
a $s+g$ symmetry. Both these gaps approximately follow a BCS-like
temperature dependence with $2\Delta/k_BT_c$ equal to 3.4 and 3.6,
respectively -- although an upward deviation of the experimental
points was observed at low temperature, as in films
Ref.\cite{bobrov05}. These results may be compatible with those by
Naidyuk \emph{et al.} \cite{naidyuk07b} in the $ab$ plane, since the
single-band fit can reasonably give an ``average'' gap with respect
to the two-band fit and is probably only weakly sensitive to the
faster depression of the small gap. Along the $a$ axis, a rather
wide distribution of gap values $\Delta_{[100]}=1.6-2.5$ meV was
observed, all disappearing at the bulk $T_c$ -- surprisingly similar
to the findings by Bashlakov \emph{et al.} \cite{bashlakov05} in
$\rm{YNi_2B_2C}$ films.

%%%%% HoNiBC %%%%%%%%%%%%%%%%%%%%%%%%%

\begin{figure}[ht]
\begin{center}
%\vspace{-6mm}
\includegraphics[width=0.8\columnwidth]{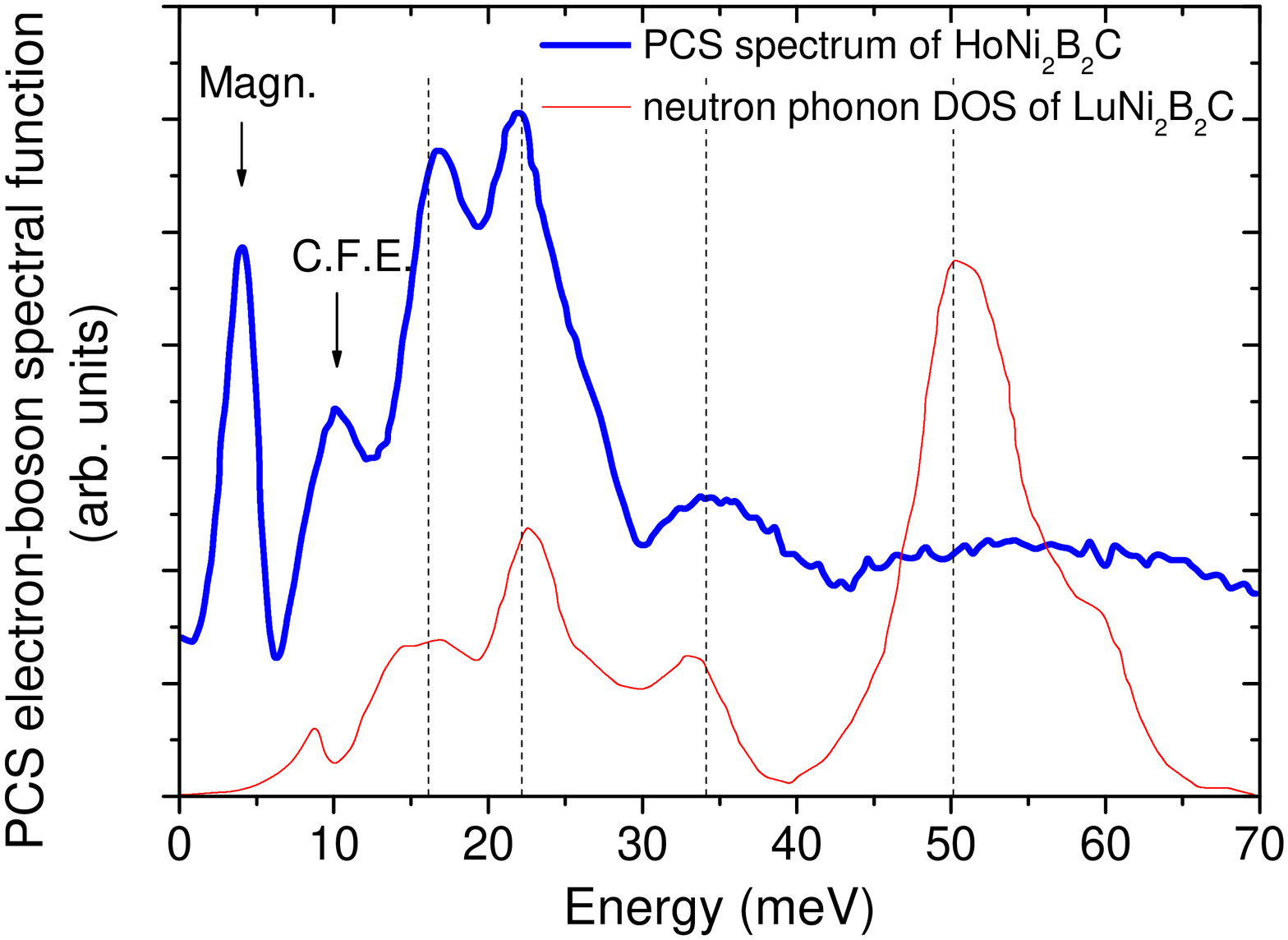}
\end{center}
\caption{Point-contact electron-boson spectral function $\alpha^2_PC
F(eV)$ of HoNi$_2$B$_2$C (thick line) compared to the neutron phonon
DOS of the isostructural compound LuNi$_2$B$_2$C (thin line). The
correspondence of the structures at 3, 10, 16, 22 and 50 mV is
clear. The additional peaks due to magnetic and crystal-field
excitations are also indicated (from
Ref.\cite{naidyuk07c}.)}\label{fig:HoNiBC_phonons}
\end{figure}

Various other compounds of the family were studied by PCS in the
normal state and PCAR, especially by the Ukrainian group, and it
would be impossible to account here for all their results. Let us
just briefly mention for its interest the magnetic compound
$\rm{HoNi_2B_2C}$ with a low-temperature commensurate AFM state
whose N\'{e}el temperature $T_N=5.3$ K is smaller than $T_c=8.5$ K.
PCS measurements of $\alpha_{PC}^2F(\omega)$ allowed identifying
structures at 16, 22, 34 and 50 meV corresponding to peaks in the
phonon DOS of isostructural LuNi$_2$B$_2$C, plus a peak at 3 meV
related to the magnetic ordered state and indeed disappearing at
$T_N$, and a peak at 10 meV possibly due to a coupling of carriers
to crystal-field excitations \cite{yanson97,naidyuk07c} (see
Fig.\ref{fig:HoNiBC_phonons}). PCAR spectra in the superconducting
state first of all showed a negligible dependence on the current
direction and admitted single-band BTK fit; the low-temperature gap
is $\Delta_0=0.95$ meV, and decreases on increasing temperature
following a BCS-like curve but disappears at $T_c^*=5.6$ K, well
below $T_c$. Between $T_c^*$ and $T_c$, an unconventional gapless
state was proposed, possibly due to the peculiar spiral magnetic
order in this temperature range. Below $T_N$, the superconducting
state coexists with an antiferromagnetic order. To explain this
phenomenon, a separation of the two phases in the $k$ space has been
proposed. In particular, superconductivity below $T_N$ should
survive only on a single (isotropic) sheet of the FS with no
contributions from the Ho $5d$ states. This picture has been
directly put in connection with the results of critical field and
anisotropy measurements as a function of temperature in the same
compound \cite{muller07}.

%%%%% ErNiBC %%%%%%%%%%%%%%%%%%%%%%%%%
More recently, PCAR measurements have been performed in the
$\rm{ErNi_2B_2C}$ compound \cite{bobrov08}, with $T_c=11$ K and a
low-temperature incommensurate antiferromagnetic order with spin
density wave between 2 K and $T_N\simeq 6$ K. To account for the
magnetic pair-breaking effect, a suitable model by Beloborod'ko
\cite{beloborodko03} was used to fit the spectra. The model contains
as adjustable parameters $\Delta$ (order parameter, OP), $Z$
(barrier height) and $\gamma$ (magnetic scattering rate). The gap
$\Delta_0$ is related to the OP by the formula
$\Delta_0=\Delta(1-\gamma^{2/3})^{3/2}$ \cite{beloborodko03}. The
single-band fit of the PCAR spectra as a function of temperature
gave $\Delta(T=2\rm{K})\simeq 1.8$ meV for both $I\parallel ab$ and
$I\parallel c$. A two-band fit was also carried out because of the
claimed unsatisfactory quality of the single-band one. This fit
gives two OPs $\Delta_1\approx 2$ meV and $\Delta_2\approx 1.2$ meV
whose temperature dependence is shown in
Fig.\ref{fig:borocarbides}(c). The anisotropy in this case is small,
but a very unconventional behavior is observed because of the AFM
order below $T_N$. The observation of one or two OPs at low
temperature clearly proves the coexistence of superconductivity and
AFM order, also observed by laser photoemission.  On heating,
$\Delta_1$ and $\Delta_2$ first increase (possibly because of the
weakening of the AFM order), which is consistent with previous
findings by tunnel and laser photoemission spectroscopy as well as
with the prediction of some theories of coexistence of
superconductivity and antiferromagnetic state \cite{chi92}. In the
paramagnetic state above 6 K, the OPs follow BCS-like curves (whose
extrapolation would give $T_{c1}'\simeq 11.3$ K and $T_{c2}'\simeq
14.5$ K) and finally abruptly disappear at $T_c=11$ K. The
extrapolated $T_c'$ values give for both OPs a ratio
$2\Delta/k_BT_c$ in the strong-coupling regime. To fit the curves,
also the weight of the larger OP $\Delta_2$ had to be varied in a
non-monotonic way, although it generally decreases on heating. The
magnetic scattering rates also evolve non-trivially as a function of
temperature, and $\gamma_2$ is always greater than $\gamma_1$. This
results were interpreted as indicating that: i) different bands are
differently affected by magnetic order; ii) the part of the FS that
develops the larger OP $\Delta_2$ tends to diminish on approaching
$T_c$; iii) there is a FS separation with distinct superconducting
and magnetic bands (or FS sheets); a more detailed analysis of the
results in Ref.\cite{bobrov08} indicates that approximately half of
the FS is nonsuperconducting.

It follows from the above that the result of PCAR measurements in
borocarbides carried out by different groups are often in
disagreement with one another. Moreover, the coexistence of magnetic
orders of some kind and superconductivity in some of this compounds
prevents any tentative description of this class of compounds as a
whole. Nevertheless, various hints strongly suggest multiband
superconductivity, with: i) a conventional phonon-mediated
superconducting coupling (at least in non-magnetic ones); ii) weak
interband coupling (in $\rm{YNi_2B_2C}$
\cite{raychaudhuri04,mukhopadhyay05} and in $\rm{LuNi_2B_2C}$
\cite{bobrov06,naidyuk07b}) with different gaps and $T_c$; iii) a
generally anisotropic distribution of gap values over the FS, with a
larger gap along the $c$ axis; iv) a separation of superconducting
and magnetic order parameters on different sheets of the FS  (in
$\rm{HoNi_2B_2C}$ \cite{naidyuk07c} and $\rm{ErNi_2B_2C}$
\cite{bobrov08}). The research in this field is still going on and
new measurements will certainly help clarifying this complex
situation.

\begin{figure}[ht]
\begin{center}
%\vspace{-6mm}
\includegraphics[width=0.9\columnwidth]{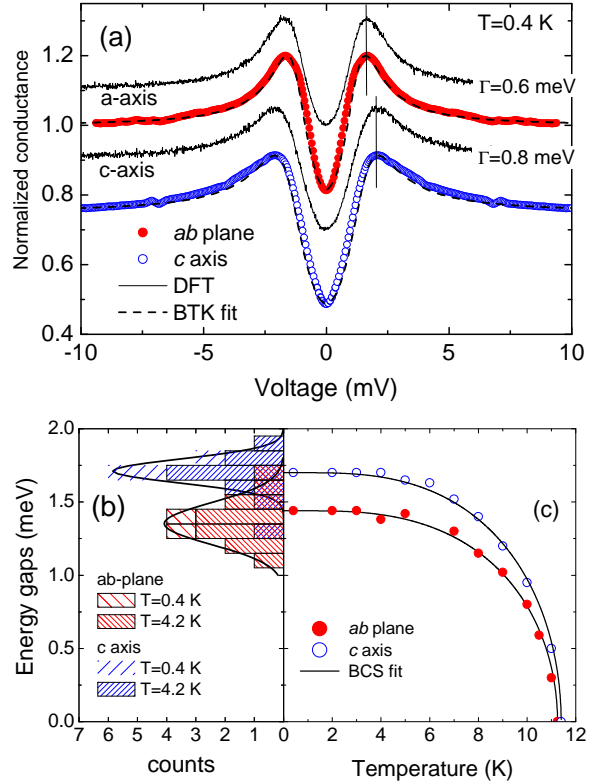}
\end{center}
\caption{(a) Low-temperature experimental PCAR spectra (circles) in
CaC$_6$ for $ab$-plane (red) and $c$-axis (blue) current injection,
superimposed to the relevant single-band BTK fit (dashed lines).
Black solid lines are the theoretical conductance curves at $T=0$
calculated from the full $k$ distribution of gap values and Fermi
velocities, and smeared with the indicated (experimental) $\Gamma$
values. (b) Distribution of measured gap values and relevant
gaussian fit. (c) Gaps vs. $T$ extracted from the fit of the
temperature dependence of the curves in (a). }\label{fig:CaC6}
\end{figure}
\subsection{PCAR in Graphite-intercalation compounds}
\label{sect:CaC6} The recent discovery of the new superconducting
Graphite Intercalation Compounds (GICs) CaC$_6$, YbC$_6$
\cite{weller05}, and SrC$_6$ \cite{kim07} has renewed the interest
for this long-known class of compounds. CaC$_6$ shows the highest
$T_c=11.5$ K among this class of compounds. Its lattice is made up
of alternating graphite layers and Ca planes, with a rhombohedral
structure, \cite{Emery05}. The similarity with MgB$_2$ is striking
and indeed the electronic band structure \cite{mazin07} includes the
$\sigma$ and $\pi$ bands, though the former are completely filled
and play no role in superconductivity. According to first-principle
calculations \cite{calandra05,kim06} superconductivity arises from
the coupling of both Ca and C phonon modes to the carriers of the
so-called interlayer band, which is formed by C and Ca orbitals
\cite{mazin07}. The Fermi surface \cite{mazin07} consists of
$\pi$-band warped cylinders parallel to the $c$ axis, and of
interlayer-band FS sheets created by the intersection of some of
these cylinders with the nearly-spherical Ca orbitals
\cite{sanna07}. Unlike in MgB$_2$ (see Fig.\ref{fig:MgB2gaps}(a))
the calculated gap \cite{sanna07} is continuously distributed in
energy between 1.1 and 2.3 meV and the mixing of C and Ca states
prevents a true multigap superconductivity. However, the gap mapping
on the FS shows that it changes from one FS sheet to the other and
is, on average, slightly larger in the 3D interlayer band (actually
depending also on the wavevector $\mathbf{k}$ within each sheet).
This suggested us to use directional soft PCAR in CaC$_6$ single
crystals to observe the predicted anisotropy. The extreme
sensitivity of CaC$_6$ to air and moisture required to cleave the
sample and make the contact in inert atmosphere, and to seal the
whole sample holder before transferring it to the cryostat. Due to
the small $T_c=11.5$ K and the smallness of the effect to be
observed, part of the measurements were carried out at 400 mK
\cite{gonnelli08}.

Fig.\ref{fig:CaC6}(a) shows a $c$-axis and a $ab$-plane PCAR
spectrum at $T=0.4$ K (symbols). The position of the conductance
peaks is clearly different and indeed the fit with a single-band 2D
BTK model gives $\Delta_{ab}=1.44$ meV and $\Delta_c=1.7$ meV. The
values of $Z$ are systematically higher in c-axis contacts
($0.74\leq Z\leq 1.01$) than in $ab$-plane ones  ($0.48\leq Z\leq
0.75$), in agreement with the different Fermi velocities in the two
directions ($v_{ab} = 0.54\cdot 10^6$ m/s, $v_{c} = 0.29\cdot 10^6$
m/s). These spectra are in excellent agrement with \emph{ab-initio}
calculations of the Andreev conductance carried out
\cite{gonnelli08} as described in Sect.\ref{sect:trueFS},
eq.\ref{eq:sigmaFS}. To allow a comparison with experiment, the
theoretical curves  at $T=0$ (shown in Fig. \ref{fig:Sanna}) were
smeared with the experimental values of $\Gamma$ (i.e. 0.6 meV for
the $ab$-plane contact, 0.8 meV for the $c$-axis one) neglecting the
(much smaller) thermal smearing. The results are shown in
Fig.\ref{fig:CaC6} as solid black lines. Fig.\ref{fig:CaC6}(b) shows
that the distributions of low-temperature gap values obtained in
several $c$-axis and $ab$-plane contacts are approximately Gaussian
and overlap only slightly. The temperature dependence of the gaps is
shown, for the spectra in Fig.\ref{fig:CaC6}(a), in panel (c). The
gap always follow a BCS-like trend with gap ratio
$2\Delta_{ab}/k_BT_c= 2.98$ and $2\Delta_{c}/k_BT_c= 3.48$. All
these results gave the first direct evidence of gap anisotropy in
CaC$_6$ and showed a very nice example of a successful feedback
between PCAR experiments and theoretical predictions from ab-initio
calculations.

\section{Conclusions}
Multiband superconductivity was theoretically investigated since the
Sixties, but it was often considered as an exotic although
interesting possibility, with little practical relevance, because of
the feebleness of its effects even in the very few cases when they
were detected, or the very small critical temperature of the
materials that displayed them \cite{binnig80}. On the other hand,
point-contact spectroscopy was invented in the mid-Seventies and,
although initially applied to the study of normal metals, it was
soon understood to be a powerful spectroscopic tool in the study of
superconductors. However, other techniques such as STM and ARPES
with better spatial and momentum resolution largely dominated, in
the Nineties, the study of cuprates. With the discovery of MgB$_2$
in 2001, multiband superconductivity suddenly became a promising and
popular field of research, and PCAR spectroscopy rapidly acquired a
great relevance thanks to its quick and successful application to
this compound. The amount of information this technique has been
able to provide, even in samples that were impossible to analyze by
STM and ARPES, was probably a surprise for most of the
superconducting community. Since then, and thanks to its
reliability, simplicity and flexibility, PCAR spectroscopy has
played an important role in the investigation of the superconducting
properties of many new (and less new) compounds.

In this review we have tried to present the recent applications of
PCS and PCAR spectroscopy to the study of multiband superconductors,
starting from MgB$_2$ (either pure or doped) to continue with
borocarbides, graphite-intercalation compounds and the recently
discovered Fe-based superconductors. We have shown that PCAR
measurements can provide information on the number, the amplitude
and the symmetry of the superconducting order parameter(s), but also
-- when integrated with first-principle calculations and with the
Eliashberg theory -- on the coupling strengths (both within and
between bands), the scattering rates, the densities of states, the
$k$ dependence of the gap within the various sheets of the Fermi
surface,  and so on. We have provided the reader with a simple
theoretical introduction to PCS and PCAR spectroscopy, showing how
the limitations of the original, pioneering Blonder-Tinkham-Klapwijk
model (often used even today to fit the point-contact spectra) can
be overcome to improve the degree of approximation to the real case
and to make PCAR a much sharper tool for the investigation of
unconventional superconductors.

\section*{Acknowledgments}
Our warmest thanks to A. Brinkman, L. Boeri, O. Dolgov, A. Plecenik,
P. Raychaudhuri for perusing the manuscript and suggesting
corrections and improvements, and to A. Sanna, S. Massidda and P.
Raychaudhuri for kindly providing original material. Special thanks
to F. Dolcini for helpful discussions on the physics of ballistic
contacts. We are indebted to all the members of our group for
contributing to the research work summarized here, in particular M.
Tortello for experimental research and G.A. Ummarino for theoretical
support. A particular acknowledgment to V. A. Stepanov for his
fundamental contribution to our research in all these years. R.S.G.
wishes to thank R.K. Kremer and the Max Planck Institute f\"{u}r
Festk\"{o}rperforschung (Stuttgart) where this review was partly
written.

\bibliographystyle{iopart-num}
\bibliography{bibliografia}

\providecommand{\newblock}{}
\begin{thebibliography}{100}
\expandafter\ifx\csname url\endcsname\relax
  \def\url#1{{\tt #1}}\fi
\expandafter\ifx\csname urlprefix\endcsname\relax\def\urlprefix{URL }\fi
\providecommand{\eprint}[2][]{\url{#2}}

\bibitem{yanson74}
Yanson I~K 1974 {\em Sov. Phys. JETP\/} {\bf 39} 506--513

\bibitem{duif89}
Duif A~M, Jansen A~G~M and Wyder P 1989 {\em J. Phys.: Condens. Matter\/} {\bf
  1} 3157--3189

\bibitem{naidyuklibro}
Naidyuk Y~G and Yanson I~K 2004 {\em Point-Contact Spectroscopy\/} ({\em
  Springer Series in Solid-State Sciences\/} vol 145) (Springer)

\bibitem{jansen80}
Jansen A~G~M, van Gelder A~P and Wyder P 1980 {\em J. Phys. C: Solid state
  Phys.\/} {\bf 13} 6073

\bibitem{deutscher05}
Deutscher G 2005 {\em Rev. Mod. Phys.\/} {\bf 77} 109

\bibitem{blonder83}
Blonder G and Tinkham M 1983 {\em Phys. Rev. B\/} {\bf 27} 112

\bibitem{srikanth92}
Srikanth H and Raychaudhuri A~K 1992 {\em Physica C\/} {\bf 190} 229

\bibitem{baltz09}
Baltz V, Naylor A~D, Seemann K~M, Elder W, Sheen S, Westerholt K, Zabel H,
  Burnell G, Marrows C~H and Hickey B~J 2009 {\em J. Phys.: Condens. Matter\/}
  {\bf 21} 095701

\bibitem{holm58}
Holm R 1958 {\em Electric Contacts Handbook\/} (Springer-Verlag)

\bibitem{sharvin65}
Sharvin Y~V 1965 {\em Zh. Eksp. Teor. Fiz.\/} {\bf 48} 984 engl. Transl. Sov.
  Phys.-JETP \textbf{21}, 655 (1965)

\bibitem{deJong94}
de~Jong M~J~M 1994 {\em Phys. Rev. B\/} {\bf 49} 7778

\bibitem{wexler66}
Wexler G 1966 {\em Proc. Phys. Soc. London\/} {\bf 89} 927

\bibitem{baranger85}
Baranger H~U, MacDonald A and Leavens C 1985 {\em Phys. Rev. B\/} {\bf 31} 6197

\bibitem{deutscher02}
Deutscher G and Maynard R 2002 {\em The Gap Symmetry and Fluctuations in
  High-T$_c$ Superconductors\/} ({\em NATO Science Series: B\/} vol 371)
  (Kluwer Academic Publishers) chap From the Andreev Reflection to the Sharvin
  Contact Conductance, pp 503--510

\bibitem{Andreev64}
Andreev A 1964 {\em Zh. Eksp. Teor. Fiz.\/} {\bf 46} 1823 engl. Transl. Sov.
  Phys.-JETP \textbf{19}, 1228 (1974)

\bibitem{degennes66}
De~Gennes P 1966 {\em Superconductivity of metals and Alloys\/} (Benjamin, New
  York)

\bibitem{saint-james64}
Saint-James D 1964 {\em J. Phys.\/} {\bf 25} 899

\bibitem{waldram96}
Waldram J~R 1996 {\em Superconductivity of metals and cuprates\/} (Insitute of
  Physics Publishing)

\bibitem{daghero06c}
Daghero D, Calzolari A, Ummarino G~A, Tortello M, Gonnelli R~S, Stepanov V~A,
  Tarantini C, Manfrinetti P and Lehmann E 2006 {\em Phys. Rev. B\/} {\bf 74}
  174519

\bibitem{BTK}
Blonder G~E, Tinkham M and Klapwijk T~M 1982 {\em Phys. Rev. B\/} {\bf 25} 4515

\bibitem{kashiwaya96}
Kashiwaya S, Tanaka Y, Koyanagi M and Kajimura K 1996 {\em Phys. Rev. B\/} {\bf
  53} 2667

\bibitem{dynes78}
Dynes R~C, Narayanamurti V and Garno J~P 1978 {\em Phys. Rev. Lett.\/} {\bf 41}
  1509

\bibitem{plecenik94}
Plecenik A, Grajcar M, Be\v{n}a\v{c}ka v, Seidel P and Pfuch A 1994 {\em Phys.
  Rev. B\/} {\bf 49} 10016

\bibitem{dolgov03}
Dolgov O~V, Gonnelli R~S, Ummarino G~A, Golubov A~A, Shulga S~V and Kortus J
  2003 {\em Phys. Rev. B\/} {\bf 68} 132503

\bibitem{ummarino10}
Ummarino G~A In preparation

\bibitem{yanson04c}
Yanson I~K, Beloborod'ko S~I, Naidyuk Y~G, Dolgov O~V and Golubov A~A 2004 {\em
  Phys. Rev. B\/} {\bf 69} 100501

\bibitem{mazin99}
Mazin I~I 1999 {\em Phys. Rev. Lett.\/} {\bf 83} 1427

\bibitem{brinkman02}
Brinkman A, Golubov A~A, Rogalla H, Dolgov O~V, Kortus J, Kong Y, Jepsen O and
  Andersen O~K 2002 {\em Phys. Rev. B\/} {\bf 65} 180517

\bibitem{golubov09}
Golubov A~A, Brinkman A, Tanaka Y, Mazin I~I and Dolgov O~V 2009 {\em Phys.
  Rev. Lett.\/} {\bf 103} 077003

\bibitem{gonnelli08}
Gonnelli R~S, Daghero D, Delaude D, Tortello M, Ummarino G~A, Stepanov V~A, Kim
  J~S, Kremer R~K, Sanna A, Profeta G and Massidda S 2008 {\em Phys. Rev.
  Lett.\/} {\bf 100} 207004

\bibitem{sanna07}
Sanna A, Profeta G, Floris A, Marini A, Gross E~K~U and Massidda S 2007 {\em
  Phys. Rev. B\/} {\bf 75} 020511(R)

\bibitem{sheet04}
Sheet G, Mukhopadhyay S and Raychaudhuri P 2004 {\em Phys. Rev. B\/} {\bf 69}
  134507

\bibitem{strijkers01}
Strijkers G~J, Ji Y, Yang F~Y, Chien C~L and Byers J~M 2001 {\em Phys. Rev.
  B\/} {\bf 67} 104510

\bibitem{mazin01}
Mazin I~I, Golubov A~A and Nadgorny B 2001 {\em J. Appl. Phys.\/} {\bf 89} 7576

\bibitem{mazin01b}
Mazin I, Golubov A~A and Nadgorny B 2001 {\em J. Appl. Phys.\/} {\bf 90} 3127

\bibitem{chalsani07}
Chalsani P, Upadhyay S~K, Ozatay O and Burham R 2007 {\em Phys. Rev. B\/} {\bf
  75} 094417

\bibitem{woods04}
Woods G~T, Soulen R~J, Mazin I~I, Nadgorny B, Osofsky M~S, Sanders J, Srikanth
  H, Egelhoff W~F and Datla R 2004 {\em Phys. Rev. B\/} {\bf 70} 054416

\bibitem{suhl59}
Suhl H, Matthias B~T and Walker L~R 1959 {\em Phys. Rev. Lett.\/} {\bf 3} 552

\bibitem{hafstrom70}
Hafstrom J~W and MacVicar M~L~A 1970 {\em Phys. Rev. B\/} {\bf 2} 4511

\bibitem{carlson70}
Carlson J~R and Satterthwaite C~B 1970 {\em Phys. Rev. Lett.\/} {\bf 24} 461

\bibitem{binnig80}
Binnig G, Baratoff A, Hoenig H~E and Bednorz J~G 1980 {\em Phys. Rev. Lett.\/}
  {\bf 45} 1352

\bibitem{buzea01}
Buzea C and Yamashita T 2001 {\em Supercond. Sci. Technol.\/} {\bf 14}
  R115--R146

\bibitem{kortus01}
Kortus J, Mazin I~I, Belashchenko K~D, Antropov V~A and Boyer L~L 2001 {\em
  Phys. Rev. Lett.\/} {\bf 86} 4656

\bibitem{kong01}
Kong Y, Dolgov O~V, Jepsen O and Andersen O~K 2001 {\em Phys. Rev. B\/} {\bf
  64} 020501(R)

\bibitem{liu01}
Liu A~Y, Mazin I~I and Kortus J 2001 {\em Phys. Rev. Lett.\/} {\bf 87} 87005

\bibitem{eliashberg60}
Eliashberg G~M 1960 {\em Sov. Phys. JETP\/} {\bf 11} 696

\bibitem{choi03}
Choi H~J, Cohen M~L and Louie S~G 2003 {\em Physica C\/} {\bf 385} 66--74

\bibitem{nicol05}
Nicol E and Carbotte J~P 2005 {\em Phys. Rev. B\/} {\bf 71} 054501

\bibitem{anderson59}
Anderson P~W 1959 {\em J. Phys. Chem. Solids\/} {\bf 11} 26

\bibitem{golubov97}
Golubov A~A and Mazin I~I 1997 {\em Phys. Rev. B\/} {\bf 55} 15146

\bibitem{choi02}
Choi H~J, Roundy D, Sun H, Cohen M~L and Louie S~G 2002 {\em Nature\/} {\bf
  418} 758--760

\bibitem{bugoslavsky02}
Bugoslavsky Y, Miyoshi Y, Perkins G~K, Berenov A~V, Lockman Z,
  MacManus-Driscoll J~L, Cohen L~F, Caplin A~D, Zhai H~Y, Paranthaman M~P, M
  C~H and Blamire M 2002 {\em Supercond. Sci. Technol.\/} {\bf 15} 526

\bibitem{szabo01}
Szab\'o P, Samuely P, Kacmarcik J, Klein T, Marcus J, Fruchart D, Miraglia S,
  Marcenat C and Jansen A~G~M 2001 {\em Phys. Rev. Lett.\/} {\bf 87} 137005

\bibitem{lee02}
Lee S, Khim Z~G, Chong Y, Moon S~H, Lee H~N, Kim H~G, Oh B and Jip~Choi E 2002
  {\em Physica C\/} {\bf 377} 202

\bibitem{gonnelli02c}
Gonnelli R~S, Daghero D, Ummarino G~A, Stepanov V~A, Jun J, Kazakov S~M and
  Karpinski J 2002 {\em Phys. Rev. Lett.\/} {\bf 89} 247004

\bibitem{kohen01}
Kohen A and Deutscher G 2001 {\em Phys. Rev. B\/} {\bf 64} 060506

\bibitem{li02}
Li Z~Z, Tao H~J, Xuan Y, Ren Z~A, Che G~C and Zhao B~R 2002 {\em Phys. Rev.
  B\/} {\bf 66} 064513

\bibitem{schmidt01}
Schmidt H, Zasadzinski J~F, Gray K~E and Hinks D~G 2001 {\em Phys. Rev. B\/}
  {\bf 63} 220504

\bibitem{Laube01}
Laube F, Goll G, Hagel J, L\"{o}hneysen H~V, Ernst D and Wolf T 2001 {\em
  Europhysics Letters\/} {\bf 56} 296

\bibitem{plecenik02}
Plecenik A, Benacka S, Kus P and Grajcar M 2002 {\em Physica C\/} {\bf 368} 251

\bibitem{gonnelli02a}
Gonnelli R~S, Calzolari A, Daghero D, Ummarino G~A, Stepanov V~A, Fino P,
  Giunchi G, Ceresara S and Ripamonti G 2002 {\em J. Phys. Chem. Solids\/} {\bf
  63} 2319

\bibitem{szabo03b}
Szab\'o P, Samuely P, Kacmarcik J, Jansen A~G~M, Klein T, Marcus J and Marcenat
  C 2003 {\em Supercond. Sci. Technol.\/} {\bf 16} 162

\bibitem{sologubenko02}
Sologubenko A, Jun J, Kazakov S~M, Karpinski J and Ott H~R 2002 {\em Phys. Rev.
  B\/} {\bf 66} 014504

\bibitem{welp02}
Welp U, Rydh A, Karapetrov G, Kwok W~K, Crabtree G~W, Marcenat C, Paulius L,
  Klein T, Marcus J, Kim K~H~P, Jung C~U, Lee H~S, Kang B and Lee S~I 2003 {\em
  Phys. Rev. B\/} {\bf 67} 012505

\bibitem{angst02}
Angst M, Puzniak R, Wisniewski A, Jun J, Kazakov S~M, Karpinski J, Roos J and
  Keller H 2002 {\em Phys. Rev. Lett.\/} {\bf 88} 167004

\bibitem{daghero03}
Daghero D, Gonnelli R~S, Ummarino G~A, Stepanov V~A, Jun J, Kazakov S~M and
  Karpinski J 2003 {\em Physica C\/} {\bf 385} 255

\bibitem{gonnelli04a}
Gonnelli R~S, Daghero D, Ummarino G~A, Dellarocca V, Calzolari A, Stepanov V~A,
  Jun J, Kazakov S~M and Karpinski J 2004 {\em Physica C\/} {\bf 408-410} 796

\bibitem{naidyuk96}
Naidyuk Y~G, H\"{a}ussler R and L\"{o}hneysen H~v 1996 {\em Physica B\/} {\bf
  218} 122

\bibitem{dahm02}
Dahm T, Graser S, Iniotakis C and Schopohl N 2002 {\em Phys. Rev. B\/} {\bf 66}
  144515

\bibitem{dahm04a}
Dahm T, Graser S and Schopohl N 2004 {\em Physica C\/} {\bf 408-410} 336

\bibitem{koshelev03}
Koshelev A~E and Golubov A~A 2003 {\em Phys. Rev. Lett.\/} {\bf 90} 177002

\bibitem{bugoslavsky04}
Bugoslavsky Y, Miyoshi Y, Perkins G~K, Caplin A~D, Cohen L~F, Pogrebnyakov A~V
  and Xi X~X 2004 {\em Phys. Rev. B\/} {\bf 69} 132508

\bibitem{naidyuk05b}
Naidyuk Y~G, Kvitnitskaya O~E, Yanson I~K, Leeb S and Tajima S 2005 {\em Solid
  State Communications\/} {\bf 133} 363

\bibitem{dahm03}
Dahm T and Schopohl N 2003 {\em Phys. Rev. Lett.\/} {\bf 91} 017001

\bibitem{gurevich03}
Gurevich A 2003 {\em Phys. Rev. B\/} {\bf 67} 184515

\bibitem{Lukic07}
Lukic V and Nicol E~J 2007 {\em Phys. Rev. B\/} {\bf 76} 144508

\bibitem{gonnelli04c}
Gonnelli R~S, Daghero D, Calzolari A, Ummarino G~A, Dellarocca V, Stepanov V~A,
  Jun J, Kazakov S~M and Karpinski J 2004 {\em Phys. Rev. B\/} {\bf 69}
  100504(R)

\bibitem{szabo03a}
Szab\'o P, Samuely P, Kacmarcik J, Klein T, Marcus J and Jansen A~G~M 2003 {\em
  Physica C\/} {\bf 388-389} 145

\bibitem{bugoslavsky05}
Bugoslavsky Y, Miyoshi Y, Perkins G~K, Caplin A~D, Cohen L~F, Pogrebnyakov A~V
  and Xi X~X 2005 {\em Phys. Rev. B\/} {\bf 72} 224506

\bibitem{miyoshi05}
Miyoshi Y, Bugoslavsky Y and Cohen L~F 2005 {\em Phys. Rev. B\/} {\bf 72}
  012502

\bibitem{gonnelli03a}
Gonnelli R~S, Daghero D, Ummarino G~A, Stepanov V~A, Jun J, Kazakov S~M and
  Karpinski J 2003 {\em Supercond. Sci. Technol.\/} {\bf 16} 171

\bibitem{yanson03}
Yanson I~K, Fisun V~V, Bobrov N~L, Naidyuk Y~G, Kang W~N, Choi E~M, Kim H~J and
  Lee S~I 2003 {\em Phys. Rev. B\/} {\bf 67} 024517

\bibitem{golubov02}
Golubov A~A, Kortus J, Dolgov O~V, Jepsen O, Kong Y, Andersen O~K, Gibson B~J,
  Ahn K and Kremer R~K 2002 {\em J. Phys.: condens. matter\/} {\bf 14} 1353

\bibitem{naidyuk03}
Naidyuk Y~G, Yanson I~K, Kvitnitskaya O~E, Lee S and Tajima S 2003 {\em Phys.
  Rev. Lett.\/} {\bf 90} 197001

\bibitem{shukla03}
Shukla A, Calandra M, d'Astuto M, Lazzeri M, Mauri F, Bellin C, Krish M,
  Karpinski J, Kazakov S~M, Jun J, Daghero D and Parlinski K 2003 {\em Phys.
  Rev. Lett.\/} {\bf 90} 095506

\bibitem{yanson04a}
Yanson I~K and Naidyuk Y~G 2004 {\em Low Temp. Phys.\/} {\bf 30} 261

\bibitem{yanson04b}
Yanson I~K and Naidyuk Y~G 2004 {\em Spectroscopy of Emerging Materials\/}
  ({\em NATO Science Series II: Mathematics, Physics and Chemistry\/} vol 165)
  (Kluwer Academic Publisher) chap Point-contact spectroscopy of two-band
  superconductor $\rm{MgB_{2}}$, p 273

\bibitem{cava03}
Cava R~J, Zandbergen H~W and Inumaru K 2003 {\em Physica C\/} {\bf 385} 8--15

\bibitem{gonnelli07}
Gonnelli R~S, Daghero D, Ummarino G~A, Tortello M, Delaude D, Stepanov V~A and
  Karpinski J 2007 {\em Physica C\/} {\bf 456} 134

\bibitem{ribeiro03}
Ribeiro R, Bud'ko S, Petrovic C and Canfield P 2003 {\em Physica C\/} {\bf 384}
  227

\bibitem{klein06}
Klein T, Lyard L, Marcus J, Marcenat C, Szab\'o P, Hol'anov\'a Z, Samuely P,
  Kang B~W, Kim H~J, Lee H~S, Lee H~K and Lee S~I 2006 {\em Phys. Rev. B\/}
  {\bf 73} 224528

\bibitem{zambano05}
Zambano A~J, Moodenbaugh A~R and Cooley L~D 2005 {\em Supercond. Sci.
  Technol.\/} {\bf 18} 1411--1420

\bibitem{putti05}
Putti M, Ferdeghini C, Monni M, Pellecchi I, Tarantini C, Manfrinetti P,
  Palenzona A, Daghero D, Gonnelli R~S and Stepanov V~A 2005 {\em Phys. Rev.
  B\/} {\bf 71} 144505

\bibitem{karpinski07}
Karpinski J, Zhigadlo N~D, Katrych S, Puzniak R, Rogacki K and Gonnelli R 2007
  {\em Physica C\/} {\bf 456} 3

\bibitem{avdeev03}
Avdeev M, Jorgensen J~D, Ribeiro R~A, Bud'ko S~L and Canfield P~C 2003 {\em
  Physica C\/} {\bf 387} 301

\bibitem{kazakov05}
Kazakov S~M, Puzniak R, Rogacki K, Mironov A~V, Zhigadlo N~D, Jun J, Soltmann
  C, Batlogg B and Karpinski J 2005 {\em Phys. Rev. B\/} {\bf 71} 024533

\bibitem{holanova04a}
Holanov\'a Z, Szab\`{o} P, Samuely P, Wilke R~H~T, Bud'ko S~L and Canfield P~C
  2004 {\em Phys. Rev. B\/} {\bf 70} 064520

\bibitem{gonnelli05}
Gonnelli R~S, Daghero D, Calzolari A, Ummarino G~A, Dellarocca V, Stepanov V~A,
  Kazakov S~M, Zhigadlo N and Karpinski J 2005 {\em Phys. Rev. B\/} {\bf 71}
  060503R

\bibitem{samuely03a}
Samuely P, Hol'anov\`{a} Z, Szab\`{o} P, Kacmarcik J, Ribeiro R~A, Bud'ko S~L
  and Canfield P~C 2003 {\em Phys. Rev. B\/} {\bf 68} 020505(R)

\bibitem{holanova04b}
Holanov\'a Z, Szab\'o P, Kacmarcik J, Samuely P, Ribeiro R~A, Bud'Ko S~L and
  Canfield P~C 2004 {\em Physica C\/} {\bf 408-410} 610

\bibitem{szabo07}
Szab\'o P, Samuely P, Pribulov\'{a} Z, Angst M, Bud'ko S, Canfield P~C and
  Marcus J 2007 {\em Phys. Rev. B\/} {\bf 75} 144507

\bibitem{daghero08}
Daghero D, Delaude D, Calzolari A, Tortello M, Ummarino G~A, Gonnelli R~S,
  Stepanov V~A, Zhigadlo N~D, Katrych S and Karpinski J 2008 {\em J. Phys.:
  Condens. Matter\/} {\bf 20} 085225

\bibitem{ummarino05a}
Ummarino G~A, Daghero D, Gonnelli R~S and Moudden A~H 2005 {\em Phys. Rev. B\/}
  {\bf 71} 134511

\bibitem{kortus05}
Kortus J, Dolgov O~V, Kremer R~K and Golubov A~A 2005 {\em Phys. Rev. Lett.\/}
  {\bf 94} 027002

\bibitem{erwin03}
Erwin S~C and Mazin I~I 2003 {\em Phys. Rev. B\/} {\bf 68} 132505

\bibitem{cooley05}
Cooley L~D, Zambano A~J, Moodenbaugh A~R, Klie R~F, Zheng J~C and Zhu Y 2005
  {\em Phys. Rev. Lett.\/} {\bf 95} 267002

\bibitem{putti03}
Putti M, Affronte M, Manfrinetti P and Palenzona A 2003 {\em Phys. Rev. B\/}
  {\bf 68} 094514

\bibitem{profeta03}
Profeta G, Continenza A and Massidda S 2003 {\em Phys. Rev. B\/} {\bf 68}
  144508

\bibitem{daghero09a}
Daghero D~Ummarino G~A, Tortello M, Delaude D, Gonnelli R~S, Stepanov V~A,
  Monni M and Palenzona A 2009 {\em Supercond. Sci. Technol.\/} {\bf 22} 025012

\bibitem{gonnelli06a}
Gonnelli R~S, Daghero D, Ummarino G~A, Calzolari A, Tortello M, Stepanov V~A,
  Zhigadlo N~D, Rogacki K, Karpinski J, Bernardini F and Massidda S 2006 {\em
  Phys. Rev. Lett.\/} {\bf 97} 037001

\bibitem{putti06}
Putti M, Affronte M, Ferdeghini C, Manfrinetti P, Tarantini C and Lehmann E
  2006 {\em Phys. Rev. Lett.\/} {\bf 96} 077003

\bibitem{rogacki06}
Rogacki K, Batlogg B, Karpinski J, Zhigadlo N~D, Schuck G, Kazakov S~M,
  W\"{a}gli P, Pu\'{z}niak R, Wi\'{s}niewski A, Carbone F, Brinkman A and
  van~der Marel D 2006 {\em Phys. Rev. B\/} {\bf 73} 174520

\bibitem{joseph07}
Joseph P~J~T and Singh P~P 2007 {\em Solid State Commun.\/} {\bf 141} 390

\bibitem{ferrando07}
Ferrando V, Affronte M, Daghero D, Di~Capua R, Tarantini C and Putti M 2007
  {\em Physica C\/} {\bf 456} 144

\bibitem{tarantini06}
Tarantini C, Aebersold H~U, Braccini V, Celentano G, Ferdeghini C, Ferrando V,
  Gambardella U, Gatti F, Lehmann E, Manfrinetti P, Marr\'{e} D, Palenzona A,
  Pallecchi I, Sheikin I, Siri A~S and Putti M 2006 {\em Phys. Rev. B\/} {\bf
  73} 134518

\bibitem{gerashenko02}
Gerashenko A~P, Mikhalev K~N, Verkhovskii S~V, Karkin A~E and Goshchitskii B~N
  2002 {\em Phys. Rev. B\/} {\bf 65} 132506

\bibitem{putti08}
Putti M, Vaglio R and Rowell J~M 2008 {\em Supercond. Sci. Technol.\/} {\bf 21}
  043001

\bibitem{physicaCFeAs}
 2009 {\em Physica C. Special Issue: Superconductivity in Iron Pnictides\/}
  {\bf 469}

\bibitem{singh08}
Singh D~J and Du M~H 2008 {\em Phys. Rev. Lett.\/} {\bf 100} 237003

\bibitem{ding08}
Ding H, Richard P, Nakayama K, Sugawara K, Arakane T, Sekiba Y, Takayama A,
  Souma S, Sato T, Takahashi T, Wang Z, Dai X, Fang Z, Chen G~F, Luo J~L and
  Wang N~L 2008 {\em Europhys. Lett.\/} {\bf 83} 47001

\bibitem{hunte08}
Hunte F, Jaroszynski J, Gurevich A, Larbalestier D~C, Jin R, Sefat A~S, McGuire
  M~A, Sales B~C, Christen D~K and Mandrus D 2008 {\em Nature (London)\/} {\bf
  453} 903

\bibitem{kohama08}
Kohama Y, Kamihara Y, Riggs S, Balakirev F~F, Atake T, Jaime M, Hirano M and
  Hosono H 2008 {\em Europhys. Letters\/} {\bf 84} 37005

\bibitem{kawasaki08}
Kawasaki S, Shimada K, Chen G~F, Luo J~L, Wang N~L and Zheng G~q 2008 {\em
  Phys. Rev. B\/} {\bf 78} 220506(R)

\bibitem{matano08}
Matano K, Ren Z~A, Dong X~L, Sun L~L, Zhao Z~X and Zheng G~q 2008 {\em
  Europhys. Lett.\/} {\bf 83} 57001

\bibitem{shan08}
Shan L, Wang Y, Zhu X, Mu G, Fang L, Ren C and Wen H~H 2004 {\em Europhys.
  Lett.\/} {\bf 83} 57004

\bibitem{wang08}
Wang Y~L, Shan L, Fang L, Cheng P, Ren C and Wen H~H 2009 {\em Supercond. Sci.
  Technol.\/} {\bf 22} 015018

\bibitem{yates08a}
Yates K~A, Cohen L~F, Ren Z~A, Yang J, Lu W, Dong X~L and Zhao Z~X 2008 {\em
  Supercond. Sci. Technol.\/} {\bf 21} 092003

\bibitem{yates09}
Yates K~A, Morrison K, Rodgers J~A, Penny G~B~S, Bos J~W~G, Attfield J~P and
  Cohen L 2009 {\em New J. Phys.\/} {\bf 11} 025015

\bibitem{samuely09a}
Samuely P, Szab\'o P, Pribulov\'a Z, Tillman M, Bud'ko S~L and Canfield P~C
  2009 {\em Supercond. Sci. Technol.\/} {\bf 22} 014003

\bibitem{chen08b}
Chen T~Y, Tesanovic Z, Liu R~H, Chen X~H and Chien C~L 2008 {\em Nature
  (London)\/} {\bf 453} 761

\bibitem{chen08a}
Chen X~H, Wu T, Wu G, Liu R~H, Chen H and Fang D~F 2008 {\em Nature (London)\/}
  {\bf 453} 1224

\bibitem{gonnelli09a}
Gonnelli R~S, Daghero D, Tortello M, Ummarino G~A, Stepanov V~A, Kim J~S and
  Kremer R~K 2009 {\em Phys. Rev. B\/} {\bf 79} 184526

\bibitem{gonnelli09b}
Gonnelli R~S, Daghero D, Tortello M, Ummarino G~A, Stepanov V~A, Kim J~S and
  Kremer R~K 2009 {\em Physica C\/} {\bf 469} 512

\bibitem{daghero09b}
Daghero D, Tortello M, Gonnelli R~S, Stepanov V~A, Zhigadlo N~D and Karpinski J
  2009 {\em Phys. Rev. B\/} {\bf 80} 060502(R)

\bibitem{chen09}
Chen T~Y, Huang S~X, Tesanovic Z, Liu R~H, Chen X~H and Chien C~L 2009 {\em
  Physica C\/} {\bf 469} 521

\bibitem{boeri08}
Boeri L, Dolgov O~V and Golubov A~A 2008 {\em Phys. Rev. Lett.\/} {\bf 101}
  026403

\bibitem{mazin08}
Mazin I~I, Singh D~J, Johannes M~D and Du M~H 2008 {\em Phys. Rev. Lett.\/}
  {\bf 101} 057003

\bibitem{lang02}
Lang K~M, Madhavan V, Hoffman J~E, Hudson E~W, Eisaki H, Uchida S and Davis J~C
  2002 {\em Nature (London)\/} {\bf 415} 412

\bibitem{mazin09}
Mazin I~I and Schmalian J 2009 {\em Physica C\/} {\bf 469} 614

\bibitem{benfatto08}
Benfatto L, Capone M, Caprara S, Castellani C and Di~Castro C 2008 {\em Phys.
  Rev. B\/} {\bf 78} 140502(R)

\bibitem{ummarino09}
Ummarino G~A, Tortello M, Daghero D and Gonnelli R~S 2009 {\em Phys. Rev. B\/}
  {\bf 80} in press

\bibitem{kondo08}
Kondo T, Santander-Syro A~F, Copie O, Liu C, Tillman M~E, Mun E~D, Schmalian J,
  Bud'ko S~L, Tanatar M~A, Canfield P~C and Kaminski A 2008 {\em Phys. Rev.
  Lett\/} {\bf 101} 147003

\bibitem{dubroka09}
Dubroka A, Kim K~W, R\"{o}ssle M, Malik V~K, Drew A~J, Liu R~H, Wu G, Chen X~H
  and Bernhard C 2008 {\em Phys. Rev. Lett.\/} {\bf 101} 097011

\bibitem{karpinski09}
Karpinski J, Zhigadlo N, Katrych S, Bukowski Z, Moll P, Weyeneth S, Keller H,
  Puzniak R, Tortello M, Daghero D, Gonnelli R~S, Maggio-Aprile I, Fasano Y,
  Fischer {\O}, Rogacki K and Batlogg B 2009 {\em Physica C\/} {\bf 469} 370

\bibitem{szabo09}
Szab\'o P, Pribulov\'{a} Z, Prist\'{a}\v{s} G, Bud'ko S~L, Canfield P~C and
  Samuely P 2009 {\em Phys. Rev. B\/} {\bf 79} 012503

\bibitem{samuely09b}
Samuely P, Pribulov\'a Z, Szab\'o P, Prist\'{a}\v{s} G, Bud'ko S~L and Canfield
  P~C 2009 {\em Physica C\/} {\bf 469} 507

\bibitem{park09}
Park J~T, Inosov D~S, Niedermayer C, Sun G~L, Haug D, Christensen N~B,
  Dinnebier R, Boris A~V, Drew A~J, Schulz L, Shapoval T, Wolff U, Neu V, Yang
  X, Lin C~T, Keimer B and Hinkov V 2009 {\em Phys. Rev. Lett.\/} {\bf 102}
  117006

\bibitem{ni08}
Ni N, Bud'ko S~L, Kreyssig A, Nandi S, Rustan G~E, Goldman A~I, Gupta S,
  Corbett J~D, Kracher A and Canfield P~C 2008 {\em Phys. Rev. B\/} {\bf 78}
  014507

\bibitem{vilmercati09}
Vilmercati P, Fedorov A, Vobornik I, Manju U, Panaccione G, Goldoni A, Sefat
  A~S, McGuire M~A, Sales B~C, Jin R, Mandrus D, Singh D~J and Mannella N 2009
  {\em Phys. Rev. B\/} {\bf 79} 220503

\bibitem{terashima09}
Terashima K, Bowen J~H, Nakayama K, Sato T, Richard P, Xu Y~M, Li L~J, Cao G~H,
  Xu Z~A, Ding H and Takahashi T 2009 {\em Proc. Natl. Acad. Sci (USA)\/} {\bf
  106} 7330

\bibitem{shulga98}
Shulga S~V, Drechsler S~L, Fuchs G, M\"{u}ller K~H, Winzer K, Heinecke M and
  Krug K 1998 {\em Phys. Rev. Lett.\/} {\bf 80} 1730

\bibitem{naidyuk07a}
Naidyuk Y~G, Bashlakov D~L, Yanson I~K, Fuchs G, Behr G, Souptel D and
  Drechsler S~L 2007 {\em Physica C\/} {\bf 460-462} 103

\bibitem{raychaudhuri04}
Raychaudhuri P, Jaiswal-Nagar D, Sheet G, Ramakrishnan S and Takeya H 2004 {\em
  Phys. rev. Lett.\/} {\bf 93} 156802

\bibitem{naidyuk07b}
Naidyuk Y~G, Bashlakov D~L, Bobrov N~L, Chernobay V~N, Kvitnitskaya O~E, Yanson
  I~K, Behr G, Drechsler S~L, Fuchs G, Souptel D, Naugle D~G, Rathnayaka K~D~D
  and Ross J~H 2007 {\em Physica C\/} {\bf 460-462} 107

\bibitem{bobrov08}
Bobrov N~L, Chernobay V~N, Naidyuk Y~G, Tyutrina L~V, Naugle D~G, Rathnayaka
  K~D~D, Bud'ko S~L, Canfield P~C and Yanson I~K 2008 {\em Europhys. Lett.\/}
  {\bf 83} 37003

\bibitem{yanson97}
Yanson I~K, Fisun V~V, Jansen A~G~M, Wyder P, Canfield P~C, Cho B~K, Tomy C~V
  and Paul D~M 1997 {\em Phys. Rev. Lett.\/} {\bf 78} 935

\bibitem{mukhopadhyay05}
Mukhopadhyay S, Sheet G, Raychaudhuri P and Takeya H 2005 {\em Phys. Rev. B\/}
  {\bf 72} 014545

\bibitem{bashlakov05}
L B~D, Naidyuk Y~G, Yanson I~K, Wimbush S~C, Holzapfel B, Fuchs G and Drechsler
  S~L 2005 {\em Supercond. Sci. Technol.\/} {\bf 18} 1094

\bibitem{mukhopadhyay09}
Mukhopadhyay S, Sheet G, Nigam A~K, Raychaudhuri P and Takeya H 2009 {\em Phys.
  Rev. B\/} {\bf 79} 132505

\bibitem{bobrov05}
Bobrov N~L, Beloborod'ko S~I, Tyutrina L~V, Yanson I~K, Naugle D~G and
  Rathnayaka K~D~D 2005 {\em Phys. Rev. B\/} {\bf 71} 014512

\bibitem{bobrov06}
Bobrov N~L, Beloborod'ko S~I, Tyutrina L~V, Chernobay V~N, Yanson I~K, Naugle
  D~G and Rathnayaka K~D~D 2006 {\em Low Temp. Phys.\/} {\bf 32} 489

\bibitem{lu08}
Lu X, Park W~K, Kim J~D, Yeo S, Lee S~I and Green L~H 2008 {\em Physica B\/}
  {\bf 403} 1098

\bibitem{naidyuk07c}
Naidyuk Y~G, Kvitnitskaya O~E, Yanson I~K, Fuchs G, Nenkov K, W\"{a}lte A, Behr
  G, Souptel D and Drechsler S 2007 {\em Phys. Rev. B\/} {\bf 76} 014520

\bibitem{muller07}
M\"{u}ller K~H, Fuchs G, Drechsler S~L, Opahle I, Eschrig H, Schultz L, Behr G,
  L\"{o}ser W, Souptel D, W\"{a}lte A, Nenkov K, Naidyuk Y and Rosner H 2007
  {\em Physica C\/} {\bf 460-462} 99

\bibitem{beloborodko03}
Beloborod'ko S~I 2003 {\em Low Temp. Phys.\/} {\bf 29} 650

\bibitem{chi92}
Chi H and Nagi A~D~S 1992 {\em J. Low Temp. Phys.\/} {\bf 86} 139

\bibitem{weller05}
Weller T~E, Ellerby M, Saxena S~S, Smith R~P and Skipper N~T 2005 {\em Nature
  Physics\/} {\bf 1} 39

\bibitem{kim07}
Kim J~S, Boeri L, O'Brien J~R, Razavi F~S and Kremer R~K 2007 {\em Phys. Rev.
  Lett.\/} {\bf 99} 027001

\bibitem{Emery05}
Emery N, H\'{e}rold C, d'Astuto M, Garcia V, Bellin C, Mar\^{e}ch\'{e} J,
  Lagrange P and Loupias G 2005 {\em Phys. Rev. Lett.\/} {\bf 95} 087003

\bibitem{mazin07}
Mazin I~I, Boeri L, Dolgov O~V, Golubov A~A, Bachelet G~B, Giantomassi M and
  Andersen O~K 2007 {\em Physica C\/} {\bf 460-462} 116

\bibitem{calandra05}
Calandra M and Mauri F 2005 {\em Phys. Rev. Lett.\/} {\bf 95} 237002

\bibitem{kim06}
Kim J~S, Boeri L, Kremer R~K and Razavi F~S 2006 {\em Phys. Rev. B\/} {\bf 74}
  214513

\end{thebibliography}

\end{document}